%% file: main.tex
\documentclass[aps, rmp, reprint]{revtex4-2}
\newcommand{\thedoctitle}{Photon localization: Local measurements, localized states, and causality}
\usepackage{etoolbox}
\newtoggle{BoxedTheorems}\settoggle{BoxedTheorems}{true}
\newtoggle{HyperrefStyling}\settoggle{HyperrefStyling}{true}
\newtoggle{PreprintStyling}\settoggle{PreprintStyling}{true}
\newtoggle{ExportTikz}\settoggle{ExportTikz}{false}
\newtoggle{FigFromPDF}\settoggle{FigFromPDF}{false}
\usepackage{setup}

\iftoggle{ExportTikz}{\tikzexternalize[prefix=tikz-export/]}{}

\begin{document}

\title{\thedoctitle}

\author{Jan Gulla}
\email{jan.gulla@fys.uio.no}
\affiliation{Department of Physics, University of Oslo, NO-0316 Oslo, Norway\looseness=-2}

\author{Johannes Skaar}
\email{johannes.skaar@fys.uio.no}
\affiliation{Department of Physics, University of Oslo, NO-0316 Oslo, Norway\looseness=-2}

\date{\today}

\begin{abstract}
We review what it means for a relativistic particle to be localized and argue that the most useful approach is through local measurement. This perspective ties localization directly to relativistic causality and measurable outcomes in spacetime. Strict localization in quantum field theory is formalized by states that give the same expectation value as vacuum for all local observables outside some spacetime region. With this formalism, single photons cannot be strictly localized, but coherent states can. The structure and interpretation of local observables and local operator algebras are discussed in connection with the Reeh--Schlieder theorem. Since probabilities of local measurement outcomes cannot vanish in vacuum, particle detection events must be phrased carefully. Key characteristics of strictly localized states are surveyed alongside concrete strategies for constructing them. The relationship between localization, causality, and on-demand sources is analyzed, finding that strictly localized states can represent localized, propagating information in quantum field theory. Explicit examples of localized photonic states are presented with an emphasis on quantum optics, including states that approximate single photons and (entangled) Bell states. 
\end{abstract}


\maketitle

\begingroup
\let\boldmath\relax
\tableofcontents
\endgroup

\input{1_introduction}

\input{2_history}

\input{3_local_measurement}

\input{4_local_operators}

\input{5_localized_states}

\input{6_localization_and_causality}

\input{7_examples_of_localized_states}

\input{8_conclusion}

\section*{Acknowledgments}
We thank Joakim Bergli, Daniel Burgarth, Kai Ryen Bush, Thiago L.\ M.\ Guedes, Robert Jonsson, Pieter Kok, Jon Magne Leinaas, Nils J.\ Mikkelsen, Jason Pye, and Sigurd S.\ Rustad for helpful discussions and comments. The authors used ChatGPT (OpenAI) as an aid in discussing scientific questions and as an additional check on some scientific arguments and derivations. The authors take full responsibility for the content of the article. This work was supported by the Research Council of Norway [grant number 354574].

\appendix
\crefalias{section}{appendix}
\crefalias{subsection}{appendix}
\crefalias{subsubsection}{appendix}
\input{9_appendix}


%

\end{document}

%% file: 1_introduction.tex

\iftoggle{ExportTikz}{%
     
}{}

\section{Introduction}
\label{c1_sec_introduction}
We begin this review with a short mathematical derivation that illustrates why the topic of photon localization in quantum field theory is fundamentally nontrivial. A more high-level introduction then follows in \cref{c1_sec_review_purpose_and_scope}. Quick answers to some common questions can be found in \cref{c8_sec_conclusion}.

\subsection{Invitation}
\label{c1_sec_invitation}
Consider the free electric field in one dimension restricted to a single linear polarization and forward propagation, meaning that the wavevector is a scalar $k > 0$. In this case there is exactly one wavevector $k$ for each frequency $\omega$, and we can express the electric field at some spatial point as a scalar function of time \cite[Ch.~6.2]{loudon2000}:
\begin{equation}\label{c1_E_t}
    E\lp(t) = \minti{\dd \omega} \lp[\mathcal{E}\lp(\omega) a\lp(\omega) e^{- i \omega t} + \mathcal{E}^*\sephn\lp(\omega) \adj{a}\sephn\lp(\omega) e^{i \omega t}].
\end{equation}
Here the frequency-mode annihilation and creation operators satisfy $\comm{a\lp(\omega)}{\adj{a}\sephn\lp(\omega')} = \delta\lp(\omega - \omega')$, and $\mathcal{E}\lp(\omega) \propto \sqrt{\omega}$ is some function of the frequency. 

How do we describe a particle (a photon) of this field? An immediate response might simply be $\ket{1_\omega} = \adj{a}\sephn\lp(\omega)  \ket{0}$, but such a monochromatic state is infinitely spread out. A better representation of some physical light pulse is achieved by superposing such basis states in a wavepacket:
\begin{equation}\label{c1_1_xi}
    \ket{1_\xi} = \adj{a}_\xi  \ket{0}, \sepc \adj{a}_\xi = \minti{\dd \omega} \xi\lp(\omega) \adj{a}\sephn\lp(\omega).
\end{equation}
Any single-photon pulse can be written in this form for some appropriate pulse spectrum $\xi\lp(\omega)$, normalized to $\minti{\dd \omega} \abs{\xi\lp(\omega)}^2 = 1$.

However, a simple calculation of the expectation value of the normal-ordered electric field squared for this state gives
\begin{equation}\label{c1_qexv_1_xi_E_t}
    \qexv{1_\xi}{\norder{E^2\sephn\lp(t)}} = 2 \, \abs{\minti{\dd \omega} \mathcal{E}\lp(\omega) \xi\lp(\omega) e^{- i \omega t}}^2.
\end{equation}
The integral in \cref{c1_qexv_1_xi_E_t} is an (inverse) Fourier transform over only positive frequencies, meaning that it cannot vanish on any interval in time (see, e.g., \cref{c2_sec_particle_and_photon_localization_in_quantum_field_theory}). Thus, for any single-photon state, the expectation value of $\norder{E^2\sephn\lp(t)}$ has nonzero tails everywhere. 

This feature turns out not to be true for all states. For example, the same calculation for a coherent state $\ket{\alpha_\xi} = \exp\lp(\alpha \adj{a}_\xi - \alpha^* a_\xi) \ket{0}$ with $\alpha \in \complexes$ gives
\begin{equation}\label{c1_qexv_alpha_xi_E_t}
    \qexv{\alpha_\xi}{\norder{E^2\sephn\lp(t)}} = \lp(\mintn{\dd \omega} \mathcal{E}\lp(\omega) Z\lp(\omega) e^{- i \omega t})^2,
\end{equation}
where we have extended $\mathcal{E}\lp(-\omega)= \mathcal{E}^*\sephn\lp(\omega)$ and
\begin{equation}\label{c1_Z_w}
    Z\lp(\omega) = 
    \begin{cases}
        \alpha \xi\lp(\omega), & \omega > 0, \\
        \alpha^* \xi^*\sephn\lp(- \omega), & \omega < 0.
    \end{cases}
\end{equation}
Thus, even though the modes we quantize are supported only for positive frequencies, coherent states combine the modes in a way that brings both positive and negative frequencies into the Fourier integral. In particular, the expectation value of $\norder{E^2\sephn\lp(t)}$ for coherent states may be compactly supported in time. 

Note that although we have here shown the simplest case of one-dimensional forward propagation ($k > 0$), it turns out that including other wavevectors $\vect{k}$ does not change the conclusion. The fundamental reason is that for every wavevector, the associated frequency is always positive, $\omega_{\vect{k}} \geq 0$, corresponding to particles always having positive energy. This absence of half the frequency spectrum for the factor $e^{- i \omega t}$ severely limits the possible time behavior of expectation values such as $\exv*{\norder{E^2\sephn\lp(t)}}$ for single particles, to the point that they cannot evolve in time with compact support. We show this in detail with a full, three-dimensional analysis based on the Reeh--Schlieder theorem in \cref{c5_sec_proof_single_particles_not_localized}.

The short derivation above raises some questions. If we send a signal by means of an optical pulse, we expect from basic causality that this signal must have a well-defined \enquote{start}---a point in time before which an observer at the other end does not see the signal. However, if the pulse consists of a single photon, then measuring $\norder{E^2\sephn\lp(t)}$ at the other end, no matter how far away, has a nonzero expectation value for all time, suggesting that single-photon signals have no start.

What is the meaning of this result? Is $\norder{E^2\sephn\lp(t)}$ actually observable? What exactly is measurable in some region of space and time in quantum field theory? Why can the expectation value of $\norder{E^2\sephn\lp(t)}$ have a start for coherent states but not for single photons? Is this related to localization of photons? What does it mean for a particle to be localized in quantum field theory? What is the exact connection between localization and causal signals? What types of quantum states may represent propagating relativistic information? These are the questions explored in this review.

\subsection{Review purpose and scope}
\label{c1_sec_review_purpose_and_scope}
Looking for answers to these questions in the literature can be confusing and quickly opens a giant can of worms, including: relativistic wavefunctions, particle localization operators, Newton--Wigner position operators, almost-localized particles, the Fermi problem, Hegerfeldt's theorem, local measurement, algebraic quantum field theory, the Reeh--Schlieder theorem, and more. Clearly, particle localization in quantum field theory is a very fundamental, well-studied topic. Yet it is also highly confusing, with a plethora of different approaches and no-go theorems, many of which can appear to contradict each other. 

The goal of this review is to make some sense of all of this. Such a goal is not small; nevertheless it is our hope to connect and understand how the various theorems, approaches, and statements related to this topic fit together. Additionally, existing results in the literature can give the impression that particle localization in quantum field theory is somehow fundamentally ill-defined. We will on the other hand argue that with a systematic treatment, particle localization in quantum field theory can be given a precise and consistent meaning. 

In \cref{c2_sec_history} we look broadly at the various approaches to photon localization. In \cref{c3_sec_what_does_it_mean_for_a_particle_to_be_localized} we argue why, among these methods, a natural starting point is to characterize local measurement. The remainder of this review then follows this approach, starting by examining what is locally measurable in quantum field theory, and formulating a coherent theory of particle localization based on it. 

One likely reason why such a theory has not fully solidified before is that conventional single-photon experiments have generally been well described without resolving these localization questions. Single-photon sources in the optical regime today typically have a high-frequency carrier modulated by a slowly varying envelope, corresponding to a narrow-bandwidth spectrum around a carrier frequency in the $\sim \! \SI{e14}{\hertz}$ range [see, e.g., \cite{eisaman2011} and references therein]. For such quasi-monochromatic signals, the difference between including and not including negative frequencies is vanishingly small. Therefore, standard treatments of quantum optics, such as \cite[Ch.~6.2]{loudon2000}, argue that we may with good accuracy replace the exact expression \eqref{c1_E_t} with an integral including negative frequencies as well. 

Within this frequency extension, we can introduce time-domain ladder operators: ${\adj{a}\sephn\lp(t) = \frac{1}{\sqrt{2\pi}} \mintn{\dd \omega} \adj{a}\sephn\lp(\omega) e^{i \omega t}}$, so that the commutator is $\comm{a\lp(t)}{\adj{a}\sephn\lp(t')} = \delta\lp(t - t')$. Single photons can then be put in arbitrary temporal modes $\mintn{\dd t} \xi\lp(t) \adj{a}\sephn\lp(t) \ket{0}$ by choosing any desired mode function $\xi\lp(t)$. Such a theory reduces in some ways to the same as nonrelativistic quantum mechanics, applied to arbitrary modes in space and time. Many works in quantum optics follow this approach in great agreement with experiments, e.g., \citetext{\citealp{gardiner1985,blow1990}; \citealp[Ch.~5.3]{gardiner2000}; \citealp[Ch.~6.2]{loudon2000}; \citealp[Ch.~7.1]{walls2008}; \citealp[Ch.~4.7]{wiseman2010}}.

However, it is not obvious exactly when this narrow-bandwidth approximation can be justified, since it involves approximating an operator. Furthermore, the approximation cannot in general be justified in situations where the bandwidth is not narrow. Surely the technological limitations of today will improve over time, making a theory for general pulse spectra increasingly relevant. Already there are proposals for methods for creating single-photon sources in the THz regime with pulse lengths on the order of a single cycle \cite{su2016}. Moreover, for laser light, ultrafast pulses of only a few cycles are now produced routinely [see, e.g., \cite{krausz2009} and references therein], meaning that the narrow-bandwidth approximation can certainly not be assumed for all types of light. 

More fundamentally, the real problem with the narrow-bandwidth approximation is that even when it is experimentally justified, there are some questions that it cannot answer. In particular, causality dictates that any signal generated on demand cannot lead to observable effects outside its lightcone. Such constraints should hold exactly, not approximately. Even if the contributions from the negative frequencies added in the approximation are vanishingly small, they are not physical degrees of freedom in quantum field theory. They can therefore not be the basis for a fully consistent discussion of localization and causality. 

In summary, what is needed is to analyze particle localization in the exact theory of quantum optics without the narrow-bandwidth approximation, retaining the positive-frequency mode spectrum. Only then do all the exciting and intriguing aspects of quantum field theory such as the Reeh--Schlieder theorem enter the picture. There are other areas of physics, such as relativistic quantum information, that do explore these aspects, but these areas tend to concentrate more on local measurements, vacuum entanglement, and related phenomena. One goal for this review is to connect ideas from relativistic quantum information to the question of photon localization and, more generally, argue for why studying localized states in quantum field theory is interesting.

\subsection{Review outline}
\label{c1_sec_review_outline}
This review is partly based on the PhD thesis of the first author \cite{gulla2024}, where the second author acted as advisor. There is therefore some overlap in the content of these two works, although large parts have been added and modified. The review is organized as follows. 

In \cref{c2_sec_history} we go through the history of studies related to photon localization, touching on topics such as particle localization operators, local observables, the Paley--Wiener criterion, the Fermi problem, Hegerfeldt's theorem, and the Reeh--Schlieder theorem. We also discuss viewpoints in the literature on what a particle is and how to interpret measurements in quantum field theory. 

In \cref{c3_sec_local_measurement} we look at what it means for a particle to be localized. In particular, we argue why local measurement is a well-suited method for formalizing localization of relativistic particles. We then discuss what is locally measurable in quantum field theory, as well as what a measurement process is physically. 

In \cref{c4_sec_local_operators} we develop the mathematical description of local operators introduced in \cref{c3_sec_local_measurement}. Concretely, we begin by constructing polynomial local operators from fields smeared over spacetime regions. We then construct bounded local algebras through spectral decomposition and discuss spectral projectors. We also introduce the Reeh--Schlieder theorem, as well as a corollary which says that the probability of any outcome of any local measurement in vacuum cannot vanish. This surprising result has been much debated in the literature, and we therefore explain in detail how to correctly interpret measurements in quantum field theory. 

In \cref{c5_sec_localized_states} we define strictly localized states as states that are exactly undetectable by any local measurement outside some sharp region in space and time. We also look at how strictly localized states can be characterized by Licht operators or unitary operators. We also discuss mixed states and the localization of post-measurement states. Finally, we demonstrate that the fact that single photons cannot be strictly localized follows from the Reeh--Schlieder theorem (generalizing the analysis in \cref{c1_sec_invitation}). 

In \cref{c6_sec_localization_and_causality} we discuss the connection between strictly localized states and causality. We first give a precise statement of relativistic causality in quantum field theory, stated in terms of external, classical control and measurement probabilities. We then look at the dynamics of strictly localized states, showing that they indeed represent localized information that stays localized within the lightcone. We then argue why any source that is operated on demand, acting on the vacuum, must in fact produce strictly localized states. 

In \cref{c7_sec_examples_of_localized_states} we give several examples of strictly localized states. First we consider coherent states and squeezed states, obtained by certain unitary operators constructed from smeared fields. Then we turn to nonunitary isometries, which can generate, e.g., strictly localized states close to single photons. These states may describe the output of on-demand sources aiming at producing single photons. Finally, we discuss strictly localized states that result from truncating a single-photon pulse by an optical shutter, consider excitations of localized modes, and give a Heisenberg-type uncertainty relation for broadband photons.

\subsection{Notation}
\label{c1_sec_notation}
We work in natural units, with $\hbar = c = \epsilon_0 = 1$, and limit the treatment to flat Minkowski space. With Fourier transforms, we use the usual physics convention of letting the sign in the exponent be different for space and time transforms. In addition, we use a unitary choice of the normalization factor $2 \pi$:
\begin{subequations}
\begin{alignat}{2}
    f\lp(\vect{k}) &= \fourier{f\lp(\vect{x})}\lp(\vect{k}) &&= \mintf{\frac{\dd[3]x}{\lp(2 \pi)^{3/2}}} f\lp(\vect{x}) e^{-i \vect{k} \cdot \vect{x}}, \label{c1_f_k} \\
    f\lp(\omega) &= \fourier{f\lp(t)}\lp(\omega) &&= \mintf{\frac{\dd t}{\sqrt{2 \pi}}} f\lp(t) e^{i \omega t}. \label{c1_f_w}
\end{alignat}
\end{subequations}

We use the same normalization factor in field expressions, so for instance a free, hermitian Klein--Gordon field at some spacetime point $x$ is written \cite[Ch.~2.4]{peskin1995} as
\begin{equation}\label{c1_phi_x}
\begin{aligned}
    \phi\lp(x) &= \mintf{\frac{\dd[3] k}{\lp(2 \pi)^{3/2}}} \frac{1}{\sqrt{2 \omega_{\vect{k}}}} \lp(\padj{a}_{\vect{k}} e^{- i k \cdot x} + \adj{a}_{\vect{k}} e^{i k \cdot x}),
\end{aligned}
\end{equation}
where $a_{\vect{k}}$ is the annihilation operator of the mode with wavevector $\vect{k}$, and $k\cdot x = k_\mu x^\mu = \omega_{\vect{k}} t - \vect{k} \cdot \vect{x}$. The frequency is given by $\omega_{\vect{k}}^2 = \vect{k}^2 + m^2$, where $m$ is the mass. With the convention in \cref{c1_phi_x}, we have $\comm[\big]{\padj{a}_{\vect{k}}}{\adj{a}_{\vect{k}'}} = \delta^{\lp(3)}\sephn\lp(\vect{k} - \vect{k}')$. Furthermore, we let
\begin{equation}\label{c1_1_k}
    \ket{1_{\vect{k}}} = \adj{a}_{\vect{k}} \ket{0},
\end{equation}
so that $\braket{1_{\vect{k}}}{1_{\vect{k}'}} = \delta^{\lp(3)}\sephn\lp(\vect{k} - \vect{k}')$. With this convention, not all $k$ integrals have factors of $2 \pi$. For instance the one-particle identity operator becomes
\begin{equation}\label{c1_identity_k}
    \identity_{\text{1 particle}} = \mint{\dd[3] k} \ket{1_{\vect{k}}} \bra{1_{\vect{k}}}.
\end{equation}

We write the quantized electromagnetic potential \citetext{\citealp[Ch.~4.8]{peskin1995}; \citealp[Ch.~7.3]{greiner1996}} as
\begin{equation}\label{c1_A_mu_x}
    A^\mu\sephn\lp(x) = \sum_s \mintf{\frac{\dd[3] k}{\lp(2 \pi)^{3/2}}} \!\frac{1}{\sqrt{2 \omega_{\vect{k}}}} \lp(e^\mu_{\vect{k}s} \padj{a}_{\vect{k}s} e^{- i k \cdot x} + e^{\mu*}_{\vect{k}s} \adj{a}_{\vect{k}s} e^{i k \cdot x}).
\end{equation}
Here $\mu$ is a spacetime index, $\omega_{\vect{k}} = \abs{\vect{k}}$, and $e^\mu_{\vect{k}s}$ are polarization vectors. \Cref{c1_A_mu_x} is valid in Coulomb gauge with the index $s$ running over two transverse polarizations, or in Lorenz gauge (with Gupta--Bleuler quantization) with $s$ running over four polarizations.

We sometimes specialize to a one-dimensional analysis in line with \cite[Ch.~6.2]{loudon2000}, by considering a single linear polarization and propagation direction for $A^\mu\sephn\lp(x)$ in Coulomb gauge. Since the forward-propagating components have a trivial spacetime dependence $t - x$, it is unnecessary to keep both $t$ and $x$. Thus we can equivalently consider a fixed spatial point, e.g., $x = 0$, expressing the potential as a function of only time:
\begin{equation}\label{c1_A_t}
    A\lp(t) = \minti{\dd \omega} \mathcal{A}\lp(\omega) \seph \lp[a\lp(\omega) e^{- i \omega t} + \adj{a}\sephn\lp(\omega) e^{i \omega t}].
\end{equation}
Here we have introduced frequency-mode ladder operators $a\lp(\omega)$ satisfying $\comm{a\lp(\omega)}{\adj{a}\sephn\lp(\omega')} = \delta\lp(\omega - \omega')$, corresponding to $a_{\vect{k}s}$ for the particular propagation and polarization direction. We have also absorbed functions of the frequency in line with \cite[Ch.~III.A]{cohen-tannoudji1997}. Specifically
\begin{equation}\label{c1_A_w}
    \mathcal{A}\lp(\omega) = \frac{1}{\sqrt{2 \pi}} \frac{1}{\sqrt{2 \omega}}.
\end{equation}
The corresponding electric field is
\begin{equation}\label{c1_E_t_notation}
\begin{aligned}
    E\lp(t) &= -\frac{\dd A\lp(t)}{\dd t} \\
    &= \minti{\dd \omega} \lp[\mathcal{E}\lp(\omega) a\lp(\omega) e^{-i\omega t}
    + \mathcal{E}^*\sephn\lp(\omega) \adj{a}\sephn\lp(\omega) e^{i\omega t}],
\end{aligned}
\end{equation}
where $\mathcal{E}\lp(\omega) = i\omega\mathcal{A}\lp(\omega)$, as in \cref{c1_E_t}.


%% file: 2_history.tex

\iftoggle{ExportTikz}{%
     
}{}

\section{History}
\label{c2_sec_history}
The issue of particle localization is a very old question that has shown up in many different areas of physics over the years. Here we are specifically interested in localization in the currently accepted framework for particle physics, namely quantum field theory. It is however not easy to get a full overview of this topic across all potentially relevant branches of physics, and some statements and no-go theorems can appear to contradict one another. 

This section aims to give a somewhat comprehensive overview of studies on localization of relativistic particles. Rather than a purely historical recollection, we aim to emphasize how the main ideas evolved and how they fit together, looking back through the lens of what we know now. Major topics we discuss include particle localization operators, local measurement, the Paley--Wiener criterion, the Fermi problem, Hegerfeldt's theorem, the Reeh--Schlieder theorem, and how to interpret measurements in quantum field theory. 

Additionally, we include some of our own analyses and interpretations, to try to connect insights that are well established in some subfields to others where they are less widely appreciated. Hopefully this makes it possible to unpack this heavily studied but confusing topic, and understand how the different aspects of localization are related. Also note that the content presented in this section incorporates parts of the historical discussion in \cite{gulla2023} written by the same authors.

\subsection{Particle and photon localization in quantum field theory}
\label{c2_sec_particle_and_photon_localization_in_quantum_field_theory}
Today, localization of photons and other relativistic particles is a relatively specialized topic. For instance in high-energy scattering calculations, it is usually sufficient to consider momentum eigenstates $\ket{\vect{p}}$ in order to predict many relevant experimental results with great accuracy \cite[Ch.~4.5]{peskin1995}. Although these states are completely delocalized in space, their usefulness reflects the fact that scattering cross sections are often insensitive to the detailed spatial profile of realistic wavepackets.

Historically, however, localization was a central consideration in the development of the theory of light. Around the early 18th century, Newton suggested that light consists of tiny, localized particles (corpuscles), which was effective at explaining sharp shadows and straight-line propagation \cite{newton1704}. Around the same time, Huygens put forth a competing view of light as a propagating wave in some underlying medium, and used it to explain reflection and refraction \cite{huygens1690}. This idea was eventually consolidated in Maxwell's synthesis of electricity, magnetism, and optics \cite{maxwell1865}, describing light as a self-sustaining electromagnetic field. 

In these wave theories, localization of light is no longer point-like but can nevertheless be formalized in terms of the support of the field disturbance. In particular we may construct wavepackets whose energy density is initially confined to a bounded region and subsequently transported causally through space (typically with spreading). For reviews on the early history of studies of light, see, e.g., \cite{sachs1973,bacry1988}.

In the early 20th century, the question about localization then reappeared during the birth of quantum theory and wave mechanics \cite{planck1901,einstein1905,bohr1913,heisenberg1925,born1926,schroedinger1926}. It became necessary to describe particles of matter by wavefunctions $\psi\lp(\vect{x})$ with a probabilistic interpretation. Still, one eventually got accustomed to particles no longer being localized to points but rather smeared out with wavefunctions. After all, the expectation value of measuring position is well-defined as the mean of the probability distribution $\abs{\psi\lp(\vect{x})}^2$, and it evolves in time in a predictable way. 

The situation would however be less straightforward when passing to quantum field theory. Incorporating special relativity into quantum physics led to quantum fields emerging as a natural framework \cite{dirac1927,heisenberg1929,heisenberg1930}, and over time this became the basis of what is today called the Standard Model of particle physics. Although popular science sometimes phrases a particle vs.\ wave debate in this construction, modern theory does not; light is simply described by states of the quantized electromagnetic field. Nevertheless, a genuine question arises: given light in some state of this quantum field, in what sense can it be said to be localized in space and time?

It was quickly noticed that the localization concepts that work well in nonrelativistic quantum mechanics do not carry over naively to relativistic particles. With no obvious wavefunction to square, how does one define \enquote{where the particle is} in quantum field theory? Over time, various analogues of the nonrelativistic localization schemes have been analyzed, including relativistic particle position operators \cite{pryce1948,moeller1949,foldy1950,jordan1963,weidlich1963,bacry1964,bacry1988a,berg1965,fleming1965,fleming1965a,fleming2000,galindo1965,barut1968,johnson1969,kalnay1971,angelopoulos1974,oconnell1978,bracken1999,kosinski2018}, relativistic particle position states \citetext{\citealp{newton1949,philips1964}; \citealp[Ch.~4]{teller1995}; \citealp{melloy2002,pavsic2018}}, and relativistic particle probability operators \cite{wightman1955,wightman1962,jauch1967,amrein1969,schlieder1971,hegerfeldt1974,hegerfeldt1985,hegerfeldt1998,hegerfeldt1998a,skagerstam1976,jancewicz1977,perez1977,kaloyerou1988,malament1996,busch1999,halvorson2002,castrigiano2015,moretti2023} (note that there is some overlap between these topics and several of the listed works deal with a combination). 

The same has been true for investigations of localization of photons specifically, with analyses of, e.g., photon wavefunctions \cite{landau1930,oppenheimer1931,good1957,moses1973,bialynicki-birula1994,bialynicki-birula1996a,bialynicki-birula1996,inagaki1994,sipe1995,hawton1999a,hawton2007,kobe1999,keller2000,chan2002,fedorov2005,muthukrishnan2005,smith2006,smith2007,tamburini2008}, photon position operators \citetext{\citealp{jadczyk1972,jordan1978,comi1980}; \citealp[Ch.~5]{bacry1981,bacry1988}; \citealp{mourad1993,skagerstam1994,shojai1997,hawton1999,hawton2001,ojima2015,babaei2017,dobrski2023}}, and spatial photon probability/number operators \cite{mandel1966,kraus1977,cook1982a,cook1982,rosewarne1992,hawton1995,holevo2006,hawton2010,terno2014,guarnieri2015} (again, note that these topics overlap). For reviews on these types of approaches to photon localization, see, e.g., \cite{bialynicki-birula1996,keller2005}.

These analyses generally find that such constructions cannot be defined for photons in a way that preserves all the desired properties familiar from nonrelativistic quantum mechanics; at least one property has to be dropped. We look at this in more detail in \cref{c3_sec_wavefunctions_position_operators_and_position_eigenstates}. Some of these works take these challenges as an indication that photons cannot be sharply localized. However, these conclusions depend on the localization criterion adopted, and this literature does not converge on a unique criterion for what it means for a photon to be localized.

On the other hand, there have been some works that have taken a different approach. One other method for studying photon localization has been looking at the (normal-ordered) energy density as a function of space and time. An early calculation of the expected energy density for a single photon \cite{amrein1969} produced a long-standing belief \cite[Ch.~12.11]{mandel1995} that a maximally localized photon has an energy density that is spread out in space $\vect{x}$ with an asymptotic falloff of $\abs{\vect{x}}^{-7}$. This belief was however later shown to be false, when methods from electromagnetic directed-energy pulse trains \cite{brittingham1983,ziolkowski1989,hellwarth1996} were used to show that solutions with arbitrarily high inverse powers of $\abs{\vect{x}}$ are possible \cite{adlard1997}. 

Finally, \cite{bialynicki-birula1998} found an almost-exponential limit to the localization of photons. The fundamental reason for this limit is due to time evolution and the frequency (energy) always being positive: $\omega_{\vect{k}} = \abs{\vect{k}} \geq 0$. The connection to localization is easiest to see in one dimension, which we looked at in \cref{c1_sec_invitation} and saw that the expected electric energy density for a single photon is proportional to the absolute square of a Fourier transform of only positive frequencies. Such Fourier amplitudes satisfy quite strict analytic properties, governed for instance by the Paley--Wiener criterion. 

\begin{theorem}\label{c2_thm_paley_wiener}\textup{(Paley--Wiener criterion).\footnote{\label{c2_foot_paley_wiener}Rephrasing of \cite{paley1934}, Theorem~XII\@. See also \cite{bialynicki-birula1998}.}}
    Let $f\lp(\omega)$ be a nonzero square-integrable function with $f\lp(\omega) = 0$ for ${\omega < 0}$. Then its (inverse) Fourier transform $f\lp(t)$ satisfies
    \begin{equation}\label{c2_paley_wiener}
        \mintn{\dd t} \frac{\abs[\big]{\log{\abs{f\lp(t)}}}}{1 + t^2} < \infty.
    \end{equation}
    Hence $f\lp(t)$ is nonzero (almost) everywhere on $\reals$ and must have an asymptotic falloff as $\abs{t} \rightarrow \infty$ that is slower than $e^{- \gamma \abs{t}}$ for any constant $\gamma > 0$. 
\end{theorem}

In three spatial dimensions, limitations imposed by energy positivity still remain but are harder to quantify. Ref.~\cite{bialynicki-birula1998} specializes to a single-photon state with a certain spherical symmetry, and shows that its expected energy density in space can again be expressed in terms of a Fourier integral over only positive arguments. Thus Bialynicki-Birula shows that the expected energy density has to be nonzero for arbitrarily large radial distances $r \rightarrow \infty$ and fall off asymptotically slower than $\sim \exp\lp(- \gamma r)$ for all $\gamma > 0$. He also gives a concrete example of a single-photon state with an expected energy density that falls off asymptotically as $\sim \exp\lp(- \gamma \sqrt{r})$ for some $\gamma > 0$. 

Later, certain cylindrically symmetric photons were investigated in \cite{saari2005,saari2006}, showing that the asymptotic energy-density falloff can be faster than exponential in some spatial directions, while remaining slower along other directions. Ref.~\cite{bialynicki-birula2009} then analyzed the energy density localization of \emph{arbitrary} single-photon states. This analysis does however not rely on the Paley--Wiener criterion. Rather, it considers the spatial localization of the electric and magnetic energy densities separately. Since the electric and magnetic fields are related to each other by time derivatives, their simultaneous localization in space is again fundamentally limited by energy positivity. In particular, Ref.~\cite{bialynicki-birula2009} uses an argument based on analyticity to show that the total expected energy density of a single photon cannot be confined to a bounded spatial region (note that for the question of who was first to prove this, see also \cref{c2_sec_knights_and_lichts_treatment_of_strictly_localized_states}). For more details on approaches to photon localization based on energy density, see \cite{bialynicki-birula1998,saari2012}.

Analyzing localization of photons based on their energy density is fundamentally different from the analyses based on nonrelativistic localization. The reason is that energy density is something that is \emph{locally measurable} in quantum field theory \cite[Ch.~III.1]{haag1996}. In \cref{c3_sec_local_measurement} we argue in detail why local measurement is a suitable method for analyzing questions about localization of relativistic particles. We thus support the conclusion from \cite{bialynicki-birula2009}. A simple proof that single photons cannot be strictly localized is given in \cref{c5_sec_proof_single_particles_not_localized}, and \cref{a1_sec_proof_single_photons_cannot_be_localized} provides an explicit energy-density analysis.

\begin{remark}\label{c2_rem_single_photons_cannot_be_localized}\remheadp{}
    Single photons cannot be strictly localized---their expected normal-ordered energy density, a locally measurable quantity, cannot be confined to a bounded spatial region!
\end{remark}

This result is counterintuitive for several reasons. First, there are numerous reported experiments producing single photons, as seen in, e.g., \cite{scheel2009,eisaman2011,senellart2017,sinha2019,wang2019}. If these infinite tails are really present, why do we not see them in experiments? 

Second, and perhaps most importantly, particles not being strictly localizable appears to be irreconcilable with causality. One person to notice this was Hegerfeldt. In \cite{hegerfeldt1974}, he argued that strictly localized quantum particles and a positive Hamiltonian are incompatible with relativistic causality, based on a quite general argument using analyticity and energy positivity. Later, Hegerfeldt connected his argument to something called \emph{the Fermi problem} in \cite{hegerfeldt1994}. Here he argued that an atom decaying to its ground state under the production of photons will instantaneously cause detectable influences at some other, spatially separated atom. Such apparent signaling across spacelike points clearly challenges the idea of relativistic causality.

\subsection{The Fermi problem}
\label{c2_sec_the_fermi_problem}
The usual way of demonstrating that quantum field theory is in accordance with relativistic causality is through the field commutator. E.g., for some hermitian scalar field $\phi\lp(x)$, it is verified that $\comm{\phi\lp(x)}{\phi\lp(\splike{x})} = 0$ for spacelike separations \cite[Ch.~2.4]{peskin1995}. Since local observables in quantum field theory are built from field polynomials (see, e.g., \cref{c3_sec_pointwise_local_observables}), this means that measurements that are spacelike separated cannot influence one another. However, it is also of interest to verify causality explicitly for the fundamental dynamical process in quantum field theory: particle production, evolution, and detection. 

An early attempt at such a calculation was made already in 1930, immediately following the inception of quantum field theory. Ref.~\cite{kikuchi1930} considered an initially excited atom decaying to its ground state under the production of electromagnetic radiation, and the time evolution of the radiated energy density was analyzed. Fermi subsequently refined this setup in what has become known as the Fermi problem by adding a second, spatially separated atom initially in its ground state acting as a detector \cite{fermi1932}. Assuming the electromagnetic field starts out in vacuum, Fermi considered the timing of the energy propagation from the first to the second atom by calculating a transition probability as a function of time. In both \cite{kikuchi1930} and \cite{fermi1932} it was found that the emitted radiation propagates at the speed of light, and their solutions were regarded as demonstrating that quantum field theory is in agreement with relativistic causality \cite[Ch.~20]{heitler1954}. 

However, \cite{shirokov1964} pointed out that the causal solutions to the Fermi problem were a result of an approximation, where a certain integral's limits were extended from $\lp(0, \infty)$ to $\lp(- \infty, \infty)$. Shirokov argues that without this approximation, there would be some nonzero probability for the second atom becoming excited before the signal has had time to propagate there \cite{shirokov1967}. Today we understand that this integral's lower limit was indeed coming from the same positive-energy restriction found later by \cite{hegerfeldt1974} and \cite{bialynicki-birula1998}. 

The contradiction with causality came from another approximation done in \cite{kikuchi1930} and \cite{fermi1932}, namely of including only resonant terms in the interaction Hamiltonian. Under this assumption, the excited atom decays under the production of a single photon, in line with the Weisskopf--Wigner approximation \cite{weisskopf1930}. Since the expected energy density of single photons must have infinite tails, this clearly leads to acausal, observable influences on the second atom. 

A resolution to this issue came in \cite{ferretti1968}. First, Ferretti pointed out that a consistent analysis of atom radiation dynamics can only be done while keeping both the resonant and anti-resonant terms in the interaction Hamiltonian. This means that the time evolution operator leads to a Dyson series. Consequently, the field state generated during the emission process will in general contain multi-photon components, rather than being restricted to the vacuum and one-photon sectors.

Second, Ferretti realized that it is necessary to specify a measurement operator that is local to the measurement region. In this case, for measurements at the second atom, an appropriate operator would for example be given by the projector for the second atom in the excited state, summed over all possible photon states and first-atom states. On the other hand, if we were to include, e.g., the electromagnetic vacuum state in the measurement projector, this would correspond to measuring the photon number globally at the observation time. Since this is not a measurement local to the second atom, the associated measurement probability could show a time variation before the arrival time of the signal at the second atom. We look at what is locally measurable in quantum field theory in detail in \cref{c4_sec_local_operators}. 

Unfortunately, Ferretti's insights went largely unnoticed, and textbooks on quantum field theory repeated Fermi's approximation years later, e.g., in \cite[Ch.~5.10]{louisell1973}. The issues and solutions concerning energy positivity and local measurements in connection with the Fermi problem were subsequently rediscovered a number of times \cite{hegerfeldt1974,haan1985,rubin1987,compagno1989,biswas1990,valentini1991,hegerfeldt1994,maddox1994,milonni1994a,compagno1995,berman1997,hegerfeldt1998}. Today, the approximation of retaining only the resonant terms in the interaction Hamiltonian, now known as the rotating-wave approximation, and its limitations for precise dynamics are well understood \cite{power1997}. The causal nature of the Fermi problem is also easily demonstrated using the Heisenberg picture \cite{milonni1995}, where the field operators obey equations with the same form as the classical equations \cite[Ch.~4.6]{milonni1993}. For other discussions of the history of the Fermi problem, see \cite{shirokov1978,dickinson2016}.

In summary, the Fermi problem literature clarifies why relativistic causality in quantum field theory at the operator level is not in contradiction with the fact that photons cannot be strictly localized; namely, a decaying atom that starts decaying at some time does not emit an exact single photon, but rather a superposition/mix of photon-number states. The Paley--Wiener criterion cannot be directly applied to states with multiple photon-number terms, so it is possible that such states can be strictly localized. In particular, it was found in \cite{bialynicki-birula1998} that a particular type of superposition, namely coherent states, can be strictly localized. We also indicated this in \cref{c1_sec_invitation} and a detailed discussion is given in \cref{c7_sec_examples_of_localized_states}.

\subsection{What is a particle?}
\label{c2_sec_what_is_a_particle}
While these insights seem to have become relatively well-known in the context of solving the Fermi problem, their implications for particle localization seem to be less widespread. In general, it appears that most of the current literature deals with the issue of localizing relativistic particles in one of three ways. 

A first category is works that give the impression that a single relativistic particle may be strictly localized. For example many of the currently popular textbooks and lecture notes on quantum field theory claim that for a field $\phi\lp(x)$, the state $\phi\lp(x)  \ket{0}$ is a single particle localized at the spacetime point $x$ \citetext{\citealp[Ch.~2.2]{roman1969}; \citealp[Ch.~3.1]{itzykson1980}; \citealp[Ch.~2.3]{peskin1995}; \citealp[Ch.~1.4]{pokorski2000}; \citealp[Ch.~2.7]{tong2006}; \citealp[Ch.~5.5]{folland2008}; \citealp[Chs.~2.3 and 8.4]{schwartz2014}; \citealp[Ch.~11.4]{anastasiou2020}}. In the same category, we can group works in quantum optics that use time-domain ladder operators of the type $\comm{a\lp(t)}{\adj{a}\sephn\lp(t')} = \delta\lp(t - t')$, as detailed in \cref{c1_sec_review_purpose_and_scope}. Both of these claims are approximations that can be used in certain situations with great accuracy; however, strictly speaking they do not hold due to the absence of negative frequencies. Unfortunately, the approximate nature of this picture is often not mentioned, and the regimes where these approximations do not hold are rarely discussed. 

A second category can be said to be works that are aware that single relativistic particles cannot be strictly localized, and phrase it as an issue. This includes many of the works based on approaches to localization inspired by nonrelativistic quantum mechanics listed in \cref{c2_sec_particle_and_photon_localization_in_quantum_field_theory}. It is a natural expectation that a theory of photons should allow us to specify where the photons are \citetext{\citealp{sachs1973}; \citealp[Ch.~5]{bacry1988}}. Some authors have debated to what degree we can impose a \enquote{particle vs.\ field ontology} in quantum field theory \cite{halvorson2002}. Others have argued that if all quantum particles are always spread out over all space, including those making up detectors and ourselves, maybe no inconsistencies with causality would arise \cite{hegerfeldt1998}. Some authors have even proposed alternative theories that introduce independent negative-frequency photon modes alongside the usual positive-frequency modes \cite{southall2021}. In general, the lack of localizability of single relativistic particles is often seen as a deep conceptual problem.

Finally, a third category is studies based on transformations of field operators. This includes works that deal with Heisenberg-picture operator dynamics used to make causal signal propagation manifest \cite{compagno1995,milonni1995}, as mentioned in connection with the Fermi problem. However, these analyses typically do not explicitly construct the propagating signals as states in a photon-number basis. Similarly there are works on localized field modes and fields restricted to bounded spacetime regions, often using Bogoliubov transformations to relate local and global (Minkowski) modes \cite{shirokov1980,martinetti2003,ida2013,su2016a,foo2020,onoe2022,camblong2024}. These approaches provide certain reduced-state characterizations, such as that the global vacuum looks thermal in the local modes. However, we show in \cref{c7_sec_localized_modes} that such localized-mode excitations do not automatically produce strictly localized states relative to the global vacuum. Therefore, neither of these approaches generally provides an explicit characterization of what types of states in the (global) particle-number basis may represent strictly localized, propagating signals.

With the resolution of the Fermi problem in mind, we argue that a crucial link connecting these viewpoints is the role of states with indefinite particle number. Namely, is it possible that what we perceive as real, localized particles in experiments are not actually exact single-particle states, but instead superpositions of number states? It might seem strange to abandon the implicit, intuitive assumption that it must be single-particle states that describe the basic, localized constituents of light. After all, quantum field theory typically formulates scattering calculations in terms of states with definite particle number, for instance within perturbative expansions using Feynman diagrams. However, quantum optics routinely employs a broader class of quantum states, needed to represent distinct forms of the electromagnetic field. Indeed, number states are only a basis in a large Hilbert space of states. 

We also want to stress that such an approach does not entail any kind of modification of quantum field theory. The idea is simply to not limit localization questions to exact single-particle states, and rather examine to what degree more general states may be localized. In fact, as we show in \cref{c7_sec_eta12}, it turns out that there exist superpositions of photon-number states that can be strictly localized while simultaneously being very near a single photon in certain situations. This motivates a picture in which \enquote{particle-like} behavior we see in experiments need not arise from exact photon-number states, but can instead arise from states with small, additional terms. Such a perspective is also compatible with characterizations of single-photon sources in experiments, since a low value of the correlation function $g^{\lp(2)}\sephn\lp(0) \ll 1$ does not require the emitted state to be an exact single-photon state. 

One potential explanation for why this perspective has been less emphasized in the literature could be the relatively large focus on localization schemes inspired by nonrelativistic methods, as mentioned in \cref{c2_sec_particle_and_photon_localization_in_quantum_field_theory}. Allowing more general states does not by itself resolve the difficulties with, e.g., constructing suitable position observables. Instead, it is by considering locally measurable field observables, such as the energy density, that superpositions of number states may be strictly localized while single photons may not. Another possible explanation is the existence of several no-go theorems concerning particle localization, the most famous perhaps being that by Hegerfeldt. These no-go theorems seem at first glance to exclude \emph{any} strictly localized states compatible with relativistic causality, including superpositions of number states. However, these theorems require careful interpretations.

\subsection{Hegerfeldt's theorem}
\label{c2_sec_hegerfeldts_theorem}
In addition to connecting his 1974 argument to the Fermi problem, Hegerfeldt subsequently refined his original idea over time \cite{hegerfeldt1974,hegerfeldt1980,hegerfeldt1985,hegerfeldt1989,hegerfeldt1994,hegerfeldt1995,hegerfeldt1998,hegerfeldt1998a,hegerfeldt2001,hegerfeldt2002} into what has become known as Hegerfeldt's theorem.

\begin{theorem}\label{c2_thm_hegerfeldts_theorem}\textup{(Hegerfeldt's theorem).\footnote{\label{c2_foot_hegerfeldts_theorem}Among the many similar versions of this theorem, the clearest is arguably \cite{hegerfeldt1998}.}}
    On a Hilbert space $H$, let the Hamiltonian $\hamil$ be bounded from below, e.g., $\hamil \geq 0$. Let $D \geq 0$ be any positive operator on $H$. For any $\ket{\psi_0} \in H$, denote its time evolution $\ket{\psi\lp(t)} = e^{- i \hamil t} \ket{\psi_0}$. Then either
    \begin{subequations}
    \begin{equation}\label{c2_hegerfeldts_theorem_eq}
        \qexv{\psi\lp(t)}{D} = 0, \sepc \forall t \alme,
    \end{equation}
    or
    \begin{equation}\label{c2_hegerfeldts_theorem_neq}
        \qexv{\psi\lp(t)}{D} > 0, \sepc \forall t \alme,
    \end{equation}
    \end{subequations}
    for (almost) all time. 
\end{theorem}

The connection to particle localization is the following. In quantum mechanics, measurement probabilities are calculated by expectation values of positive operators, either projectors or so-called POVM elements \cite[Ch.~2.2]{nielsen2010}. Therefore, Hegerfeldt's theorem implies that for any pure state and any fixed measurement, the probability of any outcome either remains 0 forever or is never 0 (up to zero-measure sets). 

This clashes with the intuitive idea of states representing causally propagating, localized signals. For such states, one would expect to find measurement probabilities that are, e.g., initially nonzero only inside some finite spatial region for some time. Then at some later time, as the signal propagates, the measurement probability at some far-away spatial region should go from zero to nonzero. Stating this might sound extremely obvious, but such zero-to-nonzero behavior is in fact forbidden by Hegerfeldt's theorem! Instead, if a measurement probability for some state is nonzero for \emph{some} time interval, then it must be nonzero for \emph{all} time intervals. This is why Hegerfeldt's theorem might seem to imply that there can be no strictly localized states in quantum field theory compatible with causality; however, let us look at it in more detail. 

First note that although Hegerfeldt's theorem is most often discussed in connection with quantum field theory, it actually does not depend on any field-theoretic ingredients. It therefore applies equally well to nonrelativistic quantum mechanics. For instance, we can choose some spatial region $V$ and let $P$ be a projector onto this region: $P = \int_V \sepia \dd[3]x \sepib \ket{\vect{x}} \bra{\vect{x}}$, where $\ket{\vect{x}}$ are position eigenstates. Hegerfeldt's theorem then implies that any wavefunction that eventually becomes nonzero in some region $V$ must actually be nonzero there for (almost) all time. If this wavefunction at the instant $t = 0$ is confined to a finite region outside of $V$, it must then develop tails to $V$ immediately for $t > 0$. It is therefore not clear how to treat propagating signals that respect relativistic causality. However, these tails may in practice be very small, and we typically do not consider wavefunctions with sharp wavefronts anyway in nonrelativistic quantum mechanics. Instead, we typically use for instance gaussian wavefunctions, which of course always have infinite, decaying tails. After all, quantum mechanics with wavefunctions does not claim to incorporate relativity, meaning that the limitations of Hegerfeldt's theorem are not really problematic (although important to be aware of). 

For quantum field theory however, which is designed to include special relativity and causality with quantum mechanics, Hegerfeldt's theorem does pose a challenge. Still, to apply the theorem, we need to find a specific positive operator that is observable. In Hegerfeldt's own analysis, he assumes the existence of a particle probability operator: an operator $N\lp(V)$ whose expectation value gives the probability of finding a particle in some spatial region $V$. However, such an ideal particle-presence operator would have zero expectation in vacuum, which is impossible for a local observable, as has been pointed out by several authors \cite{kaloyerou1988,buchholz1994,falcone2024}. We also discuss this aspect in \cref{c4_sec_probabilities_of_local_measurements_do_not_vanish}. Therefore, since $N\lp(V)$ is not locally observable, it is not suitable to use it in Hegerfeldt's theorem for establishing limitations on localization of quantum states. 

Yet, what about the energy density? As argued in \cref{c2_sec_particle_and_photon_localization_in_quantum_field_theory}, this is an observable that is local in quantum field theory, meaning that it is measurable in some bounded spacetime region. The raw version (without normal ordering) is formally a positive operator: for instance for a hermitian Klein--Gordon field, the energy density is $\frac{1}{2}\lp[\partial_t\phi\lp(x)]^2 + \frac{1}{2}\lp\big[\nabla\phi\lp(x)]^2 + \frac{1}{2} m^2 \phi^2\sephn\lp(x)$ (a sum of squares), which is clearly positive. Hegerfeldt's theorem then shows that the expectation value of the energy density at any point in space cannot vanish throughout a time interval and become nonzero later, and this applies to \emph{any} state. Does this prove that there can be no states that are strictly localized with respect to the energy density? It turns out that interpreting measurement probabilities in quantum field theory requires some nuance.

\subsection{Interpreting measurement probabilities in quantum field theory}
\label{c2_sec_interpreting_measurement_probabilities_in_quantum_field_theory}
The fact that measurements in quantum field theory need to be interpreted carefully was actually also noticed by \cite{ferretti1968} in connection with the Fermi problem. Ferretti observed that even when considering the full Hamiltonian and local measurements, a nonzero excitation probability will occur for the second atom instantaneously. The reason for this is the effect of vacuum fluctuations, where the interaction Hamiltonian allows an atom to be found in an excited state even if no photons are initially present. Ferretti suggested that the nonzero excitation probability for the second atom when no other sources are present constitutes a \enquote{background signal}. By introducing the first, excited atom into the system, we can obtain its induced \enquote{pure signal} by calculating the second atom's excitation probability and subtracting the background value. Thus the final piece of the Fermi-problem puzzle is that signal propagation is not revealed by looking at raw measurement probabilities alone; rather, we should consider \emph{changes} in the measurement probabilities. 

This point was also noted by \cite{buchholz1994}. They explain in detail why signals in quantum field theory should be interpreted in terms of differences between two states. For some situations this is obvious, for instance when comparing whether two different states produce the same measurement statistics. Crucially, this principle also applies when asking whether a field disturbance is detectable in some region; it must then be compared to a no-signal reference state, which typically is the \emph{presence of vacuum}. This idea of explicitly accounting for vacuum also underlies many of the techniques in quantum optics, where for instance light incident on a beam splitter is modeled with the vacuum state at the empty input. In conclusion:

\begin{remark}\label{c2_rem_state_detectable}\remheadp{}
    A state $\ket{\psi}$ is detectable in quantum field theory if it is distinguishable from vacuum $\ket{0}$; i.e., the probability of some measurement outcome is different for $\ket{\psi}$ than for $\ket{0}$.
\end{remark}

We can see this explicitly if we apply Hegerfeldt's theorem to the energy density operator. Computing the expectation value of the raw (not normal-ordered) energy density for most states results in the value being infinite, for all spacetime arguments. Thus, Hegerfeldt's theorem saying that it is nonzero for all times is not a surprise. What is needed is to compare this value with the expected energy density for vacuum, which also turns out to be infinite. Therefore, in line with \cref{c2_rem_state_detectable}, a nonzero value for the raw energy density is not enough to conclude that a particle is present. 

The canonical way of getting finite values is to normal-order the energy density, which amounts to the same as subtracting its vacuum value. We can also confirm that the expectation value for the normal-ordered energy density for vacuum is in fact zero. Thus, considering the presence of particles by the normal-ordered energy density, as, e.g., \cite{bialynicki-birula1998} does, is meaningful because we are (implicitly) comparing with vacuum. The reason that this new operator does not itself contradict Hegerfeldt's theorem is that the energy density is no longer positive after normal ordering \cite{epstein1965}. This means that Hegerfeldt's theorem does not apply to it, and the expected normal-ordered energy density may indeed be zero in some time interval and nonzero in some other interval. We cover this point in more detail in \cref{c5_sec_post_measurement_states_of_local_measurements}. 

It turns out that this trend of positive, nonzero local operators having nonvanishing vacuum expectation values is in fact universally valid. There is a general statement about local operators in quantum field theory called the Reeh--Schlieder theorem \cite{reeh1961}. We look at this theorem and its consequences in \cref{c4_sec_structure_of_local_observables}, and at more details on how to interpret measurement probabilities in quantum field theory in \cref{c4_sec_interpreting_local_measurement_probabilities}. A corollary of the Reeh--Schlieder theorem is that the probability of any possible outcome of any local measurement in vacuum can never vanish [see \cref{c4_cor_local_probabilities_never_vanish} or \cite{redhead1995}]. Specifically, the probability will always be strictly larger than zero and strictly less than one. This is why comparing with the vacuum value is so important when interpreting results such as Hegerfeldt's theorem. 

A similar theorem often discussed in connection with relativistic particle localization is that by \cite{malament1996}. Malament sets up a list of seemingly reasonable axioms for particle localization projectors in quantum field theory, and shows that the only solution is the zero operator. In other words, such projectors do not exist. However, as we discuss in \cref{c4_sec_connection_to_malaments_theorem}, one of Malament’s axioms requires detections in disjoint spatial regions to be mutually exclusive, whereas two detectors corresponding to local observables can both respond even in vacuum. Malament's requirement is therefore too strong for the projectors to represent measurements performed locally in the respective regions.

\subsection{Knight's and Licht's treatments of strictly localized states}
\label{c2_sec_knights_and_lichts_treatment_of_strictly_localized_states}
Lastly, there is one more work that has gotten little attention but deserves special mention: Knight's 1961 treatment of strictly localized states \cite{knight1961}. Long before Bialynicki-Birula and Hegerfeldt, Knight made the explicit connection between particle localization and analytic properties implied by energy positivity. Before Ferretti, he realized the need for local measurements, and he provided a precise definition of local observables in quantum field theory in terms of field polynomials. A similar definition is used for local operator algebras in algebraic quantum field theory \cite[Ch.~II.4]{haag1996}. We introduce this framework in \cref{c4_sec_local_operators}.

Knight also discusses that measurement probabilities should be interpreted by comparing with vacuum. He therefore defines a set of so-called \enquote{strictly localized states} to some region as states that look like vacuum with respect to all local measurements outside of that region. We develop this idea in detail in \cref{c5_sec_localized_states} and argue why this is a good way to formalize strict localization in quantum field theory.  

In Knight's analysis, he considers a free, hermitian scalar field. He then presents a proof that strictly localized pure states (except vacuum) for this field must contain terms with arbitrarily high particle number. He then shows that coherent states, which are such infinite sums of particle-number terms, can be strictly localized. We discuss optical coherent states in \cref{c7_sec_coherent_states}. Another discussion of Knight's statement can be found in \cite{bievre2006}. 

A continuation of Knight's ideas was done by \cite{licht1963}. Licht showed that for hermitian scalar fields, there is a one-to-one correspondence between strictly localized pure states and certain isometric operators. This is a major breakthrough in the study of strictly localized states, and we expand upon this analysis in \cref{c5_sec_generating_strictly_localized_states}. Licht also generalized some results to mixed states in \cite{licht1966}, which we comment on in \cref{c5_sec_mixed_localised_states}.

\subsection{Summary}
\label{c2_sec_summary}
In summary, there are three key lessons from the literature presented in this section. First, when analyzing particle localization in quantum field theory, it is useful to phrase the question in terms of quantities that are locally measurable. Using this approach, it has been shown that single photons cannot be strictly localized. This might seem problematic in connection with relativistic causality; however, a second principle is to not limit the discussion of localized states to \emph{exact} single-photon states. We will instead consider superpositions of number states, which may include small perturbations of single-photon states. There are other theorems, e.g., by Hegerfeldt, that at first glance seem to imply that strict localization in quantum field theory is incompatible with causality. However, this tension is resolved by a third principle, namely to interpret measurement probabilities by comparing to vacuum.


%% file: 3_local_measurement.tex


\newcommand{\FigThreeOne}{%
    \begin{figure}%
    \hspace*{-5mm}%
    \raisebox{0mm}{%
        \iftoggle{FigFromPDF}{%
            \includegraphics{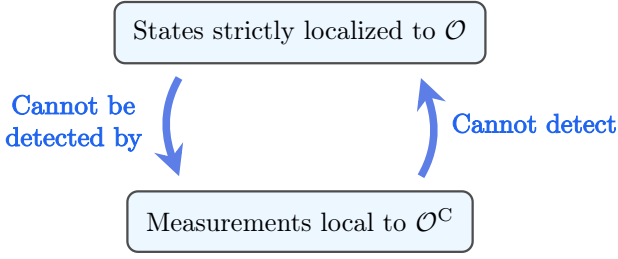}%
        }{%
            \tikzstyle{box} = [%
                rectangle,%
                line width=0.3mm,%
                rounded corners=4,%
                minimum height=0.8cm,%
                text height=0.25cm,%
                text centered,%
                draw=black!70,%
                inner sep=7pt,%
                fill=MyBlueL%
            ]%
            \tikzsetnextfilename{fig3-1}%
            \begin{tikzpicture}%
                \node (1) [box] {States strictly localized to $\mathcal{O}$};%
                \node (2) [box, yshift=-2.5cm] {Measurements local to $\compl{\mathcal{O}}$};%
                \draw [big_arrow] (-1.6, -0.6) to [out=-120, in=110] (-1.6, -1.9);%
                \draw [big_arrow] ( 1.6, -1.9) to [out=60, in=-70]   ( 1.6, -0.6);%
                \node at (-3.0, -1.2)   (3) [text width=25mm, text centered, color=MyBlueMC] {\sboldtext{Cannot be detected by}};%
                \node at ( 3.1, -1.2) (4) [color=MyBlueMC] {\sboldtext{Cannot detect}};%
            \end{tikzpicture}%
        }%
    }%
    \caption{\label{c3_fig_localized_states_and_local_measurements}Potentially circular definition of strictly localized states and local measurements.}%
    \end{figure}%
}

\newcommand{\FigThreeTwo}{%
    \begin{figure}%
    \hspace*{0mm}%
    \raisebox{0mm}{%
        \iftoggle{FigFromPDF}{%
            \includegraphics{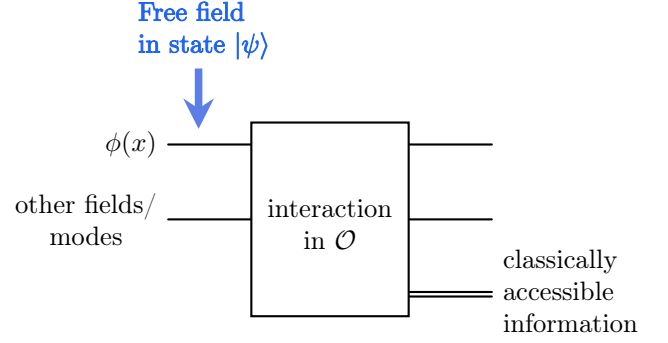}%
        }{%
            \tikzsetnextfilename{fig3-2}%
            \begin{tikzpicture}%
                \node{%
                    \begin{quantikz}[column sep=1.1cm, row sep=0.7cm]%
                        \lstick{$\phi\lp(x)$} %
                        & %
                        \gate[3]{%
                            \begin{minipage}{18mm}%
                                \centering%
                                \vspace{3mm}%
                                \smash{\parbox{18mm}{\centering interaction\\in $\mathcal{O}$}}%
                            \end{minipage}%
                        } %
                        & %
                        \\ %
                        %
                        \lstick{other fields/\\modes} %
                        & %
                        & %
                        \\ %
                        %
                        & %
                        \wireoverride{n} %
                        & %
                        \setwiretype{c} \rstick[label style={text width=20mm}]{%
                            \vspace{-5.5mm}%
                            \begin{flushleft}classically\\accessible\\information\end{flushleft}%
                        }%
                    \end{quantikz}%
                };%
                \draw[big_arrow] (-1.8, 2.1) to (-1.8, 1.3);%
                \node at (-0.6, 2.6) [color=MyBlueMC] {%
                    \parbox{40mm}{\begin{flushleft}\sboldtext{Free field\\in state $\ket{\psi}$}\end{flushleft}}%
                };%
            \end{tikzpicture}%
        }%
    }%
    \caption{\label{c3_fig_measurement_process}A quantum circuit model of a physical process that implements a measurement of a field $\phi\lp(x)$ in some spacetime region $\mathcal{O}$.}%
    \end{figure}%
}

\newcommand{\FigThreeThree}{%
    \begin{figure}%
    \hspace*{0mm}%
    \raisebox{0mm}{%
        \iftoggle{FigFromPDF}{%
            \includegraphics{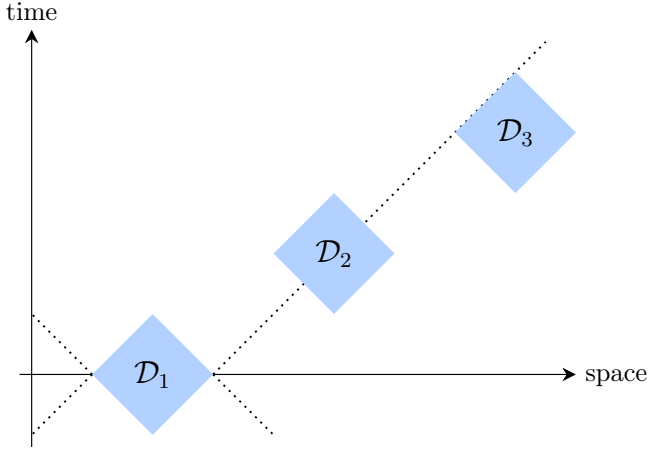}%
        }{%
            \tikzsetnextfilename{fig3-3}%
            \begin{tikzpicture}[scale=0.8]%
                \draw[thick, dotted, line width=0.7pt] (-1, 0) -- (-2,   1);%
                \draw[thick, dotted, line width=0.7pt] ( 1, 0) -- ( 6.5, 5.5);%
                \draw[thick, dotted, line width=0.7pt] (-1, 0) -- (-2,  -1);%
                \draw[thick, dotted, line width=0.7pt] ( 1, 0) -- ( 2,  -1);%
                \draw[axis_arrow] (-2.2,  0.0) -- ( 7.0, 0.0) node[right, yshift=-.4mm] {space};%
                \draw[axis_arrow] (-2.0, -1.2) -- (-2.0, 5.7) node[above] {time};%
                \fill[MyBlueML] (1, 0) -- (0, 1) -- (-1, 0) -- (0, -1) -- (1, 0);%
                \fill[MyBlueML] (7, 4) -- (6, 5) -- ( 5, 4) -- (6,  3) -- (7, 4);%
                \fill[MyBlueML] (4, 2) -- (3, 3) -- ( 2, 2) -- (3,  1) -- (4, 2);%
                \node at (0, 0) {\large$\mathcal{D}_1$};%
                \node at (6, 4) {\large$\mathcal{D}_3$};%
                \node at (3, 2) {\large$\mathcal{D}_2$};%
            \end{tikzpicture}%
        }%
    }%
    \caption{\label{c3_fig_sorkin}A Sorkin-type scenario with three local measurements in extended regions $\mathcal{D}_1$, $\mathcal{D}_2$, and $\mathcal{D}_3$. The regions $\mathcal{D}_1$ and $\mathcal{D}_3$ are spacelike separated, whereas $\mathcal{D}_2$ partially overlaps with the lightcones of both $\mathcal{D}_1$ and $\mathcal{D}_3$.}%
    \end{figure}%
}

\iftoggle{ExportTikz}{%
    \FigThreeOne
    \FigThreeTwo
    \FigThreeThree
     
}{}

\section{Local measurement}
\label{c3_sec_local_measurement}
We saw in \cref{c2_sec_history} that analyzing localization properties of relativistic particles has been attempted by many different approaches. We also claimed that methods based on \emph{locally measurable} quantities, such as the energy density, seem to be the most useful, in contrast to approaches inspired by nonrelativistic quantum mechanics. However, analyzing localization through the energy density is quite indirect and different from how localization is formulated in point-particle classical mechanics or in nonrelativistic quantum mechanics. These theories have a notion of \emph{position}, either as a simple position parameter in classical mechanics, or as a position observable in quantum mechanics. Do these traditional concepts of localization not apply to relativistic particles? What does it really mean for a particle to be localized? 

In this section we explore this question, to motivate why local measurement is a natural framework for defining localization for relativistic particles. We then discuss what is locally measurable in quantum field theory, as well as what a measurement process is physically. We take a conservative approach for characterizing the set of all physically realizable local observables through inner and outer bounding sets: we assume that it contains at least the set of all pointwise field observables (as a limit of narrowly smeared fields), and we assume that it is contained in the local operator algebra of smeared fields (see \cref{c4_sec_local_operators}), as defined in \cite{knight1961} and algebraic quantum field theory \cite[Chs.~II.4 and III.1]{haag1996}. We also discuss some limitations of measurement models based on local algebras, such as Sorkin's scenario \cite{sorkin1993}. Finally, we describe how some local measurements are treated in quantum optics.

\subsection{What does it mean for a particle to be localized?}
\label{c3_sec_what_does_it_mean_for_a_particle_to_be_localized}
In classical point-particle mechanics, a particle is assumed to have a definite position, and an idealized position measurement can be regarded as revealing this value. In that sense, different manifestations of localization---where a particle is defined to be, where an ideal position measurement would find it, and where it can undergo contact interaction---all coincide. In quantum mechanics, this automatic correspondence is broken: one must distinguish between the state of a system and the outcomes of measurements. In general, the quantum state determines only a probability distribution for the outcomes of a specified measurement procedure. It is therefore necessary to formalize the notion of localization for particles in a more nuanced way.

\subsubsection{Wavefunctions, position operators, and position eigenstates}
\label{c3_sec_wavefunctions_position_operators_and_position_eigenstates}
In the wavefunction formalism of quantum mechanics, a single particle in a pure state is characterized by a wavefunction $\Psi\lp(\vect{x}, t)$. In this representation, localization in the sense of where the particle is defined to be, is reflected in where the wavefunction $\Psi\lp(\vect{x}, t)$ has support. On the other hand, when conducting an idealized position measurement of this particle, a single value $\vect{x}$ is observed, selected at random with a probability density of $\abs{\Psi\lp(\vect{x}, t)}^2$. Similarly, the expected value of a position measurement is $\mint{\dd[3] x} \vect{x} \abs{\Psi\lp(\vect{x}, t)}^2$, and so on for higher moments. 

In general in nonrelativistic quantum mechanics, position measurements are modeled by a position observable $\hat{\vect{x}}$ [see, e.g., \cite[Ch.~1.6]{sakurai1994}]. The expected value of a position measurement for some quantum state $\ket{\psi}$ is then calculated from this observable as $\qexv{\psi}{\hat{\vect{x}}}$. The observable $\hat{\vect{x}}$ has (generalized) eigenstates $\ket{\vect{x}}$ that provide a (generalized) basis for the Hilbert space. An ideal position measurement yields an observed result $\vect{x}_0$ and leaves the particle in a position eigenstate $\ket{\vect{x}_0}$. The statistics of the same measurement performed immediately after are calculated as $\qexv{\vect{x}_0}{\hat{\vect{x}}}$, and so on, which show that the state $\ket{\vect{x}}$ corresponds well with the intuitive concept of something localized at $\vect{x}$.

We thus have several avenues for inferring localization properties of nonrelativistic quantum particles. Unlike in classical mechanics, there is a difference between where we define a particle to be and the result of measuring where it is. However, this difference is not particularly jarring. The larger the value of $\abs{\Psi\lp(\vect{x}, t)}^2$ at $\vect{x}$, the more likely the particle is to be found at $\vect{x}$. Furthermore, many classical quantities have natural quantum observables that we can measure and observe the statistics of in the quantum theory. This includes for instance position, which has a well-defined, canonical observable $\hat{\vect{x}}$. 

What is then the situation in quantum field theory? As discussed in \cref{c2_sec_history}, a large part of the history of investigations of particle localization in quantum field theory has consisted of trying to generalize the above nonrelativistic localization schemes to relativistic particles. For instance, numerous works have studied localization of photons using different types of wavefunction formalisms, or attempted various constructions of position operators or position eigenstates for relativistic quantum particles. 

The findings of these analyses are generally that for relativistic systems, these constructions do not retain all the properties we expect from nonrelativistic quantum mechanics while satisfying relativistic requirements. For instance, natural choices for sharp position eigenstates for relativistic particles either fail to be orthogonal or fail to preserve localization under Lorentz boosts \citetext{\citealp{newton1949}; \citealp[Ch.~4]{teller1995}}. Similarly, natural extensions of position operators to relativistic particles exhibit difficulties such as failure of Lorentz covariance, noncommuting $x$, $y$, and $z$ components, or noncanonical commutation relations with momentum \cite{jordan1963,mourad1993,shojai1997}. And the problems with a particle position probability interpretation of Klein--Gordon wavefunctions have been well-known for a long time \cite[Ch.~4.6]{bohm1951}.

That is not to say that these localization schemes cannot in principle be defined for relativistic quantum systems. However, in quantum field theory, wavefunctions and position operators give way to fields and spacetime parameters. From this perspective, the difficulties encountered by the nonrelativistic notions of localization suggest a limitation of their starting point, rather than a need for more elaborate constructions.

\subsubsection{Indirect indications of localization}
\label{c3_sec_indicrect_indications_of_localization}
In contrast, some authors such as \citet{knight1961,haag1965} have approached this problem from a completely different angle, and asked: what is it that is \emph{locally} measurable in quantum field theory in the first place? That is, what can we measure in some bounded region of space and time? This idea might not seem so different from the preceding approaches, but the point is that measuring, e.g., field strength or energy density yields a value of that quantity, not a particle position. Therefore, localization is characterized through the effects of the state on such measurement statistics, rather than through a particle-position observable.

This is a more indirect notion of particle localization than in classical mechanics or nonrelativistic quantum mechanics. It suggests that localization in quantum field theory is not best formulated in terms of finding \emph{the} correct particle localization operator, as particles themselves can be created and destroyed dynamically. Instead, what we have are fields and changing excitations of these fields, which show up by making some local measurement results more likely, and others less likely. It is these differences that are the observable effects of particles in quantum field theory.

One may argue that this line of reasoning underplays the existence of nonrelativistic regimes. After all, other relativistic particles such as electrons are also fundamentally described by quantum fields, but for nonrelativistic energies, their localization can nevertheless be accurately analyzed in terms of wavefunctions. However, there is a practical difference between these particles in that photons are created and destroyed even in everyday life, not only in high-energy physics. While massive particles are well described as single-particle states in many situations, photonic states routinely come in all versions: multimode coherent states, (displaced) squeezed states, Fock states, etc. A general theory of localization for light must therefore work for all such states, meaning that localization of photons must necessarily be analyzed in the field-theoretic setting. 

To be clear, the point here is not to criticize the studies based on position operators or wavefunctions, but rather point out that many approaches to relativistic particle localization have been done with a nonrelativistic starting point. This often entails starting with some familiar quantity from classical mechanics or nonrelativistic quantum mechanics and seeking an analogue that could apply to relativistic systems. Then, one tries to work out whether this new quantity has reasonable properties and whether it could still correspond to anything locally measurable in quantum field theory. Knight, on the other hand, takes the opposite approach, starting by asking what is locally measurable and then working out the consequences. The advantage of this approach is that the localization analysis gets rooted in a physically motivated assumption about what is locally measurable. Of course it might or might not turn out that some nonrelativistic operators still have close relativistic analogues under this new measurement theory. Still, the main idea is to not \emph{start out} ahead of time assuming that this is the case, but rather see what consequences a given theory of measurement has.

\subsubsection{Measurability}
\label{c3_sec_measurability}
The role of measurement for defining localization is also worth reflecting on. Imagine for example that a local decision is made to produce some particle-like excitation by acting in some localized spacetime region. What would it mean if the particle's wavefunction were found to spread out outside the future lightcone of this region? This could certainly be somewhat surprising, but unless we can connect it to observations, it does not directly violate relativistic causality. After all, there can be features in physics that exhibit superluminal behavior, such as the electromagnetic scalar potential in Coulomb gauge or a wave's group velocity in certain dispersive media, without allowing superluminal transfer of information. 

In contrast, physical features that do transmit information are according to relativity limited to the speed of light. Therefore, if a local choice of producing a particle-like excitation would lead to observable changes outside the lightcone, this would violate relativistic causality. Said in another way, purely formal localization concepts may or may not have causality as an optional desired property, whereas for approaches based on measurable outcomes, it is a necessary consistency requirement.

\subsection{What is locally measurable in quantum field theory?}
\label{c3_sec_what_is_locally_measurable_in_quantum_field_theory}
To formalize what is meant by particle localization, we therefore want to start by determining what it is that is locally measurable in quantum field theory. Note however that even with a theory of measurement in place, there could be many different ways to define what should be meant by localization for quantum states. The idea of localization that we have in mind has as a main goal to be able to describe localized \emph{information} contained in the quantum fields, that propagates according to relativistic causality. As such, the most natural definition for our purposes is to consider a state to be strictly localized to some spacetime region $\mathcal{O}$ if and only if it is \emph{undetectable} by local measurements outside $\mathcal{O}$.

\FigThreeOne

For this choice of definition, there is potentially a circular way to argue: Let $\mathcal{O}$ be some spacetime region and $\compl{\mathcal{O}}$ its complement region. We then say that states strictly localized to $\mathcal{O}$ are precisely those that cannot be detected by any local measurements in $\compl{\mathcal{O}}$. Conversely, we could try defining local measurements to $\compl{\mathcal{O}}$ as the measurements that cannot detect any states strictly localized to $\mathcal{O}$ (see \cref{c3_fig_localized_states_and_local_measurements}).

This circular definition clearly needs a starting point. However, it suggests that particle localization could potentially be approached by two different paths: A usual way is to start with axioms determining what can be locally measured and then derive results about strictly localized states. Alternatively, one could try to start with axioms specifying what states are strictly localized and use these to reconstruct the local measurement structure. This second approach could be interesting for studying localization for, e.g., fermionic fields or curved spacetime. Here we take the usual approach of starting by defining local measurement.

\subsubsection{What is a measurement?}
\label{c3_sec_what_is_a_measurement}

\FigThreeTwo

To begin, what do we really mean by a measurement? Figuring out the exact physical process that takes place during a quantum measurement is known as the measurement problem in quantum mechanics \cite{hance2022} and is unfortunately not solved today. Nevertheless, we can give a black-box type of description of a possible measurement process, as outlined in \cref{c3_fig_measurement_process}. There is some quantum field of interest $\phi\lp(x)$ that we assume is initially in a pure state $\ket{\psi}$. This represents some incoming signal that is sufficiently stable and isolated, such as a light pulse traveling through empty space. As such, we assume that interaction effects on the signal are negligible at first, so that the signal's time evolution is accurately described by a free field. It is then convenient to use the Heisenberg picture where the free time evolution is carried by the field operators. 

Located somewhere far from the incoming signal, there is some measurement apparatus that is initially decoupled and uncorrelated with the signal's degrees of freedom. This apparatus may in reality also contain excited modes of the field $\phi\lp(x)$. For instance when measuring an electromagnetic signal, we may use some electromagnetic apparatus, which consists of many electrically charged particles that themselves are interacting via the electromagnetic field. For simplicity, we assume that any such excitations occupy modes orthogonal to those carrying the incoming signal, so that the latter may be described independently by the state $\ket{\psi}$. 

In some region of space and time $\mathcal{O}$, an interaction takes place between $\phi\lp(x)$ and some other fields or modes making up the measurement apparatus. This can for example be when an incoming light pulse hits a photon detector. Leaving the fundamental account of quantum measurement open, we want to assume as little as possible about the exact time evolution of this interaction, other than that it takes place in the region $\mathcal{O}$. For example, if the dynamics predicted by the Standard Model is a complete description of physics, this time evolution must be unitary (on the whole, global system). Alternatively, the time evolution may involve some unknown, not yet understood physics, including a continuously stochastic process, nonlinear quantum effects due to gravity, or similar. Whatever this unknown time evolution is, we know that after some time, certain information about $\ket{\psi}$ ends up accessible as classical information, which is copyable and stable. This is the measurement result. 

One might question whether a measurement apparatus, consisting of elementary particles, must itself fail to be strictly localized. However, even though single particles cannot be strictly localized, there \emph{are} strictly localized states in quantum field theory (see \cref{c5_sec_localized_states}), so a strictly localized apparatus is not necessarily impossible. More importantly, we may take a pragmatic view where it is not the apparatus itself but rather its coupling to the signal that is effectively confined. In such a model, we idealize this confinement to $\mathcal{O}$ as exact.

This description is quite general, but it does motivate the use of free fields. In reality, the interactions between various quantum fields in nature are always present, and can as such never be fully \enquote{switched off}. Nevertheless, it can still make sense to describe the incoming signal by a free field, while retaining the local coupling to the apparatus as part of the measurement process. We therefore choose to focus on this simplified starting point: what is locally measurable in a free field theory, under local couplings to an apparatus? Future work may extend these results to fully interacting theories, including renormalization issues that arise in more detailed detector-field interactions.

On the other hand, since we do not assume much detail about the time evolution in $\mathcal{O}$, the model in \cref{c3_fig_measurement_process} does not determine which measurements can actually be implemented in $\mathcal{O}$. This is of course a common problem in quantum theories, in that the set of observables associated with actually implementable measurements is not fixed by the system dynamics alone, but requires additional physical input. This issue is particularly challenging in quantum field theory since there is limited direct experimental access to measurements that are simultaneously highly local and highly energetic. After all, many experiments in particle physics are described in terms of asymptotic scattering observables that do not directly reveal which field observables can be measured within small spacetime regions.

\subsubsection{Pointwise local observables}
\label{c3_sec_pointwise_local_observables}
We therefore go back to first principles. Bohr and Rosenfeld argued that the basic observable in quantum field theory is the field itself \cite{bohr1933}. Following this viewpoint, we take local observables at a spacetime point $x$ for a free, hermitian scalar field $\phi\lp(x)$ (no gauge symmetry) to be given by:
\begin{equation}\label{c3_observables}
\begin{aligned}
    &q_0, \ \phi\lp(x), \ \phi^2\sephn\lp(x), \ \dotsc \\
    &\quad = \tn{constant}, \ \tn{field strength}, \\
    &\quad \sepeq \tn{quadratic fields (energy density)}, \ \dotsc, 
\end{aligned}
\end{equation}
as well as sums, products, and derivatives of such expressions. This includes for instance time derivatives $\dot\phi\lp(x)$, space derivatives $\nabla\phi\lp(x)$, normal-ordered expressions $\norder{\phi^2\sephn\lp(x)} = \phi^2\sephn\lp(x) - \text{const.}$, and so on. 

Note that even though textbooks on quantum field theory rarely discuss explicitly what is locally measurable in detail, it often enters implicitly. Namely, most textbooks introduce correlation functions $\exv{\phi\lp(x) \phi\lp(y) \dotsb}$ as the basic building blocks for computing observable quantities [see, e.g., \cite[Ch.~4.6]{peskin1995}]. This reflects the same field-based description: expectation values of polynomial field observables, including expressions involving two or more spacetime points, are determined by these correlation functions. This was also the implicit definition of local measurement we used in \cref{c1_sec_invitation} when we demonstrated the expected electric energy density of single-photon states cannot be strictly localized. 

Furthermore, a standard calculation in most textbooks on quantum field theory is to verify that the Klein--Gordon field $\phi\lp(x)$ satisfies microcausality \cite[Ch.~2.4]{peskin1995}:
\begin{equation}\label{c3_comm_phi_x_phi_y}
    \comm{\phi\lp(x)}{\phi\lp(\splike{x})} = 0, \sepc \forall \seph \tn{spacelike $x$ and $\splike{x}$}.
\end{equation}
The argument for causality is that this implies that spacelike separated measurements cannot influence each other. This argument again clearly implicitly assumes that local observables at $x$ are constructed from the field $\phi\lp(x)$. 

The local measurability of the field also makes sense abstractly, since it is the field that coordinates the spacetime dependence of operators. In quantum mechanics, observables are represented by hermitian operators, and in gauge theories they must also be gauge invariant. Quantum field theory should then, as a refinement of quantum mechanics, tie these operators to spacetime, which is precisely what the field operator $\phi\lp(x)$ does. As such, it makes sense that the local observables at point $x$ are functions of the field $\phi\lp(x)$, whereas the choice on where and when to measure is reflected in its spacetime argument.

\subsubsection{Extended local observables}
\label{c3_sec_extended_local_observables}
One weakness of limiting ourselves to the expressions \eqref{c3_observables} is that they only involve pointwise local observables. Clearly it is possible to let the measurement apparatus interact with the system or particle in an extended region of space and time. Keeping in mind the dynamics of quantum measurement, it is conceivable that such an extended, coherent measurement process cannot be described by a sequence of individual, pointwise measurements. Rather, a local observable in some extended region of spacetime $\mathcal{O}$ is in general described by \emph{smeared} field expressions, for example
\begin{equation}\label{c3_observables_smeared}
    \int_{\mathcal{O}} \sepia \dd[4] x \sepib f\lp(x) \phi\lp(x),
\end{equation}
for some real smearing function $f\lp(x)$ that has support only in $\mathcal{O}$. In \cref{c4_sec_local_operators}, we introduce local operator algebras generated by such smeared fields, which provide a systematic framework for studying local observables.

For now, one thing we can notice from the example \eqref{c3_observables_smeared} is that the types of observables we can generate depend on what smearing functions $f\lp(x)$ we consider. For example, restricting the smearing functions to delta functions recovers the pointwise observables \eqref{c3_observables}, whereas allowing extended functions produces a larger class of possible local observables. Thus, even when assuming that pointwise observables are indeed given by field expressions \eqref{c3_observables}, a remaining question is what smearing functions $f\lp(x)$ correspond to physically realizable local observables in some extended spacetime region $\mathcal{O}$.

An early characterization of what functions $f\lp(x)$ are permitted was given by \cite{knight1961}. Knight's choice was in a way the simplest one, assuming that \emph{any} function $f\lp(x)$ with support in $\mathcal{O}$ in fact corresponds to some possible local observable. The same choice of letting $f\lp(x)$ be an arbitrary (sufficiently nice) function in $\mathcal{O}$ is also used for describing local observables through local operator algebras in algebraic quantum field theory \citetext{\citealp{haag1962}; \citealp[Chs.~II.4 and III.1]{haag1996}}. 

The logic behind this definition is that it is still true that spacelike separated measurements commute: since $\comm{\phi\lp(x)}{\phi\lp(\splike{x})}$ vanishes for spacelike $x$ and $\splike{x}$, it follows that the commutator between any smeared local observables, e.g., 
\begin{equation}\label{c3_comm_phi_f_phi_g}
\begin{aligned}
    &\comm{\int_{\mathcal{O}} \sepia \dd[4] x \sepib f\lp(x) \phi\lp(x)}{\int_{\splike{\mathcal{O}}} \sepia \dd[4] \splike{x} \sepib g\lp(\splike{x}) \phi\lp(\splike{x})} = 0, \\[1.5mm]
    &\; \forall \seph \tn{spacelike $\mathcal{O}$ and $\splike{\mathcal{O}}$}.
\end{aligned}
\end{equation}
That is, any two smeared field observables commute when their smearing functions $f\lp(x)$ and $g\lp(x)$ have spacelike separated supports. Thus, in the same way that \cref{c3_comm_phi_x_phi_y} ensures that any measurements of two pointwise local observables cannot influence each other when spacelike separated, \cref{c3_comm_phi_f_phi_g} implies that any measurements of two extended local observables cannot influence each other when spacelike separated. This is a good internal consistency check of Knight's characterization of local measurement. 

That being said, Knight's characterization has some drawbacks. For one thing, it does not describe how the measurements may actually be performed experimentally. A procedure for measuring certain simple observables is of course not difficult to imagine; e.g., we can measure a smeared electric field observable by looking at the dynamics of some electric test charge in some region. However, Knight's characterization does not give a general prescription for how to perform arbitrary local measurements of the form \eqref{c3_observables_smeared}.

\subsubsection{Sorkin's scenario}
\label{c3_sec_sorkins_scenario}
There is also a somewhat hidden conceptual difference between pointwise observables commuting and extended observables commuting. Namely, when two pointwise observables commute for spacelike separations, as, e.g., $\phi\lp(x)$ does, it means that any sequence of pointwise observables cannot exert influence outside the lightcones. However, Sorkin pointed out that the same does not seem to automatically be true for measurements associated with extended spacetime regions \cite{sorkin1993}. Specifically, what happens when we have two smeared observables in regions $\mathcal{D}_1$ and $\mathcal{D}_3$ that are spacelike separated, but we conduct a third measurement in $\mathcal{D}_2$ which partially overlaps with the lightcones of both $\mathcal{D}_1$ and $\mathcal{D}_3$, as depicted in \cref{c3_fig_sorkin}? Sorkin\footnote{\label{c3_foot_sorkins_setup}The setup in \cite{sorkin1993} assumes a local unitary process in $\mathcal{D}_1$ and local measurements in $\mathcal{D}_2$ and $\mathcal{D}_3$, and the region $\mathcal{D}_2$ is taken to be all of space (but finite in time).} argues that applying Lüders post-measurement rule to observables associated with extended spacetime regions can lead to superluminal signaling \cite{sorkin1993}. 

Note that care must be taken here to not consider signaling from $\mathcal{D}_1$ to $\mathcal{D}_3$ that depends on knowledge about specific outcomes of measurements in $\mathcal{D}_1$ or $\mathcal{D}_2$. This is because we have to assume that knowledge about such outcome values is not necessarily available to an observer in $\mathcal{D}_3$. Thus the relevant calculation is finding the nonselective post-measurement state from the interactions in $\mathcal{D}_1$ and $\mathcal{D}_2$ averaged over the measurement outcomes. Sorkin argues that this average state gives different measurement statistics in $\mathcal{D}_3$ depending on the choice of interaction in $\mathcal{D}_1$, which implies superluminal signaling from $\mathcal{D}_1$ to $\mathcal{D}_3$. 

\FigThreeThree

There is an ongoing discussion about the physical implementability and causality of measurement prescriptions of Sorkin's scenario, and about measurement in quantum field theory at large \cite{beckman2001,borsten2021,bostelmann2021,ramon2021,jubb2022,albertini2023,papageorgiou2024,oeckl2025}. To address some of these issues and to handle more general situations (such as curved spacetime), explicit system-probe models have been developed. This includes for instance Unruh--DeWitt detectors \cite{unruh1976,dewitt1979,martinmartinez2015,polo-gomez2022} and the Fewster--Verch measurement framework \cite{fewster2020,bostelmann2021,fewster2023a,mandrysch2025}. These models take the general process description in \cref{c3_fig_measurement_process}, and introduce a more fine-grained description of the external degrees of freedom, their coupling to the system, and their readout. This external system can for example be a two-level atom or a single quantum field, which is treated as a probe that extracts some information from the primary field $\phi\lp(x)$ by a local interaction.

\subsubsection{Physically realizable local measurements}
\label{c3_sec_physically_realizable_local_measurements}
The precise details of physically realizable local measurements in quantum field theory are still debated and may turn out to not be fully resolved until the quantum measurement problem is. We will not attempt to solve this issue here, but at the same time, a concrete definition of local measurement is needed for characterizing strictly localized states. We therefore take a pragmatic, conservative approach for describing the set of all physically realizable local observables by inner and outer bounding sets: We assume that the pointwise field observables \eqref{c3_observables} in $\mathcal{O}$ form an inner set, meaning that they correspond to some, but not necessarily all, local observables physically realizable in $\mathcal{O}$. We then assume that observables in the local operator algebra of smeared fields in $\mathcal{O}$, as defined in \cite{knight1961} and algebraic quantum field theory \cite[Chs.~II.4 and III.1]{haag1996}, form an outer set. In other words, we assume that the local operator algebra contains all local observables physically realizable in $\mathcal{O}$, although it might also contain observables that are not locally realizable.

Note that there is a nuance in that truly pointwise observations are not really achievable, as any measurement apparatus must have some finite extent. Nevertheless, we assume that this finite extent may be taken as small as desired by a sequence of measurements with progressively finer resolution. Therefore, if, e.g., some pointwise local observable does not vanish for any argument in some small neighborhood, there will be some smeared observable in that neighborhood that does not vanish either. 

In practice this means that for proving that some state \emph{is} strictly localized to $\mathcal{O}$, we show that it is indistinguishable from vacuum with respect to all observables in the full, local operator algebra to $\compl{\mathcal{O}}$. By our assumptions, any such state is then at least localizable with respect to all local observables physically realizable in $\compl{\mathcal{O}}$. On the other hand, for proving that some state \emph{cannot} be strictly localized to $\mathcal{O}$, we show that its expectation value differs from that of vacuum for some specific pointwise observable local to $\compl{\mathcal{O}}$. For instance for photons, this can be the normal-ordered electromagnetic energy density at some spacetime point. 

One exception is the discussion of on-demand sources and strictly localized states in \cref{c6_sec_on_demand_sources}, where we specialize to assuming that all observables in the local algebra correspond to physically realizable local measurements. This leads to the conclusion that any localized on-demand source must produce strictly localized states. Note first that this conclusion can also follow from weaker measurement assumptions or suitable locality properties of the source dynamics, as discussed there. Second, the analysis only assumes physical realizability of the local algebra in terms of measurement statistics and does not use any particular post-measurement rule. In particular, it does not assume that measuring every local observable induces the corresponding Lüders state update. Therefore, the difficulty with causality in Sorkin's scenario discussed in \cref{c3_sec_sorkins_scenario} does not itself rule out this assumption, as other, causal measurement implementations could reproduce the same measurement statistics.

\subsection{Measurements in quantum optics}
\label{c3_sec_loc_meas_quantum_optics}
As mentioned, most experiments in modern particle physics deal with scattering experiments. These experiments are modeled by cross sections and scattering matrices, which are inherently nonlocal in that they involve asymptotic in/out states. In quantum optics, on the other hand, it is more natural to consider local measurements, since they often can roughly be viewed as measurements of the electromagnetic fields, or powers or correlations of these fields.

\subsubsection{Glauber's intensity operator}
However, some commonly used observables to describe such measurements need not be strictly local. A common example is Glauber's correlation functions, which are often used to describe local optical measurements \citetext{\citealp{glauber1963}; \citealp[Ch.~4.12]{loudon2000}}. Strictly speaking, these expressions are not local \cite{bykov1989,franson2008,plimak2011}. To see this, consider the intensity operator $E^-\lp(t) E^+\lp(t)$ from Glauber's detection theory \citetext{\citealp{glauber1963}; \citealp[Ch.~4.11]{loudon2000}}, where $E^+\lp(t)$ and $E^-\lp(t)$ are the positive and negative frequency parts of the electric field $E\lp(t)$ at a fixed point in space. Being the analytic signal of the field, $E^+\lp(t)$ is determined from $E\lp(t)$ using the Hilbert transform. Similarly, $E^-\lp(t)$ is determined from the field with the Hilbert transform. Thus the intensity operator is nonlocally related to the field $E\lp(t)$. Alternatively, the nonlocality can be viewed as an artifact resulting from the use of the rotating-wave approximation \cite{franson2008}. 

The situation is completely analogous to the quasi-monochromatic assumption in the derivation of the classical intensity. Writing the electric field as
\begin{equation}
    E\lp(t)=\Re\lp[\hat E\lp(t)e^{-i\omega_0 t}] = \frac{1}{2}\lp[\hat E\lp(t)e^{-i\omega_0 t} + \hat E^*\sephn\lp(t)e^{i\omega_0 t}]
\end{equation}
with envelope $\hat E\lp(t)$ and central frequency $\omega_0$, the electric energy density is
\begin{equation}
    \frac{1}{2}E^2\lp(t) = \frac{1}{8} \lp[\hat E^2\lp(t)e^{-2i\omega_0 t} + {\hat E^{*2}}\lp(t)e^{2i\omega_0 t} + 2\hat E^*\sephn\lp(t)\hat E\lp(t) ].
\end{equation}
In the quasi-monochromatic case where the envelope $\hat E\lp(t)$ varies slowly over $2\pi/\omega_0$, we can average over time (which is a nonlocal operation) such that the first two terms are washed out. This leads to a result proportional to $\hat E^*\sephn\lp(t)\hat E\lp(t)$ or $E^-\lp(t) E^+\lp(t)$. For ultrashort pulses of the order of a few cycles, one should retain all terms, keeping the original expression $\frac{1}{2}E^2\lp(t)$, which clearly is local.

The error in approximating the local energy density by Glauber’s intensity operator depends on the exact state, since the state determines to what extent the different frequency components of the operator contribute to the expectation value. As already seen in \cref{c1_sec_invitation}, coherent states can describe compactly supported field disturbances. For the coherent state $\ket{\alpha_\xi}$ we had in \cref{c1_qexv_alpha_xi_E_t}:
\begin{equation}\label{c3_qexv_alpha_xi_E_t}
    \qexv{\alpha_\xi}{\norder{E^2\sephn\lp(t)}} = e^2\sephn\lp(t),
\end{equation}
where
\begin{equation}\label{c3_e_t_coherent}
    e\lp(t) \equiv \qexv{\alpha_\xi}{E\lp(t)}
    = \mintn{\dd \omega} \mathcal{E}\lp(\omega) Z\lp(\omega) e^{- i \omega t},
\end{equation}
and
\begin{equation}\label{c3_Z_w}
    Z\lp(\omega) = 
    \begin{cases}
        \alpha \xi\lp(\omega), & \omega > 0, \\
        \alpha^* \xi^*\sephn\lp(- \omega), & \omega < 0.
    \end{cases}
\end{equation}
Here $\xi\lp(\omega)$ may be chosen such that $e\lp(t)$ is confined to a finite time interval $\mathcal{O}$. In particular, assume that $e\lp(t)=0$ outside $\mathcal{O}$, where $\mathcal{O}$ has duration $T$. 

For this coherent state, we may quantify the nonlocality of the Glauber intensity operator by the fraction of signal predicted outside the actual pulse,
\begin{equation}\label{c3_epsilon_glauber}
    \varepsilon_{\mathrm{Gl}}
    =
    \frac{\displaystyle \int_{\compl{\mathcal{O}}} \sepia \dd t \sepib
    \qexv{\alpha_\xi}{E^-\sephn\lp(t) E^+\sephn\lp(t)}}
    {\displaystyle \mintn{\dd t} 
    \qexv{\alpha_\xi}{E^-\sephn\lp(t) E^+\sephn\lp(t)}}.
\end{equation}
To calculate this leakage, we use
\begin{equation}\label{c3_qexv_glauber_coherent}
    \qexv{\alpha_\xi}{E^-\sephn\lp(t) E^+\sephn\lp(t)}
    = \abs{e^+\sephn\lp(t)}^2,
\end{equation}
where
\begin{equation}\label{c3_eplus}
    e^+\sephn\lp(t)
    = \alpha\minti{\dd \omega} \mathcal{E}\lp(\omega) \xi\lp(\omega) e^{- i \omega t}
\end{equation}
is the positive-frequency part of $e\lp(t)$. Although $e\lp(t)$ vanishes outside $\mathcal{O}$, the positive-frequency part $e^+\sephn\lp(t)$ has infinite tails, since the negative frequencies are removed in \cref{c3_eplus}.

For concreteness, consider a compact pulse of the form $e\lp(t) = w\lp(t/T)\exp\lp(-i\omega_0 t) + w^*\sephn\lp(t/T)\exp\lp(i\omega_0 t)$, where $w\lp(t/T)$ is an envelope chosen such that the resulting coherent state is normalized, and $T$ sets the pulse duration. Its spectrum consists of two shifted envelope spectra, centered at $\pm\omega_0$. Since the spectral width of the envelope is of order $1/T$, the relevant parameter is $\omega_0 T$, which measures the number of optical cycles in the pulse. For $\omega_0 T \gg 1$, the two spectral components are concentrated well away from $\omega=0$. The deviation of $e^+\lp(t)$ from $w\lp(t/T)\exp\lp(-i\omega_0 t)$ is due to the small parts of the two shifted spectra that extend across zero frequency. Since $w\lp(t/T)\exp\lp(-i\omega_0 t)$ vanishes outside $\mathcal O$, any nonzero values of $e^+\lp(t)$ outside $\mathcal O$ are entirely due to this deviation. The leakage $\varepsilon_{\mathrm{Gl}}$ is therefore small in the quasi-monochromatic limit $\omega_0 T \gg 1$. Its exact dependence on $\omega_0 T$ is easily calculated from \cref{c3_epsilon_glauber,c3_qexv_glauber_coherent,c3_eplus}, and depends on the envelope $w$, in particular on the decay of its spectral tails.

\subsubsection{Artificial photon-number operator}
As discussed in \cref{c1_sec_review_purpose_and_scope}, it is also common to introduce time-domain ladder operators $a\lp(t)$ satisfying $\comm{a\lp(t)}{a^\dagger\sephn\lp(t')}=\delta\lp(t-t')$ by introducing artificial negative-frequency modes. The quantity $\adj{a}\sephn\lp(t)a\lp(t)$ may then be interpreted as a local photon-number density within the resulting narrow-bandwidth approximation. It is important to distinguish this from the Glauber intensity: the latter is defined in the exact theory but is nonlocal, whereas $\adj{a}\sephn\lp(t)a\lp(t)$ is local only as a consequence of the artificial frequency extension. Nevertheless, the corresponding errors turn out to be mathematically similar.

Again, we must specify the state before evaluating the error of the approximation. In analogy with the Glauber case, we define the leakage
\begin{equation}\label{c3_epsilon_art}
    \varepsilon_{\mathrm{art}}
    =
    \frac{\displaystyle \int_{\compl{\mathcal{O}}} \sepia \dd t \sepib
    \qexv{\alpha_\xi}{\adj{a}\sephn\lp(t)a\lp(t)}}
    {\displaystyle \mintn{\dd t} 
    \qexv{\alpha_\xi}{\adj{a}\sephn\lp(t)a\lp(t)}},
\end{equation}
where
\begin{equation}\label{c3_art_number_density}
    \qexv{\alpha_\xi}{\adj{a}\sephn\lp(t)a\lp(t)}
    =
    \abs{f^+\sephn\lp(t)}^2,
\end{equation}
and
\begin{equation}\label{c3_fplus}
    f^+\sephn\lp(t)
    =
    \frac{\alpha}{\sqrt{2\pi}}
    \minti{\dd \omega} \xi\lp(\omega) e^{-i\omega t}.
\end{equation}
In obtaining \cref{c3_art_number_density}, we have taken the artificial negative-frequency modes to be in their \enquote{vacuum state}. Again, although the physical field disturbance $e\lp(t)$ is confined to $\mathcal{O}$, the artificial photon-number signal \eqref{c3_art_number_density} has infinite tails. Comparing \cref{c3_eplus,c3_fplus} we find that $e^+\sephn\lp(t)$ and $f^+\sephn\lp(t)$ contain the same spectrum apart from the factor $\mathcal{E}\lp(\omega)$. For a quasi-monochromatic pulse this factor is approximately constant over the relevant bandwidth, so $\varepsilon_{\mathrm{art}}\approx\varepsilon_{\mathrm{Gl}}$, whereas for broadband pulses the two errors generally differ.


%% file: 4_local_operators.tex


\newcommand{\FigFourOne}{%
    \begin{figure}%
    \hspace*{-5mm}%
    \raisebox{0mm}{%
        \iftoggle{FigFromPDF}{%
            \includegraphics{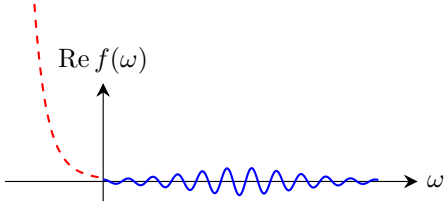}%
        }{%
            \tikzstyle{box} = [%
                rectangle,%
                line width=0.3mm,%
                rounded corners=4,%
                minimum height=0.8cm,%
                text height=0.25cm,%
                text centered,%
                draw=black!70,%
                inner sep=7pt,%
                fill=MyBlueL%
            ]%
            \tikzsetnextfilename{fig4-1}%
            \begin{tikzpicture}[domain=-1:3,scale=1.3]%
                \draw[axis_arrow] (-1,0) -- (3.2,0) node[right] {$\omega$};%
                \draw[axis_arrow] (0,-0.2) -- (0,1) node[above] {$\Re f\lp(\omega)$};%
                \draw[blue,thick,smooth] plot[domain=0:4,samples=200,scale=0.7] (\x,{0.2*cos(1000*\x)*exp(-1.5*(\x-2)^2/(1+abs(\x-2)))});%
                \draw[red,dashed,thick,smooth] plot[domain=-1:0,samples=100,scale=0.7] (\x,{0.02*exp(1.5*(\x-0.8)^2)});%
            \end{tikzpicture}%
        }%
    }%
    \caption{\label{c4_fig_superoscillations}Illustrative sketch of a function with superoscillations for $\omega>0$ at the expense of a large norm for $\omega<0$.}%
    \end{figure}%
}

\iftoggle{ExportTikz}{%
    \FigFourOne
     
}{}

\section{Local operators}
\label{c4_sec_local_operators}
In \cref{c3_sec_local_measurement} we discussed various descriptions of local measurement in quantum field theory, and settled on a pragmatic approach of assuming that physically realizable local observables contain pointwise field observables and are contained in the local operator algebra. Thus, even if the local operator algebra is \enquote{too large}, properties applying to all operators in the local algebra should by our assumption also apply to any operator describing a physically realizable local observable. In this section, we therefore look at local algebras and characterize the properties of operators in these algebras. 

We first introduce a method for generating all polynomial local operators for hermitian scalar fields and for the electromagnetic field. These operators are generally unbounded, and they form the polynomial local algebra of the field. Local observables are self-adjoint operators in this algebra. These observables can be spectrally decomposed by the spectral theorem for unbounded operators, which yields the associated spectral projectors. These projectors allow us to predict the full statistics of measuring a local observable, rather than just the expectation value. This also gives us the concept of the bounded local algebra. 

We then introduce the Reeh--Schlieder theorem, and discuss why this theorem does not violate causality, despite some discussion to that effect. We give a simple proof in a special case, which captures the main idea. The theorem is then used to establish some interesting properties of local operators. In particular, we show in \cref{c4_cor_local_probabilities_never_vanish} that the probability of any outcome of any local measurement in vacuum cannot vanish. This surprising result has been the subject of some discussion in connection with causality in the literature. We therefore look in detail at how to interpret measurements in quantum field theory, with implications for Malament's theorem and for particle localization.

\subsection{Polynomial local algebra}
\label{c4_sec_polynomial_local_algebra}
Consider first a free, hermitian scalar quantum field $\phi\lp(x)$. In \cref{c3_sec_what_is_locally_measurable_in_quantum_field_theory} we introduced pointwise local observables for $\phi\lp(x)$ as $q_0$, $\phi\lp(x)$, $\phi^2\sephn\lp(x)$, etc.\ (where $q_0$ is a constant), as well as sums and derivatives of such expressions. We then argued that in order to describe all possible measurements that can be done coherently in a \emph{region} of spacetime $\mathcal{O}$, we must in general consider smeared observables such as
\begin{equation}\label{c4_phi_f}
    \phi_f = \int_{\mathcal{O}} \sepia \dd[4] x \sepib f\lp(x) \phi\lp(x),
\end{equation}
where $f\lp(x)$ is some real and suitably nice function supported in $\mathcal{O}$. 

This smearing is however not only of physical importance but is mathematically motivated as well. The reason is that the quantum field $\phi\lp(x)$ is technically not an operator but rather an operator-valued distribution. I.e., it is to operators what the Dirac delta distribution is to functions. By smearing it out as in \cref{c4_phi_f}, we obtain an actual operator $\phi_f$.

For gauge fields there is an additional subtlety. Even after smearing $A^\mu\sephn\lp(x)$, the resulting operators $A_f$ need not be well defined on the physical Hilbert space. For example in Lorenz gauge, one first works with states that have negative norm, associated with unphysical modes of timelike and longitudinal polarization, and only afterwards passes to the Hilbert space of physical photons. Fortunately, operators that are gauge-invariant preserve this physical subspace and therefore define operators on the physical Hilbert space \cite[Ch.~7.3]{strocchi2013}. We therefore restrict the local operator algebra to gauge-invariant operators.

\subsubsection{Polynomial local operators}
\label{c4_sec_polynomial_local_operators}
We are ready to define the polynomial local algebra. Recall that an (associative) \emph{algebra} is a vector space over $\complexes$ together with an associative, bilinear multiplication operation (in this case operator composition) \cite[Ch.~1.4]{schmuedgen1990}. When the algebra is closed under taking adjoints, we call it a $\ast$\nobreakdash-algebra. We always include the identity operator, so the algebras considered here are unital. The terminology \enquote{operator rings} is also used in the older literature, for example by Licht \cite{licht1963}.

\begin{definition}\label{c4_def_polynomial_local_algebra_phi}\textup{(Polynomial local algebra for  $\phi\lp(x)$).\footnote{\label{c4_foot_polynomial_local_algebra_phi}See, e.g., \cite{haag1996}, Ch.~III.1 and \cite{fewster2015}.}}
    For a free, hermitian scalar field $\phi\lp(x)$ and some region $\mathcal{O}$, consider all self-adjoint smeared operators of the form
    \begin{equation}\label{c4_phi_f_pol_algebra}
    \begin{aligned}
        &\phi_f = \int_{\mathcal{O}} \sepia \dd[4]x \sepib f\lp(x) \phi\lp(x), \\
        &\,\forall f\lp(x) \tn{ real and supported in $\mathcal{O}$},
    \end{aligned}
    \end{equation}
    where $f\lp(x)$ are smearing functions: $\reals^4 \rightarrow \reals$ supported in $\mathcal{O}$. Form an algebra by taking all possible finite products and linear combinations (over $\complexes$) of all possible smeared fields \eqref{c4_phi_f_pol_algebra} and the identity. This is the unital $\ast$\nobreakdash-algebra generated by the smeared fields \eqref{c4_phi_f_pol_algebra}; we call it the polynomial local algebra associated with the region $\mathcal{O}$. 
\end{definition}

A closely related formulation is used in \cite{knight1961}, Eq.~(1), and \cite{haag1996}, Eq.~(II.4.1), where field polynomials are written in the form
\begin{equation}
\begin{aligned}
    \label{c4_L_G_phi}
    &\sum_n \int_{\mathcal{O}} \sepia \dd[4] x_1 \dotsm \dd[4] x_n \sepib  f_n\lp(x_1, \dotsc, x_n) \phi\lp(x_1) \dotsm \phi\lp(x_n) \\
    &\quad = l_0 + \int_{\mathcal{O}} \sepia \dd[4] x_1 \sepib f_1\lp(x_1) \phi\lp(x_1) \\
    &\quad \sepeq + \int_{\mathcal{O}} \sepia \dd[4] x_1 \dd[4] x_2 \sepib f_2\lp(x_1, x_2) \phi\lp(x_1) \phi\lp(x_2) + \dotsb.
\end{aligned}
\end{equation}
Here $f_n\lp(\cdot)$ are a sequence of complex smearing functions with all arguments supported in $\mathcal{O}$, and $l_0$ is a complex constant. The sum is assumed to be a finite polynomial of fields. The algebras defined by \cref{c4_L_G_phi} and \cref{c4_def_polynomial_local_algebra_phi} are essentially the same, since products of smeared fields correspond to factorized smearing functions in \cref{c4_L_G_phi}, while more general smearing functions $f_n\lp(x_1, \dotsc, x_n)$ can be approximated by finite sums of such factorized functions [see, e.g., \cite[Example~11.7.17]{lebl2026}].

For electrodynamics, the primary field of interest is the potential $A^\mu\sephn\lp(x)$. Since this is gauge dependent, we restrict the algebra to gauge-invariant operators. 

\begin{definition}\label{c4_def_polynomial_local_algebra_A}\textup{(Polynomial local algebra for $A^\mu\sephn\lp(x)$).\footnote{\label{c4_foot_polynomial_local_algebra_A}See \cite{haag1996}, Eq.~(II.1.7) and Ch.~III.1, \cite{strocchi2013}, Ch.~7.3, and \cite{finster2015}.}}
    For the free electromagnetic potential $A^\mu\sephn\lp(x)$ and some region $\mathcal{O}$, consider all self-adjoint smeared operators of the form
    \begin{equation}\label{c4_A_f_pol_algebra}
    \begin{aligned}
        &A_f = \int_{\mathcal{O}} \sepia \dd[4] x \sepib f^\mu\sephn\lp(x) A_\mu\lp(x), \\
        &\,\forall f^\mu\sephn\lp(x) \tn{ real and supported in $\mathcal{O}$},
    \end{aligned}
    \end{equation}
    restricted to those smearing functions that make $A_f$ gauge invariant and hence well defined on the physical Hilbert space. Here $f^\mu\sephn\lp(x)$ are functions: $\reals^4 \rightarrow \reals^4$ supported in $\mathcal{O}$. The polynomial algebra associated with $\mathcal{O}$ is now defined as the unital $\ast$\nobreakdash-algebra generated by smeared fields of the form \eqref{c4_A_f_pol_algebra}.
\end{definition}

Clearly, complex-valued smearings are included by adding fields smeared with the real and imaginary parts. Moreover, derivatives of the field $\partial_\nu \phi\lp(x)$ or $\partial_\nu A^\mu\sephn\lp(x)$ are unnecessary to add separately since they are also already implicitly included in the definition, as can be seen with integration by parts.

Note that the role of the smearing in \cref{c4_phi_f} is somewhat parallel to how the Dirac delta distribution is often used in physics. When we need an actual operator, for instance when we wish to use the spectral theorem (see \cref{c4_sec_spectral_decomposition}), the smearing is necessary. Other times we may use the usual physics shorthand of considering simpler, unsmeared operators, in the same way that we use the Dirac delta distribution in calculus. An unsmeared local operator, e.g., $\phi\lp(x)$ or $\norder{\phi^2\lp(x)}$, can then be thought of as realized by fields smeared out with some narrow smearing function centered around the point $x$.

An important property of the polynomial local algebra is its spacelike commutation property. Microcausality requires gauge-invariant operators local to spacelike separated regions $\mathcal{O}$ and $\splike{\mathcal{O}}$ to commute. For hermitian scalar fields $\phi\lp(x)$, this comes from the field itself commuting at spacelike separations: $\comm{\phi\lp(x)}{\phi\lp(\splike{x})} = 0$ for all points $x \in \mathcal{O}$ and $\splike{x} \in \splike{\mathcal{O}}$. For gauge fields such as $A^\mu\sephn\lp(x)$, the field itself may or may not commute with itself at spacelike separations, depending on the gauge choice. For example in Coulomb gauge, $A^\mu\sephn\lp(x)$ does not commute with itself for spacelike separations \cite[Ch.~7.7]{greiner1996}. Nevertheless, gauge-invariant local operators, such as the electric and magnetic fields, must by microcausality commute for spacelike separations. 

\begin{remark}\label{c4_rem_spacelike_commute_polynomial}\textup{(Spacelike polynomial operators commute).\footnote{\label{c4_foot_spacelike_commute_polynomial}See \cite{strocchi2013}, Ch.~3.1.}}
    For two regions $\mathcal{O}$ and $\splike{\mathcal{O}}$ that are spacelike separated:
    \begin{equation}
        \comm{L}{\splike{L}} = 0,
    \end{equation}
    for all operators $L$ and $\splike{L}$ in the polynomial algebras associated with $\mathcal{O}$ and $\splike{\mathcal{O}}$, respectively. 
\end{remark}

According to the usual rules of quantum mechanics, local observables are defined as those local operators that are self-adjoint. 

\begin{definition}\label{c4_def_polynomial_local_observables}\textup{(Polynomial local observables).}
    Define polynomial observables local to a region $\mathcal{O}$ as the set of all operators $L$ local to $\mathcal{O}$ such that $L$ is self-adjoint.
\end{definition}

Note that the set of local observables is a subset of the polynomial local operators, but not a sub\emph{algebra} since it is not closed under multiplication. For instance, the product of two hermitian smeared fields $\phi_f$ and $\phi_g$ of the form \eqref{c4_phi_f} is not necessarily hermitian:
\begin{equation}
    \adj{\lp(\phi_f \phi_g)} = \phi_g \phi_f \neq \phi_f \phi_g.
\end{equation}

\subsubsection{Gauge invariance, unboundedness, self-adjointness, and choice of spacetime regions}
\label{c4_sec_gauge_invariance_unboundedness_self_adjointness_and_choice_of_spacetime_regions}
There are a few points to note about \cref{c4_def_polynomial_local_algebra_phi,c4_def_polynomial_local_algebra_A}. First, concerning gauge invariance, we will not need an explicit characterization of the full set of gauge-invariant local operators. When we need concrete examples, we use familiar gauge-invariant operators, such as smeared electric and magnetic fields. For arguments that involve all possible gauge-invariant polynomial local operators, we can often make use of \cref{c4_rem_spacelike_commute_polynomial}.

The second subtlety is the fact that quantum fields, e.g., $\phi\lp(x)$, are not just operator-valued distributions, but \emph{unbounded} operator-valued distributions. That means that even after smearing as in \cref{c4_phi_f_pol_algebra}, the polynomial local operators $L$ are generally unbounded \cite[Ch.~6.1]{strocchi2013}. I.e., there exists a sequence of states $\ket{\psi_n}$ of unit norm $\norm{\ket{\psi_n}} = 1$ such that $\norm{L \ket{\psi_n}}$ grows without a bound \cite[Ch.~1.1]{schmuedgen2012}. To see this, consider the usual expansion of a hermitian Klein--Gordon field
\begin{equation}\label{c4_phi}
    \phi\lp(x) = \frac{1}{\lp(2\pi)^{3/2}} \mintf{\frac{\dd[3]k}{\sqrt{2\omega_{\vect{k}}}}}  \lp(\padj{a}_{\vect{k}} e^{i \vect{k} \cdot \vect{x} - i \omega_{\vect{k}} t} + \adj{a}_{\vect{k}} e^{- i \vect{k} \cdot \vect{x} + i \omega_{\vect{k}} t}).
\end{equation}
In this case, it suffices to smear over space at a fixed time to obtain a well-defined operator:
\begin{equation}\label{c4_phi_f_d3x}
    \phi_f = \mint{\dd[3]x} f\lp(\vect{x})\phi\lp(\vect{x}) = \padj{a}_f + \adj{a}_f.
\end{equation}
Here $\phi\lp(\vect{x})$ is the field evaluated at some fixed time $t=0$, and the smearing function $f\lp(\vect{x})$ has been assumed real. Furthermore, in the second equality we have substituted the field expression \eqref{c4_phi}, defining
\begin{equation}\label{c4_ad_f}
    \adj{a}_f = \mint{\dd[3]k} \frac{1}{\sqrt{2 \omega_{\vect{k}}}} f\lp(\vect{k}) \adj{a}_{\vect{k}},
\end{equation}
where $f\lp(\vect{k})$ is the Fourier transform of $f\lp(\vect{x})$. With normalization
\begin{equation}\label{c4_f_x_normalization}
    \mint{\dd[3]k} \frac{1}{2 \omega_{\vect{k}}} \abs{f\lp(\vect{k})}^2 = 1,
\end{equation}
we get $\comm[\big]{\padj{a}_f}{\adj{a}_f} = 1$, so we identify associated number states
\begin{equation}
    \ket{n_f} = \frac{1}{\sqrt{n!}} {a^{\dagger n}_f} \ket{0}.
\end{equation}
This leads to
\begin{equation}
    \norm{\phi_f \ket{n_f}}^2 = 2n + 1.
\end{equation}
Thus for a sequence of states $\ket{n_f} = \ket{1_f}, \ \ket{2_f}, \ \dotsc$ (all of which have unit norm), $\norm{\phi_f \ket{n_f}}$ grows without a bound. Thus $\phi_f$ is an unbounded operator. 

The tricky thing about these unbounded operators is that they are never defined on the whole Hilbert space---only on a subset $\domain \lp(L)$ called the domain of $L$ \cite[Ch.~1.1]{schmuedgen2012}. For example consider
\begin{equation}
    \ket{\psi} = \sum_{n = 1}^\infty \frac{1}{n} \ket{n_f}.
\end{equation}
Its norm is finite,
\begin{equation}
    \norm{\ket{\psi}}^2 = \sum_{n = 1}^\infty \frac{1}{n^2} = \frac{\pi^2}{6} < \infty,
\end{equation}
meaning that $\ket{\psi}$ belongs to the Hilbert space. However, when the operator $\phi_f$ from \cref{c4_phi_f_d3x} acts on $\ket{\psi}$,
\begin{equation}
\begin{aligned}
    \norm{\phi_f \ket{\psi}}^2 &= \sum_{n, m = 1}^\infty \frac{1}{n} \frac{1}{m} \Big[\sqrt{n} \seph \bra{{n - 1}_f} + \sqrt{n + 1} \seph \bra{{n + 1}_f}\Big] \\
    &\hspace{4em} \cdot \Big[\sqrt m \seph \ket{{m - 1}_f} + \sqrt{m + 1} \seph \ket{{m + 1}_f}\Big].
\end{aligned}
\end{equation}
Every term in this double sum is non-negative, meaning that we can get a lower bound by keeping only the $\braket{{n - 1}_f}{{m - 1}_f}$ terms:
\begin{equation}
    \norm{\phi_f \ket{\psi}}^2 \geq \sum_{n = 1}^\infty \frac{1}{n^2} \sqrt{n}\sqrt{n} \rightarrow \infty.
\end{equation}
Thus we might in physics say that the operator $\phi_f$ acting on $\ket{\psi}$ gives a state of infinite norm (an unnormalizable state). However, mathematically this is not allowed, so we must instead say that $\ket{\psi} \notin \domain\lp(\phi_f)$.

Importantly, although the domain of these unbounded operators cannot be the whole Hilbert space, the domain of \emph{self-adjoint} unbounded operators is always \emph{dense} in the Hilbert space (by definition) \cite[Ch.~1.2]{schmuedgen2012}. In this work, we will for the most part not keep track of the domains of various operators. We will however use some of the mathematical machinery for unbounded operators, such as the spectral theorem. When applying such machinery to a polynomial local operator, we implicitly mean the closure of the operator initially defined on the usual common dense invariant domain of the smeared fields.

Another potential complication coming from the fact that the polynomial local operators are unbounded is the difference between hermitian (symmetric) and self-adjoint operators. The difference is that self-adjointness also requires an equality of domains of the operator and its adjoint \cite[Ch.~1.2]{schmuedgen2012}. A standard assumption for the field $\phi\lp(x)$ is that when smeared with a real smearing function, it yields an operator that is self-adjoint (after closure) \citetext{\citealp[Ch.~III.1]{haag1996}; \citealp[Ch.~8.8]{derezinski2022}}. Similarly, we assume that when $A^\mu\sephn\lp(x)$ is smeared with a real smearing function that yields a gauge-invariant operator, the resulting operator (after closure) is also self-adjoint \citetext{\citealp[Ch.~7.3]{strocchi2013}; \citealp{buchholz2015}}. For a general polynomial local operator that is symmetric, self-adjointness is not automatic and must in principle be individually checked for it to qualify as an observable according to \cref{c4_def_polynomial_local_observables}.

We also ignore mathematical restrictions on the spacetime regions $\mathcal{O}$. Strictly speaking, many definitions and theorems are typically stated for spacetime regions $\mathcal{O}$ that are bounded and open \cite[Ch.~III.1]{haag1996}. We will consider more general regions, including unbounded ones, where appropriate. We generally leave mathematical details concerning the regions implicit, while noting geometric restrictions when they matter.

\subsection{Spectral decomposition}
\label{c4_sec_spectral_decomposition}
The previous section characterizes the set of polynomial local observables. For some state $\ket{\psi}$, the expected value of measuring such an observable $L$ can be computed as $\qexv{\psi}{L}$. On the other hand, to obtain the set of possible outcomes and associated measurement probabilities for a given observable $L$, we can use spectral decomposition. Since the operators are generally unbounded, we then need to look at the spectral theorem for unbounded operators. Throughout this review, the Hilbert space is the usual vacuum Fock representation space associated with a free quantum field, with only physical photon states in the electromagnetic case. This space is separable, meaning that it has a countable orthonormal basis.

\subsubsection{The spectral theorem}
\label{c4_sec_the_spectral_theorem}
\begin{theorem}\label{c4_thm_spectral}\textup{(Spectral theorem for self-adjoint, unbounded operators).\footnote{\label{c4_foot_spectral_theorem}See \cite{schmuedgen2012}, Theorem~5.7.}}
    On a Hilbert space $H$, any self-adjoint (possibly unbounded\hspace{1pt}\footnote{\label{c4_foot_spectral_theorem_unbounded}Note that the spectral theorem applies also if the operator $L$ is bounded. In that case, the domain is the whole Hilbert space, $\domain\lp(L) = H$. Also note that any bounded hermitian operator is self-adjoint.}\hspace{-1pt}) operator $L$ defined on some domain $\domain\lp(L) \subset H$ can be spectrally decomposed as
    \begin{equation}\label{c4_spectral_decomposition_general}
        L = \int_{\sigma\lp(L)} \hspace{-0.9em} \lambda \sepim \dd \mu_L\lp(\lambda).
    \end{equation}
    Here $\sigma\lp(L) \subseteq \reals$ is the spectrum of $L$, and $\mu_L$ is a projection-valued measure. Functions $h\lp(x)$ defined for real $x$ can then be extended to the operator $L$ in the usual fashion:
    \begin{equation}\label{c4_operator_functions}
        h\lp(L) = \int_{\sigma\lp(L)} \hspace{-0.9em} h\lp(\lambda) \sepim \dd \mu_L\lp(\lambda).
    \end{equation}
\end{theorem}

A measurement of $L$ results in some outcome $\lambda \in \sigma\lp(L)$, similar to eigenvalues in the finite-dimensional case. To find measurement projectors, we must in general choose some subset of outcomes $\chi \subseteq \sigma\lp(L)$. The projector associated with obtaining an outcome $\lambda \in \chi$ is found by integrating the projection-valued measure $\mu_L$ over $\chi$:
\begin{equation}\label{c4_P_chi}
    P_\chi = \int_{\chi} \hspace{-0.1em} \dd \mu_L\lp(\lambda).
\end{equation}
In this review we mostly need only two common cases of the spectral theorem: the discrete case where the spectral resolution is a countable sum, and the continuous case where the spectral resolution is an integral.

\subsubsection{Discrete spectra}
\label{c4_sec_discrete_spectra}
For self-adjoint operators with a pure point spectrum, meaning that
they have a complete orthonormal basis of eigenvectors, we can write \cite[Ch.~3.5]{madrid2001}:
\begin{equation}\label{c4_dd_mu_L_discrete}
    \dd \mu_L\lp(\lambda) = \sum_n P_n \delta\lp(\lambda - l_n) \dd \lambda,
\end{equation}
where $l_n \in \reals$ are the eigenvalues of $L$, and $P_n$ are the corresponding projectors. In this case, \cref{c4_spectral_decomposition_general} becomes a (countable) sum over the eigenvalues:
\begin{equation}\label{c4_L_spectral_discrete}
    L = \sum_n l_n P_n.
\end{equation}
Therefore, when measuring $L$, the eigenvalues $\lp{l_n}$ are the possible outcomes, each associated with a spectral projector $P_n$. Given some normalized state $\ket{\psi}$, the probability of obtaining the outcome $\lambda = l_n$ is
\begin{equation}
    \prob\lp(\lambda \probsymb{=} l_n) = \qexv{\psi}{P_n}.
\end{equation}

For each eigenvalue $l_n$, we can find a set of eigenstates $\ket{u_{ni}}$:
\begin{equation}
    L \ket{u_{ni}} = l_n \ket{u_{ni}},
\end{equation}
that are orthonormal
\begin{equation}\label{c4_u_n_i}
    \braket{u_{ni}}{u_{mj}} = \delta_{nm} \delta_{ij},
\end{equation}
so that
\begin{equation}
    P_n = \sum_i \ket{u_{ni}} \bra{u_{ni}},
\end{equation}
where the sum is over eigenstates corresponding to the eigenvalue $l_n$.

\subsubsection{Continuous spectra}
\label{c4_sec_continuous_spectra}
For self-adjoint operators with a purely continuous spectrum, the spectrum may in principle contain an absolutely continuous part and a singular continuous part. However, in this review and in many physics contexts, we assume that the singular continuous part is empty. We can then write \cite[Ch.~3.5]{madrid2001}:
\begin{equation}\label{c4_dd_mu_A}
    \dd \mu_L\lp(\lambda) = P\lp(\lambda) \sepib \dd \lambda,
\end{equation}
where $P\lp(\lambda)$ is a projector density. In this case, \cref{c4_spectral_decomposition_general} becomes an integral over the spectral values:
\begin{equation}\label{c4_spectral_decomposition_lambda}
    L = \int_{\sigma\lp(L)} \hspace{-0.9em} \dd \lambda \sepib \lambda \seph P\lp(\lambda).
\end{equation}
Therefore, when measuring $L$, the spectrum $\sigma\lp(L) \subseteq \reals$ is the sample space of possible outcomes, for which we can associate a probability \emph{density}
\begin{equation}
    p\lp(\lambda) = \qexv{\psi}{P\lp(\lambda)}.
\end{equation}

To form actual projectors (not projector densities), we can for example consider intervals $\lp[a, b] \subseteq \sigma\lp(L)$ for $a, b \in \reals$, with associated spectral projectors
\begin{equation}
    P_{\lp[a, b]} = \int_a^b \hspace{-0.3em} \dd \lambda \sepib P\lp(\lambda).
\end{equation}
Given some normalized state $\ket{\psi}$, the probability of obtaining an outcome $\lambda \in \lp[a, b]$ is
\begin{equation}
    \prob\lp(\lambda \probsymb{\in} \lp[a, b]) = \qexv{\psi}{P_{\lp[a, b]}} = \int_a^b \hspace{-0.3em} \dd \lambda \sepib p\lp(\lambda).
\end{equation}

With continuous spectra we can in many cases find a set of generalized eigenstates $\ket{u_{\lambda i}}$ \cite{gadella2003}:
\begin{equation}
    L \ket{u_{\lambda i}} = \lambda \ket{u_{\lambda i}},
\end{equation}
that are delta-function normalized
\begin{equation}\label{c4_u_lambda_j}
    \braket{u_{\lambda i}}{u_{\lambda' j}} = \delta_{ij} \delta\lp(\lambda - \lambda'),
\end{equation}
so that
\begin{equation}
    P\lp(\lambda) = \sum_i \ket{u_{\lambda i}} \bra{u_{\lambda i}},
\end{equation}
where the sum is over eigenstates corresponding to the same spectral value $\lambda$.

This situation is of course familiar from nonrelativistic quantum mechanics, as we are used to for instance writing the position operator as
\begin{equation}\label{c4_hat_x}
    \hat{x} = \mintn{\dd x} x \seph \ket{x} \bra{x},
\end{equation}
where $\ket{x}$ are generalized eigenstates satisfying $\braket*{x}{x'} = \delta\lp(x - x')$. The spectral theorem for unbounded operators is part of the mathematical details going on behind the scenes. These details can sometimes be important and nontrivial, for instance for operators with a spectrum containing both discrete and continuous parts. 

We will however take a middle ground. It is good to be aware of some of the mathematical subtleties and formal definitions underpinning these analyses. On the other hand, the observables $L$ we analyze in this review are typically smeared field expressions, which we know are related to harmonic-oscillator modes. We can therefore try relating an observable $L$ by unitary transformations to some known harmonic-oscillator operator to find the spectrum $\sigma\lp(L)$, and then look for (generalized) eigenstates to use the expressions in \cref{c4_sec_discrete_spectra,c4_sec_continuous_spectra}.

\subsection{Bounded local algebra}
\label{c4_sec_bounded_local_algebra}
The mathematics of unbounded operator-valued distributions is heavy, and formally treating technical details such as domains, self-adjointness, and convergence of operator sequences is difficult. To handle these types of questions, it has therefore been suggested to go over to algebras of \emph{bounded} operators instead \cite[Ch.~III.1]{haag1996}.

\subsubsection{Bounded local operators}
\label{c4_sec_bounded_local_operators}
The algebra of bounded local operators can be constructed using the spectral theorem in the following way.

\begin{definition}\label{c4_def_bounded_local_algebra_phi}\textup{(Bounded local algebra $\balg{\mathcal{O}}$ for $\phi\lp(x)$).\footnote{\label{c4_foot_bounded_local_algebra_phi}See \cite{araki1964}. See also \cite{derezinski2022}, Ch.~2.6 or \cite{strohmaier2024}.}}
    For a free, hermitian scalar field $\phi\lp(x)$ and some region $\mathcal{O}$, consider all self-adjoint smeared operators of the form
    \begin{equation}\label{c4_phi_f_bounded_algebra}
    \begin{aligned}
        &\phi_f = \int_{\mathcal{O}} \sepia \dd[4]x \sepib f\lp(x) \phi\lp(x), \\
        &\,\forall f\lp(x) \tn{ real and supported in $\mathcal{O}$},
    \end{aligned}
    \end{equation}
    where $f\lp(x)$ are smearing functions: $\reals^4 \rightarrow \reals$ supported in $\mathcal{O}$. Since $f\lp(x)$ is real, $\phi_f$ is self-adjoint and can be spectrally decomposed according to \cref{c4_spectral_decomposition_general} as
    \begin{equation}
        \phi_f = \int_{\sigma\lp(\phi_f)} \hspace{-0.9em} \lambda \sepim \dd \mu_{\phi_f}\sephn\lp(\lambda).
    \end{equation}
    Consider then all bounded operator functions of $\phi_f$:
    \begin{equation}
        h\lp(\phi_f) = \int_{\sigma\lp(\phi_f)} \hspace{-0.9em} h\lp(\lambda) \sepim \dd \mu_{\phi_f}\sephn\lp(\lambda), \sepc \forall h\lp(x) \tn{ bounded},
    \end{equation}
    where $h\lp(x)$ are functions: $\reals \rightarrow \reals$. Since $h\lp(x)$ is a bounded function, the spectrum of $h\lp(\phi_f)$ is bounded, meaning that $h\lp(\phi_f)$ is a bounded operator. Finally, form a $\ast$\nobreakdash-algebra by taking all possible finite products and linear combinations (over $\complexes$) of all possible $h\lp(\phi_f)$. It is also useful to have a closed set containing this algebra, in order to handle infinite sequences of operators. One therefore takes the closure in the weak operator topology, resulting in a so-called von Neumann algebra. We label it the bounded local algebra $\balg{\mathcal{O}}$ generated by $\phi\lp(x)$. 
\end{definition}

The bounded local algebra for electrodynamics is constructed in the same way except we must smear over all components of $A^\mu\sephn\lp(x)$, and we require gauge invariance. 

\begin{definition}\label{c4_def_bounded_local_algebra_A}\textup{(Bounded local algebra $\balg{\mathcal{O}}$ for $A^\mu\sephn\lp(x)$).\footnote{\label{c4_foot_bounded_local_algebra_A}See \cite{grundling2000}, \cite{finster2015}, and \cite{strohmaier2024}.}}
    For the free electromagnetic potential $A^\mu\sephn\lp(x)$ and some region $\mathcal{O}$, consider all self-adjoint smeared operators of the form
    \begin{equation}\label{c4_A_f_bounded_algebra}
    \begin{aligned}
        &A_f = \int_{\mathcal{O}} \sepia \dd[4] x \sepib f^\mu\sephn\lp(x) A_\mu\lp(x), \\
        &\,\forall f^\mu\sephn\lp(x) \tn{ real and supported in $\mathcal{O}$},
    \end{aligned}
    \end{equation}
    restricted to those smearing functions that make $A_f$ gauge invariant. Here $f^\mu\sephn\lp(x)$ are functions: $\reals^4 \rightarrow \reals^4$ supported in $\mathcal{O}$. The construction then proceeds identically with \cref{c4_def_bounded_local_algebra_phi} by spectrally decomposing all such self-adjoint, gauge-invariant smeared operators $A_f$ and building an algebra from their spectral projectors. This defines the bounded, gauge-invariant local algebra $\balg{\mathcal{O}}$ generated by $A^\mu\sephn\lp(x)$.
\end{definition}

Mathematically the bounded algebra $\balg{\mathcal{O}}$ is much better behaved than the polynomial algebra. All operators $L \in \balg{\mathcal{O}}$ are bounded and defined on the whole Hilbert space, and are therefore closed. This means that there is no distinction between hermitian and self-adjoint operators in this algebra, since the domain is always the whole Hilbert space. Additionally, it is easier to treat infinite sequences of operators in a systematic way. 

Similar to \cref{c4_rem_spacelike_commute_polynomial}, bounded local operators must satisfy microcausality.

\begin{remark}\label{c4_rem_spacelike_commute_bounded}\textup{(Spacelike bounded operators commute).\footnote{\label{c4_foot_spacelike_commute_bounded}See \cite{derezinski2022}, Ch.~2.7.}}
    For two regions $\mathcal{O}$ and $\splike{\mathcal{O}}$ that are spacelike separated:
    \begin{equation}
        \comm{L}{\splike{L}} = 0, \sepc \forall L \in \balg{\mathcal{O}}, \; \forall \splike{L} \in \balg*{\splike{\mathcal{O}}}.
    \end{equation}
\end{remark}

Additionally, for the bounded local algebra, there is a deep structural fact available known as the Schlieder property. 

\begin{remark}\label{c4_rem_schlieder_property}\textup{(Schlieder property).\footnote{\label{c4_foot_schlieder_property}Original version for projectors by \cite{schlieder1969}. This version is by \cite{roos1970}. See also \cite{horuzhy1990}, Lemma~1.3.33 and Proposition~1.3.34.}}
    For two regions $\mathcal{O}$ and $\splike{\mathcal{O}}$ that are spacelike separated with a finite gap (not touching), any two nonzero operators $L \in \balg{\mathcal{O}}$ and $\splike{L} \in \balg{\splike{\mathcal{O}}}$ satisfy
    \begin{equation}
        L \, \splike{L} \neq 0.
    \end{equation}
\end{remark}

In the same way as for the polynomial local algebra, restricting $\balg{\mathcal{O}}$ to hermitian operators then yields the subset of bounded local observables.

\begin{definition}\label{c4_def_bounded_local_observables}\textup{(Bounded local observables).}
    Define bounded local observables local to a region $\mathcal{O}$ as the set of all $L \in \balg{\mathcal{O}}$ such that $L$ is hermitian.
\end{definition}

Note that this is a subset but not a subalgebra of $\balg{\mathcal{O}}$ (see discussion after \cref{c4_def_polynomial_local_observables}).

\subsubsection{Polynomial vs.\ bounded algebra}
\label{c4_sec_polynomial_vs_bounded_algebra}
At this point, a natural question might be which of the two local algebras, the polynomial or the bounded one, represents a more appropriate characterization of physically realizable local operators associated with a spacetime region $\mathcal{O}$. In the case of observables specifically, one may argue that the bounded local observables are a physically more realistic definition. For a measurement to truly be described by, e.g., an unbounded operator $\phi_f$, the measurement apparatus must be able to measure arbitrarily large values. For any real, physical measurement apparatus, this is arguably not possible. One way to model this limitation is by a version of $\phi_f$ that saturates (or drops) measurement outcomes beyond some maximal and minimal values. This is precisely what the bounded local observables $h\lp(\phi_f)$ allow us to model: a suitable function $h\lp(x)$ acting on the spectrum sets some maximum and minimum value of outcomes. 

However, by measuring these bounded observables $h\lp(\phi_f)$ for an expanding range over which $h\lp(l) = l$, one can approach the statistics predicted by the unbounded operator $\phi_f$. In this way, one may ask whether local observables from either algebra may be considered equally physical. The observable $h\lp(L)$ can be measured using the same apparatus as $L$, simply by reporting $h\lp(l)$ for each measured value $l$ \cite[Ch.~1.3]{araki1999}. This idea is also discussed in \cite[Ch.~III.1]{haag1996}, where it is argued that one may in general pass between the polynomial and bounded algebras for most purposes. We therefore take this approach, meaning that we will use whichever of the two algebras is most convenient for analyzing the question at hand.

Due to its simplicity and explicit construction, we often use the polynomial algebra. This is especially true for questions where unsmeared operators are suitable, as it is simple to compute expressions such as $\qexv{\psi}{\norder{\phi^2\sephn\lp(x)}}$ for some state $\ket{\psi}$. Other times, for questions where more mathematical nuance is needed, we may consider the bounded algebra $\balg{\mathcal{O}}$ instead. Throughout this review, $\balg{\mathcal{O}}$ will, strictly speaking, denote the bounded version of the algebra, although for most purposes either version may be kept in mind. This is of course while keeping in mind the very different mathematical properties that operators in the two algebras have.

\subsubsection{Spectral projectors}
\label{c4_sec_spectral_projectors}
One difference between the polynomial and bounded algebras is the role of projectors. From \cref{c4_P_chi}, it is clear that the spectral projectors $P_\chi$ of local observables $L$ are bounded operator functions of $L$, obtained by applying the indicator functions\footnote{An indicator function $\indicator_{\chi}\lp(\lambda)$ equals 1 if $\lambda \in \chi$ and 0 otherwise, serving to test whether an element belongs to a subset.} $\indicator_\chi$ of the set $\chi$:
\begin{equation}
    P_\chi = \int_{\chi} \hspace{-0.1em} \dd \mu_L\lp(\lambda) = \int_{\sigma\lp(L)} \hspace{-0.9em} \indicator_\chi\lp(\lambda) \sepim \dd \mu_L\lp(\lambda) = \indicator_\chi\lp(L),
\end{equation}
with the operator function defined in \cref{c4_operator_functions}. 

For bounded local observables $L \in \balg{\mathcal{O}}$, it follows by using von Neumann's bicommutant theorem that spectral projectors $P_\chi$ of $L$ are also in the bounded local algebra $\balg{\mathcal{O}}$ \citetext{\citealp[Ch.~III.2]{haag1996}; \citealp[Ch.~2.6]{derezinski2022}}. Furthermore, $P_\chi$ is hermitian. Thus the spectral projectors are themselves bounded local observables. In other words, the set of bounded local observables is closed with respect to taking spectral projectors. Because of this, the set is also closed with respect to taking arbitrary real-valued operator functions that are bounded on the observable's spectrum. 

For polynomial local observables $L$, the situation is different. First, note that the polynomial local algebra is closed under polynomial operator functions $p\lp(L)$, but need not contain the spectral projectors $P_\chi$ of $L$, which are obtained using indicator functions. This means that the polynomial local algebra is in general \emph{not} closed with respect to taking spectral projectors. One can however ask whether the projectors $P_\chi$ are in the corresponding bounded local algebra $\balg{\mathcal{O}}$. For a smeared field of the form \eqref{c4_phi_f_bounded_algebra} or \eqref{c4_A_f_bounded_algebra}, its projectors are by construction in $\balg{\mathcal{O}}$.

\begin{remark}\label{c4_rem_spectral_projectors_local}\textup{(Spectral projectors are local).}
    For some region $\mathcal{O}$, let $L$ be a local observable in $\balg{\mathcal{O}}$ or a smeared single field of the form \eqref{c4_phi_f_bounded_algebra} or \eqref{c4_A_f_bounded_algebra} local to $\mathcal{O}$. Then the spectral projectors $P_\chi$ of $L$ are in $\balg{\mathcal{O}}$.
\end{remark}

The fact that the polynomial local algebra is not closed with respect to spectral projectors raises some questions. In quantum mechanics, a standard axiom connecting operators to measurable statistics is that the result of measuring some quantum observable $L$ gives a random variable outcome $\lambda \in \sigma\lp(L)$, with a mean given by
\begin{equation}\label{exv_L_qexv_L}
    \exv{\lambda} = \qexv{\psi}{L},
\end{equation}
where $\ket{\psi}$ is the state of the system. However, measuring $L$ does of course not only produce an average value, but rather a whole probability distribution $\prob\lp(\lambda \probsymb{\in} \chi)$ over outcome sets $\chi \subseteq \sigma\lp(L)$, as discussed in \cref{c4_sec_spectral_decomposition}. With the usual interpretation of operator functions, measuring $P_\chi = \indicator_\chi\lp(L)$ amounts to reporting 1 if $\lambda \in \chi$ and 0 otherwise. Thus no separate probability axiom is needed, since we can use \cref{exv_L_qexv_L} again:
\begin{equation}
    \prob\lp(\lambda \probsymb{\in} \chi) = \qexv{\psi}{P_\chi}.
\end{equation}
Crucially, this satisfies the same axiom because the projectors $P_\chi$ are themselves observables, since they are hermitian. However, for polynomial local observables, this argument can take us outside the polynomial local algebra, since their spectral projectors need not belong to it.

\subsubsection{Other local operators}
\label{c4_sec_other_local_operators}
We have thus introduced interpretations of two classes of the operators $L \in \balg{\mathcal{O}}$: hermitian $L$ correspond to \emph{local observables} and projective $L$ describe \emph{outcomes of projective local measurements}. One might ask whether these interpretations are part of the general classification scheme of operators in $\balg{\mathcal{O}}$. For instance:
\begin{itemize}
    \item Unitary operators describe closed-system time evolution.

    \item Kraus operators describe branches of quantum operations.

    \item Hermitian operators describe observable quantities.

    \item Projectors describe outcomes of projective measurements.

    \item POVM elements describe outcomes of general measurements.
\end{itemize}

The idea of the local algebra $\balg{\mathcal{O}}$ is to connect quantum theory to the underlying spacetime through the quantum fields. Specifically, the algebra $\balg{\mathcal{O}}$ declares which operators on the Hilbert space are attached to any given spacetime region $\mathcal{O}$. This raises the question of whether all operators in $\balg{\mathcal{O}}$ correspond to operations that are physically attainable by working only inside $\mathcal{O}$. One could then expect the interpretations from the underlying theory of quantum mechanics to apply within the local algebra:
\begin{itemize}
    \item Do unitary operators in $\balg{\mathcal{O}}$ describe closed-system time evolution in $\mathcal{O}$?

    \item Do Kraus operators in $\balg{\mathcal{O}}$ describe branches of quantum operations in $\mathcal{O}$?

    \item Do hermitian operators in $\balg{\mathcal{O}}$ describe observable quantities in $\mathcal{O}$?

    \item Do projectors in $\balg{\mathcal{O}}$ describe outcomes of projective measurements in $\mathcal{O}$?

    \item Do POVM elements in $\balg{\mathcal{O}}$ describe outcomes of general measurements in $\mathcal{O}$?
\end{itemize}

However, as discussed in \cref{c3_sec_sorkins_scenario}, there is an ongoing discussion about which local observables actually correspond to physically realizable measurements in some region $\mathcal{O}$. The same distinction between algebraic localization and physical realizability arises for other types of local operators, so many of the above interpretations are not fully determined at this time. It is for instance not clear whether a local unitary operator necessarily corresponds to a closed-system time evolution that can be physically realized by acting only within $\mathcal{O}$. In general, the precise characterization of physically realizable local measurements and operations remains an active topic of discussion \cite{jubb2022,albertini2023,fewster2023a}.

\subsection{Structure of local observables}
\label{c4_sec_structure_of_local_observables}
Having defined algebras of local operators $L$ in the previous sections, one might ask whether it is possible to say anything general about their structure or properties. For instance, if we act on vacuum as $L \ket{0}$ with an operator local to $\mathcal{O}$, do we always get a state that is strictly localized to $\mathcal{O}$? Or can we construct a local operator by finding an appropriate superposition of annihilation operators $a_{\vect{k}}$? It turns out that the answer to both of these questions is in general no, and the reason lies in the Reeh--Schlieder theorem.

\subsubsection{The Reeh--Schlieder theorem}
\label{c4_sec_the_reeh_schlieder_theorem}
\begin{theorem}\label{c4_thm_reeh_schlieder}\textup{(Reeh--Schlieder theorem).\footnote{\label{c4_foot_reeh_schlieder}Original proof by \cite{reeh1961}. This version is close to \cite{licht1963}. See also \cite{streater1964}, Theorem~4.2, \cite{fleming2000}, \cite{witten2018}, or \cite{haag1996}, Theorem~5.3.1.}}
    For any nonempty, open region of spacetime $\mathcal{O}$, the set of states $L \ket{0}$, where $L$ is a local operator, is dense in the Hilbert space $H$. In other words, \begin{equation}\label{c4_closure_L_0_bounded}
        \closure\lp\Big[\set[\big]{L \ket{0}}{L \in \balg{\mathcal{O}}}] = H.
    \end{equation}
    \Cref{c4_closure_L_0_bounded} also  applies if $\balg{\mathcal{O}}$ is replaced by the corresponding polynomial algebra, and/or
    if $\ket{0}$ is replaced by any state $\ket{\psi}$ of bounded energy spectrum.
\end{theorem}

A state of bounded energy spectrum means that there is some maximum energy value $E_{\text{max}}$, so that the state has no content above $E_{\text{max}}$. This restricts both the frequency content of the modes and the allowed multiparticle content of the state. Clearly there are many states of interest that do not satisfy this condition; however, states of bounded energy spectrum are still dense in the Hilbert space. The condition of bounded energy spectrum is sufficient but not necessary for the Reeh--Schlieder theorem; more general conditions can be formulated in terms of analyticity under translations \cite{witten2018,redei2002}.

Intuitively, the Reeh--Schlieder theorem says that any state in the Hilbert space can be approximated to arbitrary precision (in norm) by acting on the vacuum $\ket{0}$ by operators in the local algebra. If this were stated in terms of all operators on the Hilbert space, the theorem would not be very surprising; we can make any state by finding some appropriate operator acting on vacuum. However, the Reeh--Schlieder theorem says that this remains true even if we no longer can access \emph{all} operators, but rather only operators local to some region $\mathcal{O}$---for \emph{any} region $\mathcal{O}$, no matter how small.

\subsubsection{Why doesn't the Reeh--Schlieder theorem violate causality?}
\label{c4_sec_why_doesnt_the_reeh_schlieder_theorem_violate_causality}
That the local algebra can generate any state in the Hilbert space (to arbitrary precision) is quite surprising, as the local algebra is intended to contain operators that describe physically realizable operations in the region $\mathcal{O}$. It might therefore seem that by acting only inside $\mathcal{O}$, we could in principle create any state $\ket{\psi}$, including states that are localized very far away and spacelike from $\mathcal{O}$. For example, acting only with physical operations on the Earth, we could create a photon on the Moon out of vacuum. At first glance this might feel impossible and seem to allow superluminal signaling, as some kind of extreme version of quantum entanglement. This apparent tension has led to considerable discussion of the physical interpretation of the Reeh--Schlieder theorem.

However, the reason the Reeh--Schlieder theorem does not imply superluminal signaling is a matter of physics, as explained in, e.g., \cite{fredenhagen2015,witten2018}. To reliably create a photon on the Moon, we do not need just any operator but rather an isometric (or unitary) operator. The Reeh--Schlieder theorem does not guarantee that the approximating local operators can be chosen isometric. 

In the most general case, we could consider whether $L$ could represent one branch in some quantum operation creating a photon on the Moon. We would then need $L$ to be a Kraus operator: $\adj{L} L \leq \identity$. This requires $L$ to be bounded and have operator norm $\norm{L} \leq 1$. However, given a state $\ket{\psi}$ with unit norm and $\ket{\psi} = L \ket{0}$, it follows that $\norm{L} \geq 1$. Still, if we take $L \in \balg{\mathcal{O}}$, then $L$ is bounded, i.e., there exists some $M$ such that $\norm{L} = M$. We can rescale it as
\begin{equation}\label{c4_L_M}
    L_M = \frac{1}{M} L.
\end{equation}
This implies that $\adj{L}_M \padj{L}_M \leq \identity$, meaning that $L_M$ is in fact a valid Kraus operator.

The caveat is that this rescaling changes the state norm $\norm{L_M \ket{0}}^2 = 1 / M^2 \leq 1$. Taking $L_M$ as a single Kraus operator in a local quantum operation, the postselected outcome state $\ket{\psi} = L_M \ket{0} / \norm{L_M \ket{0}}$ occurs with probability $\norm{L_M \ket{0}}^2$ \cite[Ch.~8.2]{nielsen2010}. As such, it is not difficult to see why the Reeh--Schlieder theorem can be consistent with relativistic causality. Operators $L$ local to $\mathcal{O}$ that create excitations near $\mathcal{O}$ can have a high probability $\norm{L_M  \ket{0}}^2$. For states $\ket{\psi}$ that are localized to inside $\mathcal{O}$ this probability may even be unity: $\norm{L_M  \ket{0}}^2 = 1$. On the other hand, local operators $L$ that create excitations away from $\mathcal{O}$ will generally have lower probability. In the extreme case of $\ket{\psi} = L \ket{0}$ being localized very far away from $\mathcal{O}$, we expect the probability of obtaining $\ket{\psi}$ with a Kraus operator local to $\mathcal{O}$ will be extremely small: $\norm{L_M  \ket{0}}^2 \ll 1$.

When averaging over all measurement outcomes, the resulting nonselective state remains indistinguishable from vacuum by measurements at spacelike separation from $\mathcal O$ [see, e.g., \cite{fewster2020, rustad2026} and the example in \cref{c5_sec_post_measurement_states_of_local_measurements}]. Thus the measurement cannot be used for superluminal signaling, even though individual postselected states may differ from vacuum in those regions.

\subsubsection{Simple proof and explicit protocol}
\label{c4_sec_simple_proof_special_case}
We now give a very simple proof of the Reeh--Schlieder theorem in a special case limited to the one-particle subspace associated with 1D modes in the $+x$-direction of a massless, hermitian scalar field. The proof is inspired by the connection to superoscillations in \cite{ber2015}, and, to some extent, by the proof in \cite{witten2018}.

In line with \cref{c1_A_t}, we can consider a fixed point $x=0$ since the forward-propagating field has a trivial spacetime dependence, writing the field in this case as
\begin{equation}\label{c4_phi_t}
    \phi\lp(t) = \frac{1}{\sqrt{2\pi}} \minti{\dd \omega} \frac{1}{\sqrt{2\omega}} \seph \lp[a\lp(\omega) e^{- i \omega t} + \adj{a}\sephn\lp(\omega) e^{i \omega t}].
\end{equation}
Let $f\lp(t)$ be a square-integrable function supported in $\lp(-T,T)$. We will prove that a smeared field acting on the vacuum,
\begin{equation}
    \mint{\dd t} f\lp(t)\phi\lp(t) \ket{0} 
\end{equation}
can approach any given particle
\begin{equation}\label{c4_target_particle}
    \minti{\dd\omega} \xi\lp(\omega) a^\dagger\sephn\lp(\omega) \ket{0},
\end{equation}
for any square-integrable $\xi\lp(\omega)$. 

This is a very surprising result: the field in a bounded interval is capable of creating one-particle wavepackets concentrated far from the interval! Furthermore, note that 
\begin{equation}\label{c4_f_t_omega}
    \mint{\dd t} f\lp(t)\phi\lp(t) \ket{0} = \minti{\dd\omega} \frac{1}{\sqrt{2\omega}} f\lp(\omega) a^\dagger\sephn\lp(\omega) \ket{0}, 
\end{equation}
where $f\lp(\omega)$ is the Fourier transform of $f\lp(t)$. Since $f\lp(\omega)$ is essentially the spectrum of the produced particle\footnote{The factor $1/\sqrt{2\omega}$ in \cref{c4_f_t_omega} does not change the density result. To see this, take $f\lp(t)=g\lp(t)-g\lp(t-\tau)$, with $g\lp(t)$ supported in $\lp(-T/2,T/2)$ and $0<\tau<T/2$. The resulting spectrum is $m\lp(\omega)g\lp(\omega)$, where $m\lp(\omega)=\lp(1-e^{i\omega\tau})/\sqrt{2\omega}$ is bounded and nonzero almost everywhere. To show that the set of $m\lp(\omega)g\lp(\omega)$'s is dense, it suffices to show that no nonzero $h\lp(\omega)$ is orthogonal to all of them \cite[Lemma~3.3-7]{kreyszig1978}. If $h\lp(\omega)$ is orthogonal to $m\lp(\omega)g\lp(\omega)$, then $m^*\sephn\lp(\omega)h\lp(\omega)$ is orthogonal to $g\lp(\omega)$. Since the set of $g\lp(\omega)$'s forms a dense subspace by \cref{c4_rem_superoscillations}, it follows that $m^*\sephn\lp(\omega)h\lp(\omega)=0$. As $m\lp(\omega)\neq0$ almost everywhere, $h\lp(\omega)=0$, so $m\lp(\omega)g\lp(\omega)$ is again dense.}, and since the inverse Fourier transform of $f\lp(\omega)$ has compact support, we would expect that rapid oscillations in $f\lp(\omega)$ are impossible. But this is not the case! The reason is that we use only the spectrum for $\omega>0$ in \cref{c4_f_t_omega}, allowing superoscillatory behavior: on the positive-frequency axis, $f\lp(\omega)$ can mimic oscillations corresponding to time scales or locations outside the interval $\lp(-T,T)$, at the expense of a very large norm for negative frequencies (see \cref{c4_fig_superoscillations}).

\begin{remark}\label{c4_rem_superoscillations}\textup{(Superoscillations).\footnote{See, e.g., \cite{katsnelson2012}, Lemma~3.}}
    Let $f\lp(t)$ be a square-integrable function supported in $\lp(-T,T)$. Then the Fourier transform $f\lp(\omega)$ restricted to $\lp(0,\infty)$ can approximate any square-integrable function $\xi\lp(\omega)$ supported in $\lp(0,\infty)$ (in $L^2$ norm).\hfill\null
\end{remark}

The proof is by contradiction. We use the standard result that a subspace is dense in a Hilbert space iff there are no nonzero elements orthogonal to the subspace \cite[Lemma~3.3-7]{kreyszig1978}. Suppose there is a nonzero $\xi\lp(\omega)$ such that the $L^2$ inner product $\lp\big(\xi\lp(\omega),f\lp(\omega))=0$ for all $f\lp(t)$ supported in $\lp(-T,T)$. Define $\xi\lp(t)$ as the inverse Fourier transform of $\xi\lp(\omega)$. From the Plancherel theorem (or unitarity of the Fourier transform) we have
\begin{equation}
    0 = \lp\big(\xi\lp(\omega),f\lp(\omega)) = \lp\big(\xi\lp(t),f\lp(t)) = \int_{-T}^{T} \sepia \dd t \sepib \xi^*\sephn\lp(t)f\lp(t).
\end{equation}
Since this is valid for any $f\lp(t)$ we obtain $\xi\lp(t)=0$ on $\lp(-T,T)$. From the Paley--Wiener criterion (\cref{c2_thm_paley_wiener}) this is impossible, since $\xi\lp(\omega)=0$ for $\omega<0$.

\FigFourOne

An explicit protocol for remote-state generation \cite{ber2015} follows immediately. Choose $f\lp(t)$, supported in $\lp(-T,T)$, such that $\int \sepia \dd t \seph f\lp(t)\phi\lp(t) \ket{0}$ approximates the desired one-particle state. Couple the field locally to a two-level ancilla initially in an excited state $\ket{\mathrm{e}}$, through the interaction-picture Hamiltonian (possibly with suitable regularization to avoid divergences)
\begin{equation}
    H_{\mathrm{I}}\lp(t)=\lambda\lp\big[ f\lp(t)\sigma_- + f^*\sephn\lp(t)\sigma_+ ] \phi\lp(t).
\end{equation}
Here we have defined raising and lowering operators $\sigma_+ = \ket{\mathrm{e}} \bra{\mathrm{g}}$ and $\sigma_- = \ket{\mathrm{g}}\bra{\mathrm{e}}$, where $\ket{\mathrm{g}}$ is the ground state. After the interaction, measure the ancilla and postselect the outcome corresponding to the ground state $\ket{\mathrm{g}}$. This leads to the unnormalized conditional field state
\begin{equation}\label{c4_reeh_schlieder_postselected_state}
    -i\lambda \int \sepia \dd t \sepib f\lp(t)\phi\lp(t)\ket{0} + O\lp(\lambda^3),    
\end{equation}
whose squared norm gives the probability of success. For sufficiently small $\lambda$, the normalized field state can thus approximate any desired one-particle state, even one concentrated far from the interaction region. The corresponding success probability may, however, be extremely small. While the first term in expression \eqref{c4_reeh_schlieder_postselected_state} depends only on $f\lp(\omega)$ for $\omega>0$, the higher-order terms can also probe the negative-frequency part, whose norm can be enormous for a remote target. Keeping these terms negligible may therefore require an extremely small $\lambda$.

\subsubsection{No local operator annihilates vacuum}
\label{c4_sec_no_local_operator_annihilates_vacuum}
Although the Reeh--Schlieder theorem does not necessarily violate relativistic causality, it does have some important implications. Consider any bounded region of spacetime $\mathcal{O}$ (we will later generalize this slightly in \cref{c4_rem_spacelike_complement_open}). Now, suppose that there exists an $L \in \balg{\mathcal{O}}$ such that $L \ket{0} = 0$. Since $\mathcal{O}$ is bounded, we can find an open region $\splike{\mathcal{O}}$ that is spacelike separated from $\mathcal{O}$. By the Reeh--Schlieder theorem, any state $\ket{\psi}$ in a dense subspace of the Hilbert space can then be expressed as $\ket{\psi} = \splike{L}  \ket{0}$ using some operator $\splike{L} \in \balg{\splike{\mathcal{O}}}$. Since $\mathcal{O}$ and $\splike{\mathcal{O}}$ are spacelike, the operators $L$ and $\splike{L}$ commute. Thus
\begin{equation}
    L \ket{\psi} = L \, \splike{L} \ket{0} = \splike{L} L \ket{0} = 0.
\end{equation}
Thus $L$ vanishes on a dense set of states $\ket{\psi}$ in the Hilbert space. Since $L$ is bounded, it follows that $L$ is the zero operator: $L = 0$.

\begin{corollary}\label{c4_cor_local_operators_no_zero}\thmheadp{}\footnote{\label{c4_foot_local_operators_annihilation}See also \cite{haag1996}, Theorem~5.3.2.}
    For any bounded spacetime region $\mathcal{O}$, there are no (nonzero) local operators $L \in \balg{\mathcal{O}}$ that satisfy $L  \ket{0} = 0$.
\end{corollary}

Another way to think about this is that local operators can never consist only of annihilation operators $a_{\vect{k}}$, as this would clearly annihilate vacuum; they must contain some creation operators $\adj{a}_{\vect{k}}$ too. Similarly, since local algebras are closed under adjoints, local operators cannot consist of only creation operators either.

\begin{corollary}\label{c4_cor_local_operators_annihilation}\thmheadp{}
    For any bounded spacetime region $\mathcal{O}$, there are no (nonzero) local operators $L \in \balg{\mathcal{O}}$ that consist only of creation operators or only of annihilation operators. 
\end{corollary}

As mentioned in the Reeh--Schlieder theorem (\cref{c4_thm_reeh_schlieder}), we may replace $\balg{\mathcal{O}}$ by the corresponding polynomial algebra, which means that \cref{c4_cor_local_operators_no_zero,c4_cor_local_operators_annihilation} are valid for the polynomial algebra as well.

\makeshortsubsubsection
\subsubsection{Probabilities of local measurements do not vanish!}
\label{c4_sec_probabilities_of_local_measurements_do_not_vanish}
\Cref{c4_cor_local_operators_no_zero} has some interesting consequences for the probability of local measurement outcomes. Suppose we make a local measurement in a bounded spacetime region $\mathcal{O}$. For a projective measurement, the probability of any outcome of this measurement is given by some local projector $P \in \balg{\mathcal{O}}$. The implications of \cref{c4_cor_local_operators_no_zero} for this situation are analyzed in, e.g., \cite{redhead1995}. Here we want to be more general. 

Consider therefore a general local measurement, which may also include interactions of the primary system with other systems, prior to a (projective) measurement. In line with the discussion in \cref{c4_sec_other_local_operators}, we assume that the probability of any outcome of such a general local measurement is given by some local POVM element $D$. That is, a bounded local operator $D \in \balg{\mathcal{O}}$ satisfying
\begin{equation}
    0 \leq D \leq \identity.
\end{equation}
Furthermore, if $D = 0$ then the corresponding probability $\qexv{\psi}{D}$ vanishes for all possible states $\ket{\psi}$. We therefore exclude this trivial case.

Since $D$ is hermitian, it can be spectrally decomposed as $D = \int \lambda \sepim \dd \mu_D\lp(\lambda)$. The positive square root of the positive operator $D$ is then uniquely given by $\sqrt{D} = \int \sqrt{\lambda}\seph \sepim \dd \mu_D\lp(\lambda)$. Since $D$ is bounded, the function $\sqrt{\lambda}$ is bounded on the spectrum of $D$. As discussed in \cref{c4_sec_spectral_projectors}, $\balg{\mathcal{O}}$ is closed with respect to taking bounded operator functions of its hermitian elements, meaning that $\sqrt{D}\in\balg{\mathcal O}$. Assume that the probability of this particular measurement outcome in vacuum is zero:
\begin{equation}\label{c4_D_0_0}
    0 = p = \qexv{0}{D} = \norm[\big]{\sqrt{D}\seph \ket{0}}^2 \sepc \Rightarrow \sqrt{D}\seph  \ket{0} = 0.
\end{equation}
However, since $\sqrt{D} \in \balg{\mathcal{O}}$ and we assume $D \neq 0$, \cref{c4_cor_local_operators_no_zero} says that this is impossible. Thus the probability $p$ must in fact be positive!

Similarly, if any measurement operator $D$ has probability of exactly one, we can use an identical argument for $\identity - D$ to get the same contradiction. Finally note that the Reeh--Schlieder theorem applies not only to vacuum but to any state of bounded energy spectrum. Therefore \cref{c4_cor_local_operators_no_zero} applies to all such states as well. 

\begin{corollary}\label{c4_cor_local_probabilities_never_vanish}\textup{(Probabilities of local measurements never vanish).\footnote{\label{c4_foot_generalization_redhead}Generalization of \cite{redhead1995}, Theorem~2.}}
    For any bounded spacetime region $\mathcal{O}$, let the probability of any outcome of any (general) local measurement be given by some local POVM element $D$:
    \begin{equation}
        0 \leq D \leq \identity, \sepc D \in \balg{\mathcal{O}},
    \end{equation}
    and assume $D$ is not the zero or the identity operator. This probability satisfies
    \begin{equation}\label{c4_Pr_E_bounds}
        0 < \qexv{\psi}{D} < 1 \sepc \tn{(equality is not possible)},
    \end{equation}
    for any state $\ket{\psi}$ of bounded energy spectrum.
\end{corollary}

Note that all local projectors $P \in \balg{\mathcal{O}}$ (that are not zero or the identity) are also POVM elements satisfying \cref{c4_cor_local_probabilities_never_vanish}. The physics of \cref{c4_cor_local_probabilities_never_vanish} for vacuum can be succinctly summarized as:

\begin{remark}\label{c4_rem_local_probabilities_never_vanish_vacuum}\remheadp{}
    Any possible outcome of any possible measurement local to some bounded spacetime region will occur with nonvanishing probability in vacuum.
\end{remark}

A related consequence is \cite{redhead1995}:
\begin{remark}\label{c4_rem_local_projectors_inf_dimensional}\remheadp{}
    Nonzero local projectors have infinite rank.
\end{remark}
Indeed, if a local projector $P$ had finite rank $N$, one could choose $N+1$ linearly independent states with energy spectra with a common bound. Some nonzero linear combination $\ket{\psi}$, which itself would have bounded energy spectrum, would then satisfy $P\ket{\psi}=0$, contradicting \cref{c4_cor_local_probabilities_never_vanish}.

Finally, note that in the above corollaries, we assume that the region $\mathcal{O}$ is bounded in order to use the Reeh--Schlieder theorem on some open region $\splike{\mathcal{O}}$ that is spacelike separated from $\mathcal{O}$. Clearly, $\mathcal{O}$ being bounded is sufficient to find such a region $\splike{\mathcal{O}}$; however it is not actually necessary. In general, there may be many infinite or half-infinite regions of spacetime $\mathcal{O}$, whose spacelike region contains some open subregion. As long as such a spacelike, open subregion exists, we can invoke the Reeh--Schlieder theorem as before, and the corollaries still apply. 

\begin{remark}\label{c4_rem_spacelike_complement_open}\remheadp{}
    \Cref{c4_cor_local_operators_no_zero,,c4_cor_local_operators_annihilation,,c4_cor_local_probabilities_never_vanish} apply not only to spacetime regions $\mathcal{O}$ that are bounded, but to any region $\mathcal{O}$ whose spacelike region contains some nonempty open subregion. 
\end{remark}

Note that one region for which this is not true is when $\mathcal{O}$ is all of spacetime. In that case, one can construct a projector $P_0 = \ket{0}\bra{0}$ with the associated probability in vacuum $\qexv{0}{P_0} = 1$, violating the bounds \eqref{c4_Pr_E_bounds}. This global vacuum projector is in $\balg{\mathcal{O}}$, because the local algebra $\balg{\mathcal{O}}$ when $\mathcal{O}$ is all of spacetime is the set of all bounded operators on the Hilbert space \cite{yngvason2005}.

\subsection{Interpreting local measurement probabilities}
\label{c4_sec_interpreting_local_measurement_probabilities}
\Cref{c4_cor_local_probabilities_never_vanish} implies that no matter what we measure locally, all possible outcomes must have a nonvanishing chance of occurring even in vacuum. How do we make sense of this result? To best understand the implications, it can be useful to compare with how measurement results are interpreted in nonrelativistic quantum mechanics.

\subsubsection{Nonrelativistic quantum mechanics}
\label{c4_sec_nonrelativistic_quantum_mechanics}
Consider a qubit $\ket{\chi_s} \in \complexes^2$, which can represent, e.g., the spin degree of freedom of a spin-$\tfrac{1}{2}$ particle. Assume further that the qubit is prepared in one of two orthonormal basis states $\ket{\chi_0}$ (spin up) and $\ket{\chi_1}$ (spin down) with equal prior probabilities
\begin{equation}
    \prob\lp\big(\ket{\chi_s} \probsymb{=\!} \ket{\chi_0}) \probrel{=} \prob\lp\big(\ket{\chi_s} \probsymb{=\!} \ket{\chi_1}) \probrel{=} \frac{1}{2}.
\end{equation}
Suppose then that we perform a measurement in the $\{\ket{\chi_0}, \ket{\chi_1} \}$ basis. The outcome $l$ of this measurement is completely determined in each case, since obtaining, e.g., $l = 0$ occurs with probability
\begin{equation}
    \prob\lp\big(l \probsymb{=} 0 \mid \ket{\chi_s} \probsymb{=\!} \ket{\chi_0}) \probrel{=} 1,
    \sepc
    \prob\lp\big(l \probsymb{=} 0 \mid \ket{\chi_s} \probsymb{=\!} \ket{\chi_1}) \probrel{=} 0.
\end{equation}
One way to formalize how measurements are interpreted is to look at the update of information caused by the measurement. Whereas the prior probability was 50\% for each preparation $s$, our knowledge changes based on the result of the measurement. In particular we use Bayes' rule to calculate the updated, posterior probability of the state $\ket{\chi_s}$ given some observed measurement outcome, e.g., $l = 0$:
\begin{equation}
    \prob\lp\big(\ket{\chi_s} \probsymb{=\!} \ket{\chi_0} \mid l \probsymb{=} 0) \probrel{=} 1,
    \sepc
    \prob\lp\big(\ket{\chi_s} \probsymb{=\!} \ket{\chi_1} \mid l \probsymb{=} 0) \probrel{=} 0.
\end{equation}
Thus obtaining the outcome $l = 0$ gives us perfect knowledge about what the prior state $\ket{\chi_s}$ was. This works because the two possible prior states $\ket{\chi_0}$ and $\ket{\chi_1}$ are orthogonal. 

Of course, imperfections in the physical realization of qubit states and of measurement devices can result in observed frequencies that do not match these calculated probabilities exactly. However, in the ideal formalism, orthogonal states are perfectly distinguishable by a projective measurement in an appropriate basis.

\subsubsection{Quantum field theory}
\label{c4_sec_quantum_field_theory}
In contrast, under the conditions of \cref{c4_cor_local_probabilities_never_vanish}, the outcome of a local measurement is never 100\% certain. As an example, suppose we are given an unknown state among two possible candidate states with bounded energy spectrum $\ket{\psi_s} \in \{\ket{\psi_0}, \ket{\psi_1}\}$, and that we conduct a local measurement with POVM elements $E_0, E_1 \in \balg{\mathcal{O}}$ (possibly with additional POVM elements). By \cref{c4_cor_local_probabilities_never_vanish}, the probability of an outcome $l$ of this measurement always satisfies
\begin{equation}\label{c4_prob_l_0_gt_0}
    \prob\lp\big(l \probsymb{=} 0 \mid \ket{\psi_s} \probsymb{=\!} \ket{\psi_0}) \probrel{>} 0,
    \sepc
    \prob\lp\big(l \probsymb{=} 0 \mid \ket{\psi_s} \probsymb{=\!} \ket{\psi_1}) \probrel{>} 0.
\end{equation}
The measurement outcome does not make it possible to determine for certain which state $\ket{\psi_s}$ the system was in. For example, given the outcome $l = 0$, Bayes' rule implies that both posterior probabilities must be positive:
\begin{equation}\label{c4_prob_psi_psi1_gt_0}
    \prob\lp\big(\ket{\psi_s} \probsymb{=\!} \ket{\psi_0} \mid l \probsymb{=} 0) \probrel{>} 0,
    \sepc
    \prob\lp\big(\ket{\psi_s} \probsymb{=\!} \ket{\psi_1} \mid l \probsymb{=} 0) \probrel{>} 0.
\end{equation}
Crucially, this is true for any states of bounded energy spectrum, even if they are orthogonal! Thus any single local measurement can never perfectly distinguish $\ket{\psi_0}$ and $\ket{\psi_1}$.

That being said, the probability for obtaining $l = 0$ may be \emph{different} for the two states $\ket{\psi_0}$ and $\ket{\psi_1}$:
\begin{equation}
    \prob\lp\big(l \probsymb{=} 0 \mid \ket{\psi_s} \probsymb{=\!} \ket{\psi_0}) \probrel{\neq} \prob\lp\big(l \probsymb{=} 0 \mid \ket{\psi_s} \probsymb{=\!} \ket{\psi_1}).
\end{equation}
This in turn means that our knowledge about the prior state $\ket{\psi_s}$ can \emph{change} based on the measurement result. For instance, assume again an equal prior probability
\begin{equation}
    \prob\lp\big(\ket{\psi_s} \probsymb{=\!} \ket{\psi_0}) \probrel{=} \prob\lp\big(\ket{\psi_s} \probsymb{=\!} \ket{\psi_1}) \probrel{=} \frac{1}{2}.
\end{equation}
Given the measurement outcome $l = 0$, the bounds \eqref{c4_prob_psi_psi1_gt_0} show that we cannot \emph{rule out} that the prior state was $\ket{\psi_0}$ or $\ket{\psi_1}$. Nevertheless, the measurement result may reveal \emph{some} information about the prior state $\ket{\psi_s}$, in the sense of the posterior probabilities no longer being equal:
\begin{equation}\label{c4_prob_psi_0_l_0_neq_psi_1_l_0}
    \prob\lp\big(\ket{\psi_s} \probsymb{=\!} \ket{\psi_0} \mid l \probsymb{=} 0) \probrel{\neq} \prob\lp\big(\ket{\psi_s} \probsymb{=\!} \ket{\psi_1} \mid l \probsymb{=} 0).
\end{equation}

Furthermore, even though the bounds \eqref{c4_prob_l_0_gt_0} imply that local measurement probabilities are never exactly 0, they do not forbid them from being very small, e.g., one could in principle have
\begin{equation}\label{c4_prob_l_psi_001}
\begin{aligned}
    &\prob\lp\big(l \probsymb{=} 0 \mid \ket{\psi_s} \probsymb{=\!} \ket{\psi_0}) \probrel{=} 0.01, \\
    &\prob\lp\big(l \probsymb{=} 0 \mid \ket{\psi_s} \probsymb{=\!} \ket{\psi_1}) \probrel{=} 0.99.
\end{aligned}
\end{equation}
In such a case, the information revealed by the measurement can be very high. In particular, with equal priors, Bayes' rule gives
\begin{equation}\label{c4_prob_psi_psi1_001}
\begin{aligned}
    &\prob\lp\big(\ket{\psi_s} \probsymb{=\!} \ket{\psi_0} \mid l \probsymb{=} 0) \probrel{=} 0.01, \\
    &\prob\lp\big(\ket{\psi_s} \probsymb{=\!} \ket{\psi_1} \mid l \probsymb{=} 0) \probrel{=} 0.99.
\end{aligned}
\end{equation}
Thus, although a single local measurement can never identify the prior state with complete certainty, it can still make one of the candidate states overwhelmingly more likely.

\subsubsection{Connection to Malament's theorem}
\label{c4_sec_connection_to_malaments_theorem}
In light of \cref{c4_prob_psi_psi1_001}, one may ask what the significance of \cref{c4_cor_local_probabilities_never_vanish} even is. What does it matter if the knowledge we obtain from a local measurement can never make the posterior probability exactly 1, if it could be very close to 1, say $1 - 10^{-100}$? The answer seems to be twofold.

First, even though the previous example shows how a local measurement could make one candidate state much more likely, not every situation is of this kind. The measurement probabilities will depend on the specific states and local measurement, and need not be close to 0 or 1. Generally, the degree to which two states can be distinguished by a local measurement in some region $\mathcal{O}$ is determined by how differently the states are seen by the local operator algebra in $\mathcal{O}$. In other words, \cref{c4_prob_psi_psi1_001} illustrates an additional limitation of local information retrieval on top of the already stochastic nature of quantum measurement: even orthogonal states may only be partially distinguished by a local measurement. 

Second, even though the difference between $1$ and $1 - 10^{-100}$ is negligible physically, mathematics can sometimes be sensitive to such differences. An example of this is Malament's theorem \cite{malament1996}, which we mentioned in \cref{c2_sec_interpreting_measurement_probabilities_in_quantum_field_theory}. Malament sets up a list of seemingly reasonable axioms for a set of \enquote{particle detection} projectors for relativistic particles, and shows that these projectors must be the zero operator. In other words, no such projector set exists. 

However, let us look carefully at Malament's localizability axiom. It says that any projectors $P_{\mathcal{O}}$ and $P_{\splike{\mathcal{O}}}$ corresponding to detections in two disjoint spatial regions at some fixed time must be exactly orthogonal:
\begin{equation}\label{c4_P_G_P_Gp_0}
    P_{\mathcal{O}} P_{\splike{\mathcal{O}}} = P_{\splike{\mathcal{O}}} P_{\mathcal{O}} = 0.
\end{equation}
The question is whether such projectors can in fact correspond to anything locally measurable in quantum field theory. By our assumptions, any measurement projector local to a region $\mathcal{O}$ must be contained in the local operator algebra $\balg{\mathcal{O}}$. Spacelike commutativity of such projectors, $P_{\mathcal{O}} P_{\splike{\mathcal{O}}} = P_{\splike{\mathcal{O}}} P_{\mathcal{O}}$, is itself fine and implied by \cref{c4_rem_spacelike_commute_bounded}, but the last equality forces $P_{\mathcal{O}} = 0$ or $P_{\splike{\mathcal{O}}} = 0$ by \cref{c4_rem_schlieder_property}, since the two detectors are spacelike. Thus \cref{c4_P_G_P_Gp_0} stated as an exact equality with zero is too strict to describe any local measurement projectors. 

Note that our conclusion is similar to Malament's: two nonzero local projectors associated with spacelike regions cannot satisfy \cref{c4_P_G_P_Gp_0}. However, Malament's conclusion has sometimes been seen as denying the existence of reasonable notions of localization in quantum field theory. On the other hand, if one relaxes \cref{c4_P_G_P_Gp_0} so that $P_{\mathcal{O}} P_{\splike{\mathcal{O}}}$ is not exactly zero---its expectation value may still be very small---then this contradiction disappears.

\subsubsection{Implications for particle localization}
\label{c4_sec_implications_for_particle_localization}
Yet, if local spacelike projectors satisfy $P_{\mathcal{O}} P_{\splike{\mathcal{O}}} \neq 0$, this does raise some questions. Consider some single-particle state $\ket{1}$ of bounded energy spectrum, and suppose it is possible to conduct a projective local measurement in some region $\mathcal{O}$ that detects whether the particle $\ket{1}$ is present. The probability of detecting this particle should then be characterized by some local projector $P_{\mathcal{O}} \in \balg{\mathcal{O}}$. Suppose that we also do another projective measurement at the same time in a far-away spatial region. This other region $\splike{\mathcal{O}}$ is then spacelike separated from $\mathcal{O}$, and the probability of detecting the particle there is characterized by some other projector $P_{\splike{\mathcal{O}}} \in \balg{\splike{\mathcal{O}}}$. In this situation, \cref{c4_rem_schlieder_property} implies that $P_{\mathcal{O}} P_{\splike{\mathcal{O}}} \neq 0$. Furthermore, since $P_{\mathcal{O}}$ and $P_{\splike{\mathcal{O}}}$ commute, $P_{\mathcal{O}} P_{\splike{\mathcal{O}}} \in \balg{\mathcal{O} \cup \splike{\mathcal{O}}}$ is a projector. Assuming the regions $\mathcal{O}$ and $\splike{\mathcal{O}}$ are both bounded, \cref{c4_cor_local_probabilities_never_vanish} implies that
\begin{equation}
    \qexv{1}{P_{\mathcal{O}} P_{\splike{\mathcal{O}}}} > 0.
\end{equation}

Physically, we might be tempted to interpret this as there always being some nonzero chance of detecting the same particle in $\mathcal{O}$ and in $\splike{\mathcal{O}}$ at two different places at the same time. Even if this probability may turn out to be very small, the fact that it cannot be exactly zero seems troubling. Does this mean that there is some small but nonzero chance of the particle instantaneously traveling between the spacelike regions $\mathcal{O}$ and $\splike{\mathcal{O}}$ \cite{malament1996}? Would this not be a superluminal process in violation of relativistic causality?

The answer is: not necessarily. The reason lies in the difference between the probabilities \eqref{c4_prob_l_0_gt_0} and \eqref{c4_prob_psi_psi1_gt_0}. Suppose we observe a click by the local measurement $P_{\splike{\mathcal{O}}}$. This probability satisfies
\begin{equation}\label{c4_prob_click_psi_1}
    \prob\lp\big(\tn{click in $\splike{\mathcal{O}}$} \mid \ket{\psi} \probsymb{=\!} \ket{1}) \probrel{>} 0.
\end{equation}
However, what we really want to determine is the likelihood that there is a particle present, based on this measurement:
\begin{equation}\label{c4_prob_psi_1_click_G}
    \prob\lp\big(\ket{\psi} \probsymb{=\!} \ket{1} \mid \tn{click in $\splike{\mathcal{O}}$}) \probrel{=} {?}
\end{equation}
And this probability is not determined by the conditional detection probability \eqref{c4_prob_click_psi_1} alone, but depends on all situations that could result in a click. For instance the chance of obtaining a click in $\splike{\mathcal{O}}$ in pure vacuum also satisfies
\begin{equation}
    \prob\lp\big(\tn{click in $\splike{\mathcal{O}}$} \mid \ket{\psi} \probsymb{=\!} \ket{0}) \probrel{>} 0.
\end{equation}
In fact, any state of bounded energy spectrum gives some chance of a click in the detector $\splike{\mathcal{O}}$. Assuming a prior probability of several states being possible, the posterior probability \eqref{c4_prob_psi_1_click_G} will always be strictly less than 1:
\begin{equation}\label{c4_prob_psi_1_click_G_lt_1}
    \prob\lp\big(\ket{\psi} \probsymb{=\!} \ket{1} \mid \tn{click in $\splike{\mathcal{O}}$}) \probrel{<} 1.
\end{equation}

In other words, if we are standing in $\splike{\mathcal{O}}$ and observe a click, there is no way of knowing for sure what caused it. It could be a particle, or it could be vacuum, or it could be any number of things. Thus, just because there is a chance of the detector in $\splike{\mathcal{O}}$ giving a click does not mean that observing a click gives us perfect knowledge about what the state is. In particular, a click does not necessarily mean that a particle has propagated from $\mathcal{O}$ to $\splike{\mathcal{O}}$.

Of course, the click probability of a local measurement may be much higher for some particle state than for vacuum with a well-chosen measurement. Thus the posterior probability of a particle being present, given the measurement result, may increase (or decrease) based on observing detection clicks, in line with the inequality \eqref{c4_prob_psi_0_l_0_neq_psi_1_l_0}. In this way our knowledge about the state of the system, even though never reaching 100\%, increases or decreases based on repeated measurements (keeping in mind that measurements change the state).

\begin{remark}\label{c4_rem_presence_of_particles_qft}\remheadp{}
    In general, the presence of particles in quantum field theory can never be determined with 100\% accuracy based on a single local measurement; rather, the likelihood of their presence increases or decreases with measurements. \hfill\null
\end{remark}

This principle enables a meaningful interpretation of measurement results in quantum field theory consistent with relativistic causality. In addition, the difference between the probability statements \eqref{c4_prob_click_psi_1} and \eqref{c4_prob_psi_1_click_G} is important for how to define strictly localized states, which is the topic of the next section.


%% file: 5_localized_states.tex


\newcommand{\FigFiveOne}{%
    \begin{figure}%
    \hspace*{0mm}%
    \raisebox{0mm}{%
        \iftoggle{FigFromPDF}{%
            \includegraphics{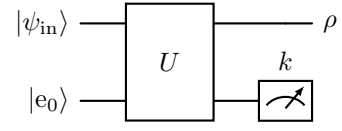}%
        }{%
            \tikzsetnextfilename{fig5-1}%
            \begin{quantikz}[column sep=0.6cm]%
                \lstick{$\ket{\psi_{\tn{in}}}$} %
                & %
                \gate[2]{U} \hphantom{\hspace{0.6cm}} %
                & %
                &[-0.6cm] %
                \rstick{$\rho$} %
                \\ %
                %
                \lstick{$\ket{\mathrm{e}_0}$} %
                & %
                & %
                \meter{k} %
            \end{quantikz}%
        }%
    }%
    \caption{\label{c5_fig_general_quantum_operation}A general quantum operation.}%
    \end{figure}%
}

\iftoggle{ExportTikz}{%
    \FigFiveOne
     
}{}

\section{Strictly localized states}
\label{c5_sec_localized_states}
Having characterized local measurements in quantum field theory, we can now turn to defining what should be meant by strictly localized states. In line with \cref{c3_fig_localized_states_and_local_measurements}, a natural requirement for states strictly localized to some spacetime region $\mathcal{O}$ is that they should not be detectable by any local measurement outside $\mathcal{O}$. However, due to the Reeh--Schlieder theorem and \cref{c4_cor_local_probabilities_never_vanish}, what is meant by \emph{detectable} needs some care. 

In this section we introduce the concept of local equivalence, which makes it possible to define so-called strictly localized states. Such states are exactly undetectable by any local measurement outside some sharp spacetime region $\mathcal{O}$. We then look at how strictly localized states can be generated by Licht or unitary operators. Furthermore, we analyze some key properties of strictly localized states, compare with nonrelativistic quantum mechanics, and discuss strictly localized mixed states including averaged post-measurement states of local measurements. Finally we give a proof of Knight's theorem \cite{knight1961} for hermitian scalar fields and the electromagnetic field, based on the Reeh--Schlieder theorem. This generalizes \cref{c1_sec_invitation} to three dimensions, showing that single photons, or any finite superposition of number states, cannot be strictly localized.

\subsection{What is a strictly localized state?}
\label{c5_sec_what_is_a_localized_state}
Our aim is to define localized states as states that are undetectable outside some spacetime region $\mathcal{O}$, but what does undetectable mean? As discussed in \cref{c2_sec_history}, it could perhaps be natural to think of states being locally detectable somewhere when there is some local measurement outcome that has nonvanishing probability there. This can however not be a good indicator because all local measurement outcomes occur with nonvanishing probability even in vacuum, as shown in \cref{c4_cor_local_probabilities_never_vanish}. 

What is needed is to use the distinction between probability of detection given a particle, and the probability of a particle given a detection, as discussed in \cref{c4_sec_implications_for_particle_localization}. Practically, one detects a particle by performing a measurement and updating the posterior probabilities of candidate prior states using observed data. For example in a repeated experiment of measuring a local observable $L$ for some unknown state $\ket{\psi}$ prepared identically every time, an observed frequency distribution $p\lp(l)$ of outcomes $l$ is produced. 

It is then easy to see that if two candidate states assign the same likelihood to every possible measurement outcome, Bayes' rule leaves their posterior odds unchanged. In other words, if two prior states $\ket{\psi_1}$ and $\ket{\psi_2}$ give the exact same statistics
\begin{equation}\label{c5_prob_l_psi_psi1}
    \prob\lp\big(l \mid \ket{\psi} \probsymb{=} \ket{\psi_1}) \probrel{=} \prob\lp\big(l \mid \ket{\psi} \probsymb{=} \ket{\psi_2}), \sepc \forall l
\end{equation}
for all possible local measurements in $\mathcal{O}$, then there is no way to distinguish between $\ket{\psi_1}$ and $\ket{\psi_2}$ within the region $\mathcal{O}$.

\subsubsection{Strictly localized pure states}
\label{c5_sec_strictly_localized_pure_states}
With this in mind we can formalize what is meant by a state being locally undetectable. The question to ask is what the possible prior states are. This is also the lesson from the Fermi-problem analysis in \cite{ferretti1968} and from the general discussion in \cite{buchholz1994}, as mentioned in \cref{c2_sec_interpreting_measurement_probabilities_in_quantum_field_theory}. In the most typical scenario, the most appropriate reference state to compare with is \emph{the vacuum}, in the sense that we want to know whether there is any other state present than just vacuum. In this situation, a state is undetectable exactly when it is locally \emph{indistinguishable} from the vacuum by any local measurement. 

\begin{definition}\label{c5_def_localized_states}\textup{(Strictly localized pure states).}
    We say that a pure state $\ket{\psi}$ is strictly localized to a region $\mathcal{O}$ if
    \begin{equation}\label{c5_qexv_psi_L_0_L}
        \qexv{\psi}{L} = \qexv{0}{L}, \sepc \forall L \in \balg{\compl{\mathcal{O}}},
    \end{equation}
    where $\compl{\mathcal{O}}$ is the complement region of $\mathcal{O}$.
\end{definition}

In line with the discussion in \cref{c3_sec_physically_realizable_local_measurements}, the set of local observables that are physically realizable as a measurement in $\compl{\mathcal{O}}$ might be strictly smaller than the hermitian operators in $\balg{\compl{\mathcal{O}}}$. In other words, \cref{c5_def_localized_states} might in fact be too strict. However, assuming that physically realizable local observables are contained in $\balg{\compl{\mathcal{O}}}$, \cref{c5_def_localized_states} is at least sufficient for ensuring that such states are indistinguishable in $\compl{\mathcal{O}}$.

Note that a state being strictly localized to $\mathcal{O}$ does not mean that expectation values of all local observables in $\compl{\mathcal{O}}$ vanish. Rather, they equal the corresponding vacuum values, which are generally nonzero. Still, a strictly localized state is perfectly undetectable outside some sharp region $\mathcal{O}$, because everywhere outside $\mathcal{O}$ it looks exactly like the vacuum. 

It might be confusing why it is the full, set-theoretic complement $\compl{\mathcal{O}}$ that is used in \cref{c5_def_localized_states}, and not the spacelike complement. After all, a state that is detectable in some spatial region will generally remain detectable in the region's future lightcone. However, it is important to remember that $\mathcal{O}$ is a region of spacetime, so the description of $\mathcal{O}$ already takes the dynamics of where the state is strictly localized over time into account. The connection between \cref{c5_def_localized_states} and how the spatial localization volume changes with time is discussed in \cref{c6_sec_localization_and_causality}.

It is also possible to consider alternatives to \cref{c5_def_localized_states} that do not demand exact equality, but rather that the difference has to go to zero for measurements taken at some fixed time and increasingly far away in space. This allows states that are \emph{concentrated} in some spatial region, with \enquote{tails} that decay at large distances but need not vanish. Such types of states are usually denoted weakly or essentially localized \cite{haag1965,schlieder1965}. This review primarily focuses on strictly localized states, but we comment on weakly localized photonic states in \cref{c7_sec_weakly_localized_photons}.

\subsubsection{Strictly localized mixed states}
\label{c5_sec_mixed_localised_states}
We may extend the definition of strictly localized states such that it applies to mixed states \cite{licht1966,hellwig1970a}:

\begin{definition}\label{c5_def_localized_mixed_states}\textup{(Strictly localized mixed states).}
    We say that a state with density operator $\rho$ is strictly localized to a region $\mathcal{O}$ if
    \begin{equation}\label{c5_qexv_rho_L_0_L}
        \tr\lp(\rho L) = \qexv{0}{L}, \sepc \forall L \in \balg{\compl{\mathcal{O}}},
    \end{equation}
    where $\compl{\mathcal{O}}$ is the complement region of $\mathcal{O}$.
\end{definition}

This definition is useful for example when the quantum field degrees of freedom have been entangled with a source or another auxiliary quantum system. After tracing out the auxiliary system, we are left with a density operator $\rho$ for the quantum field, which may correspond to a mixed state. Further properties of strictly localized mixed states are discussed in \cite{rustad2026}; in particular, it turns out that they cannot always be written as mixtures of pure states strictly localized to the same region.

More generally, we can replace the comparison with vacuum by comparison between any two states. This leads to the notion of local equivalence:

\begin{definition}\label{c5_def_local_equivalence}\textup{(Local equivalence).}
    We say that two states with density operators $\rho_1$ and $\rho_2$ are locally equivalent in a region $\mathcal{O}$ if
    \begin{equation}\label{c5_local_equivalence}
        \tr\lp(\rho_1 L) = \tr\lp(\rho_2 L), \sepc \forall L \in \balg{\mathcal{O}}.
    \end{equation}
\end{definition}

This definition covers both pure and mixed states. Locally equivalent states are indistinguishable by all physically realizable local measurements in $\mathcal{O}$, under our assumption that these observables belong to $\balg{\mathcal{O}}$. In particular, a state is strictly localized to $\mathcal{O}$ if and only if it is locally equivalent to vacuum in $\compl{\mathcal{O}}$. 

This review primarily focuses on strictly localized pure states. We only consider mixed states where needed, particularly when discussing post-measurement states averaged over measurement outcomes in \cref{c5_sec_post_measurement_states_of_local_measurements}.

\subsubsection{Hilbert space does not factor}
\label{c5_sec_hilbert_space_does_not_factor}
At this point it might seem like all this terminology would simply lead to a tensor-product picture in which states strictly localized to $\mathcal{O}$ could be easily constructed in the form $\ket{\psi}_{\mathcal{O}} \otimes \ket{0}_{\compl{\mathcal{O}}}$, where the two factors belong to Hilbert spaces associated with $\mathcal{O}$ and $\compl{\mathcal{O}}$, respectively. After all, this is how independent subsystems in nonrelativistic quantum mechanics are treated. 

However, for spacetime degrees of freedom, such a picture is in general \emph{not available}. In quantum field theory, there is no tensor-product factorization $H = H_{\mathcal{O}} \otimes H_{\compl{\mathcal{O}}}$ for which all operators in $\balg{\mathcal{O}}$ and $\balg{\compl{\mathcal{O}}}$ act only on their respective tensor factors \cite{witten2018}. There is therefore no tensor-product factorization for a sharp region and its complement. This is related to the fact that local algebras of bounded spacetime diamonds in quantum field theory are so-called von Neumann type III \cite{witten2018}, with the consequence that there is in general no partial trace available. Strictly localized states are therefore always a class of states of the entire system---they are \enquote{global states}. 

A possible source of confusion here is the so-called \emph{split property} discussed in algebraic quantum field theory. This asserts that a factorization structure is possible, but it applies to regions that are spacelike (and separated by some finite gap). This does therefore not apply to two regions that are complementary. The failure of factorization for complementary spacetime regions can be seen directly from the field commutator. If such a decomposition were true, every operator local to one region would necessarily commute with every operator local to the other. However, complementary regions contain points that are timelike or lightlike to each other, and the field does in general not commute within/on the lightcone. Hence one can find local operators built from smeared fields that do not commute, meaning that the tensor-product decomposition is not possible. 

We therefore cannot define strictly localized states to be states that \emph{are equal to} vacuum in a Hilbert space tensor factor associated with $\compl{\mathcal{O}}$; such a factor is not available. Instead, we must find other ways to formalize it, such as \cref{c5_def_localized_states}, which says that strictly localized states are those that \emph{look like} vacuum in $\compl{\mathcal{O}}$. 

It is also possible to understand this in connection with \cref{c4_cor_local_probabilities_never_vanish}. For some bounded spacetime region $\mathcal{O}$, imagine that a decomposition such as $H = H_{\mathcal{O}} \otimes H_{\compl{\mathcal{O}}}$ was possible, so that local operators to $\mathcal{O}$ would be the operators acting on $H_{\mathcal{O}}$. In that case, one local operator that should be available would be a local vacuum projector $P_{\tn{vacuum in $\mathcal{O}$}} = \ket{0}_{\mathcal{O}}\bra{0} \otimes \identity_{\compl{\mathcal{O}}}$. However, assuming that the global vacuum state would factor similarly as $\ket{0} = \ket{0}_{\mathcal{O}} \otimes \ket{0}_{\compl{\mathcal{O}}}$, measuring this projector would clearly result in $\qexv{0}{P_{\tn{vacuum in }\mathcal{O}}} = 1$. This is impossible according to \cref{c4_cor_local_probabilities_never_vanish}. Thus, this situation is not correct, and there is in general no local projector available in $\mathcal{O}$ that can check if the state \emph{is} vacuum in $\mathcal{O}$. The only thing we can do is check whether a state \emph{looks like} vacuum in the sense of \cref{c5_qexv_psi_L_0_L}. 

In summary, we are from nonrelativistic quantum mechanics used to dividing different degrees of freedom of quantum systems into relevant ones, and tracing over irrelevant ones. However, these results show that complementary regions of spacetime cannot be traced out.

\subsubsection{Localization with respect to observables vs.\ the full algebra}
\label{c5_sec_localization_with_respect_to_observables_vs_the_full_algebra}
As noted in \cref{c4_sec_bounded_local_operators}, the set of bounded local observables is a subset of local operators $\balg{\mathcal{O}}$ but not a subalgebra. One question is whether phrasing \cref{c5_def_local_equivalence} in terms of all operators in $\balg{\mathcal{O}}$ or just the hermitian ones gives equivalent definitions. The nuance is whether the additional requirement of observables $L$ being hermitian allows more solutions.

It turns out that there is no difference. Assume for contradiction that the set of pairs of states satisfying \cref{c5_def_local_equivalence} for all hermitian operators in $\balg{\mathcal{O}}$ is larger than the set satisfying it for all operators in $\balg{\mathcal{O}}$. That would imply there exist a (non-hermitian) operator $L \in \balg{\mathcal{O}}$ and two states $\rho_1$ and $\rho_2$ such that
\begin{equation}\label{c5_psi1_L_psi2_L_l}
    \tr\lp(\rho_1 L) - \tr\lp(\rho_2 L) = l \neq 0,
\end{equation}
for some nonzero complex constant $l$, while this difference must be zero for all $L$ that are hermitian. However, any $L \in \balg{\mathcal{O}}$ can be written as a complex linear combination of two hermitian operators in the same algebra, $L = L_1 / 2 - i L_2 / 2$, where
\begin{equation}
    L_1 = L + \adj{L}, \sepc L_2 = i L - i \adj{L}.
\end{equation}
Thus
\begin{align}
    l &= \frac{1}{2} \lp\big[\tr\lp(\rho_1 L_1) - \tr\lp(\rho_2 L_1)] - \frac{i}{2} \lp\big[\tr\lp(\rho_1 L_2) - \tr\lp(\rho_2 L_2)] \nonumber \\
      &= 0,
\end{align}
which is a contradiction. Hence we could equally phrase \cref{c5_def_local_equivalence} in terms of all hermitian $L \in \balg{\mathcal{O}}$.

\subsection{Generating strictly localized states}
\label{c5_sec_generating_strictly_localized_states}
An important continuation of Knight's work was done in \cite{licht1963,licht1966}. Licht found a way to characterize and generate all strictly localized states from the vacuum through a class of isometric operators, which we label \emph{Licht operators}.

\subsubsection{Licht operators}
\label{c5_sec_licht_operators}
\begin{definition}\label{c5_def_licht_operators}\textup{(Licht operators).\footnote{\label{c5_foot_licht_operators}See \cite{licht1963}.}}
    An operator $W$ is called a Licht operator associated with a region $\mathcal{O}$ if it satisfies
    \begin{subequations}\label{c5_licht_W}
    \begin{align}
        \sepc\sepc \comm{W}{L} &= 0, \sepc \forall L \in \balg{\compl{\mathcal{O}}}, \label{c5_comm_W_L} \\
        \adj{W} W &= \identity. \label{c5_Wd_W_I}
    \end{align}
    \end{subequations}
    I.e., $W$ commutes with all local operators \emph{outside} of $\mathcal{O}$, and it is an isometry. Note that since we have an infinite-dimensional Hilbert space, $W$ may fail to be unitary, $W \adj{W} \neq \identity$, even though it satisfies \cref{c5_Wd_W_I}. For electromagnetism, we consider $W$ on the physical photon Hilbert space.
\end{definition}

For any operator $W$ satisfying \cref{c5_licht_W}, it is clear that
\begin{equation}
    \qexv{0}{\adj{W} L W} = \qexv{0}{L}, \sepc \forall L \in \balg{\compl{\mathcal{O}}},
\end{equation}
meaning that the state $\ket{\psi} = W \ket{0}$ is strictly localized to $\mathcal{O}$.

\begin{remark}\remheadp{}
    For every Licht operator $W$ associated with a region $\mathcal{O}$, the state $\ket{\psi} = W \ket{0}$ is strictly localized to $\mathcal{O}$.
\end{remark}

Less obvious is the converse, i.e., that for every state $\ket{\psi}$ strictly localized to $\mathcal{O}$, there is a Licht operator $W$ associated with $\mathcal{O}$. Licht was able to prove this in \cite{licht1963}, and further that this operator is in fact unique. 

\begin{theorem}\label{c5_thm_lichts_theorem}\textup{(Licht's theorem).\footnote{\label{c5_foot_lichts_theorem}See \cite{licht1963}, Theorem~III and Corollary~1.}}
    Let $\mathcal{O}$ be a region whose complement $\compl{\mathcal{O}}$ contains a nonempty open subregion. For every state $\ket{\psi}$ strictly localized to $\mathcal{O}$, there is a corresponding Licht operator $W$ associated with $\mathcal{O}$ such that $\ket{\psi} = W \ket{0}$. This operator is the unique operator $W$ satisfying \cref{c5_comm_W_L} that produces the state from vacuum $\ket{\psi} = W \ket{0}$.
\end{theorem}

Thus \cref{c5_qexv_psi_L_0_L,c5_licht_W} are equivalent in the following sense: For every state $\ket{\psi}$ strictly localized to $\mathcal{O}$, there exists a unique operator $W$ satisfying \cref{c5_licht_W} such that $\ket{\psi} = W \ket{0}$. Conversely, for every operator $W$ satisfying \cref{c5_licht_W}, the state $\ket{\psi} = W \ket{0}$ is strictly localized to $\mathcal{O}$.

\subsubsection{Finding Licht operators}
\label{c5_sec_finding_licht_operators}
Licht's theorem is very powerful and contains a lot of information if one is given a valid Licht operator $W$. On the other hand, obtaining an explicit expression for $W$ from a given strictly localized state $\ket{\psi}$ can still be difficult. What's worse, even if we are given a candidate isometry $W$, it may be difficult to establish that it commutes with the entire algebra $\balg{\compl{\mathcal{O}}}$. Here we introduce a simpler, sufficient condition.

\begin{remark}\label{c5_rem_licht_operators_phi}\remheadp{}
    For a hermitian scalar field $\phi\lp(x)$, let $W$ be an operator that satisfies
    \begin{subequations}\label{c5_licht_W_phi}
    \begin{align}
        \sepc\sepc \comm{W}{\phi\lp(x)} &= 0, \sepc \forall x \in \compl{\mathcal{O}}, \label{c5_comm_W_phi} \\
        \adj{W} W &= \identity, \label{c5_Wd_W_I_phi}
    \end{align}
    \end{subequations}
    for some region $\mathcal{O}$. Then $W$ is a Licht operator associated with $\mathcal{O}$.
\end{remark}

For electromagnetism, we can use the exact same method, where we require \cref{c5_comm_W_phi} to hold component-wise.

\begin{remark}\label{c5_rem_licht_operators_A}\remheadp{}
    For the electromagnetic potential $A^\mu\sephn\lp(x)$, let $W$ be an operator on the physical photon Hilbert space that satisfies
    \begin{subequations}\label{c5_licht_W_A}
    \begin{align}
        \sepc\sepc \comm{W}{A^\mu\sephn\lp(x)} &= 0, \sepc \forall x \in \compl{\mathcal{O}}, \label{c5_comm_W_A} \\
        \adj{W} W &= \identity, \label{c5_Wd_W_I_A}
    \end{align}
    \end{subequations}
    for some region $\mathcal{O}$. Then $W$ is a Licht operator associated with $\mathcal{O}$.
\end{remark}

The sufficiency of \cref{c5_rem_licht_operators_phi,c5_rem_licht_operators_A} follows by noting that such $W$ must commute with all self-adjoint smeared fields, $\phi_f$ or $A_f$, respectively, with suitable functions $f\lp(x)$ supported in $\compl{\mathcal{O}}$. By a standard result in functional analysis, it then follows that $W$ must commute with every spectral projector of these smeared fields \cite[Theorem~13.33]{rudin1973}. Thus $W$ commutes with all finite linear combinations and products of these spectral projectors too, and this also holds for weak operator limits. By construction of the bounded algebra, it therefore follows that $W$ commutes with all $L \in \balg{\compl{\mathcal{O}}}$. Together with $\adj{W}W = \identity$, this shows that $W$ satisfies \cref{c5_def_licht_operators} and is a Licht operator associated with $\mathcal{O}$, meaning that $W \ket{0}$ is a state strictly localized to $\mathcal{O}$.

It is also useful to note that compositions of Licht operators are themselves Licht operators associated with the union of the regions:

\begin{remark}\label{c5_rem_licht_operators_comp}\remheadp{}
    Let $W_1$ be a Licht operator associated with a region $\mathcal{O}_1$, and $W_2$ be a Licht operator associated with a region $\mathcal{O}_2$. Then $W = W_1 W_2$ is a Licht operator associated with $\mathcal{O}_1 \cup \mathcal{O}_2$.
\end{remark}

This follows from \cref{c5_def_licht_operators}, because $\compl{\lp(\mathcal{O}_1 \cup \mathcal{O}_2)} = \compl{\mathcal{O}_1} \cap \compl{\mathcal{O}_2}$. Therefore any $L \in \balg*{\compl{\lp(\mathcal{O}_1 \cup \mathcal{O}_2)}}$ satisfies $\comm{W_1 W_2}{L} = \comm{W_1}{L} W_2 + W_1 \comm{W_2}{L} = 0$. The operator $W$ is also isometric since $W^\dagger W = W_2^\dagger W_1^\dagger \padj{W}_1 \padj{W}_2 = \identity$.

On the other hand, superpositions of states strictly localized to some region are generally not strictly localized to that region \cite{licht1963}. In other words, the states strictly localized to a region $\mathcal{O}$ do not form a subspace of the Hilbert space. We discuss this surprising fact in more detail in \cref{c5_sec_diff_nonrel_qm}.

\begin{remark}\label{c5_rem_licht_operators_superposition}\remheadp{}
    Let $W_1$ and $W_2$ be Licht operators associated with the same region $\mathcal{O}$. Then every nonzero superposition $c_1 W_1 + c_2 W_2$ can be normalized to a Licht operator for all $c_1, c_2 \in \complexes$ if and only if $W_2^\dagger \padj{W}_1$ is proportional to the identity operator.
\end{remark}

It is important to note that Licht operators are not the only operators creating some given strictly localized state from the vacuum. In other words, if an operator $W$ does \emph{not} satisfy \cref{c5_licht_W}, it does not follow immediately whether the state $W \ket{0}$ is strictly localized or not. This is because Licht's theorem only guarantees that for any strictly localized state $\ket{\psi}$, there exists an operator $W$ satisfying \cref{c5_licht_W} such that $\ket{\psi} = W \ket{0}$. There are generally other operators $W$ such that $\ket{\psi} = W  \ket{0}$ but that do not satisfy \cref{c5_licht_W}.

For showing that some state is \emph{not} strictly localized to a region $\mathcal{O}$, the most straightforward method is picking some specific local observable $L$ associated with region $\compl{\mathcal{O}}$ and showing that \cref{c5_qexv_psi_L_0_L} does not hold for that particular $L$. In line with \cref{c3_sec_physically_realizable_local_measurements}, to be safe that the state really is distinguishable from the vacuum by some physically realizable local measurement, we should show it for some pointwise observable $L$. For example for a hermitian scalar field $\phi\lp(x)$, this observable can typically be the normal-ordered squared field $\norder{\phi^2\sephn\lp(x)}$, and for electromagnetism it can typically be the normal-ordered energy density $\frac{1}{2}\norder{\vect{E}^2\sephn\lp(x)} + \frac{1}{2}\norder{\vect{B}^2\sephn\lp(x)}$. Since the expectation values of these observables vanish in vacuum, showing that they do not vanish for some state $\ket{\psi}$ in a region $\compl{\mathcal{O}}$ is sufficient to show that $\ket{\psi}$ cannot be strictly localized to $\mathcal{O}$.

\subsubsection{Unitary operators}
\label{c5_sec_unitary_operators}
It can sometimes be useful to consider unitary operators $U$ instead of Licht operators $W$, which are only isometric. Given any Licht operator $W$ satisfying \cref{c5_licht_W_phi} or \cref{c5_licht_W_A}, we can always obtain a unitary operator by extending the Hilbert space $H$ of the field to include a two-dimensional source space $\complexes^2$ as well \cite{gulla2023}. Let a basis for this source space be a ground state $\ket{\mathrm{g}}$ and an excited state $\ket{\mathrm{e}}$, and define corresponding raising and lowering operators $\sigma_+ = \ket{\mathrm{e}} \bra{\mathrm{g}}$ and $\sigma_- = \ket{\mathrm{g}}\bra{\mathrm{e}}$. We can then construct an operator $U$ acting on $H \otimes \complexes^2$ based on the Licht operator $W$ acting on $H$ as
\begin{equation}\label{c5_U_W}
    U = W \otimes \sigma_- + \adj{W} \otimes \sigma_+ + \lp(\identity - W \adj{W}) \otimes \ket{\mathrm{g}} \bra{\mathrm{g}}.
\end{equation}
Furthermore, this operator $U$ is unitary.

The commutation relations \eqref{c5_comm_W_phi} and \eqref{c5_comm_W_A} also carry over to $U$ defined by \cref{c5_U_W}. In the scalar case, since $\phi\lp(x)$ is hermitian, $\comm{W}{\phi\lp(x)}=0$ implies $\comm{\adj{W}}{\phi\lp(x)}=0$. A similar argument holds for $A^\mu\lp(x)$, with the adjoint interpreted with extra nuance in Lorenz gauge. Thus, every term in \cref{c5_U_W} commutes with the field outside $\mathcal{O}$, meaning that $U$ inherits the same commutation property as $W$. Since $U$ is unitary, we may rewrite this property in the case of $\phi\lp(x)$ as
\begin{equation}\label{c5_Ud_phi_U}
    \adj{U} \phi\lp(x) U = \phi\lp(x), \sepc \forall x \in \compl{\mathcal{O}},
\end{equation}
and in the case of $A^\mu\lp(x)$ as
\begin{equation}\label{c5_Ud_A_U}
    \adj{U} A^\mu\sephn\lp(x) U = A^\mu\sephn\lp(x), \sepc \forall x \in \compl{\mathcal{O}}.
\end{equation}

Given such an operator $U$, the corresponding strictly localized state $\ket{\psi} = W  \ket{0}$ can be expressed by letting $U$ act on the initial state $\ket{0} \otimes \ket{\mathrm{e}}$, and tracing out the source space:
\begin{equation}\label{c5_psi_tr_src}
    \ket{\psi} \bra{\psi} = \tr_{\tn{src}} \lp\Big[U \lp\big(\ket{0}\bra{0} \otimes \ket{\mathrm{e}}\bra{\mathrm{e}}) \seph \adj{U}].
\end{equation}
We can write it directly in this form because the reduced state is pure. Thus, given a Licht operator $W$ satisfying \cref{c5_licht_W_phi} or \cref{c5_licht_W_A}, we can generate the same strictly localized state using a unitary operator $U$ on $H \otimes \complexes^2$ and tracing out the source, as in \cref{c5_psi_tr_src}. This unitary leaves the field outside $\mathcal{O}$ invariant, as expressed by \cref{c5_Ud_phi_U} or \cref{c5_Ud_A_U}, respectively.

More generally, we may start directly with an arbitrary unitary operator $U$ on a combined field-source space satisfying \cref{c5_Ud_phi_U} or \cref{c5_Ud_A_U}. By the same argument as in \cref{c5_sec_finding_licht_operators}, $U$ then commutes with every $L \in \balg{\compl{\mathcal{O}}}$. Consequently, letting $U$ act on $\ket{0} \otimes \ket{\mathrm{e}}$ and tracing out the source gives a reduced field state $\rho$ that satisfies $\tr\lp(\rho L) = \qexv{0}{L}, \sepc \forall L \in \balg{\compl{\mathcal{O}}}$. Thus $\rho$ is strictly localized to $\mathcal{O}$, although it need not be pure. If it is pure, we can compute it using \cref{c5_psi_tr_src}.

\FigFiveOne

The fact that we get a pure state in \cref{c5_psi_tr_src} after interacting with the source warrants a comment. Consider a general quantum system in some arbitrary state $\ket{\psi_{\tn{in}}}$ interacting with some environment under an arbitrary unitary operator $U$ on the total space. After the interaction, label the system's reduced state by $\rho$ and let the environment be measured in some basis, as depicted in \cref{c5_fig_general_quantum_operation}. Assuming the environment input is some fixed state $\ket{\mathrm{e}_0}$, an interesting question is whether it is possible to write
\begin{equation}\label{c5_U_Us_Ue}
    U \lp(\ket{\psi_{\tn{in}}} \otimes \ket{\mathrm{e}_0}) \overset{\displaystyle ?}{=} \lp(U_{\tn{sys}} \otimes U_{\tn{env}}) \, \lp(\ket{\psi_{\tn{in}}} \otimes \ket{\mathrm{e}_0}), \sepc \forall \ket{\psi_{\tn{in}}},
\end{equation}
for some fixed unitary operators $U_{\tn{sys}}$ and $U_{\tn{env}}$ acting on each subsystem alone. When such a decomposition is possible, the system's input-output evolution for this environment can be described by the unitary operator $U_{\tn{sys}}$ alone, without including the environment.

When \cref{c5_U_Us_Ue} holds, the system's reduced state $\rho$ will always be pure:
\begin{equation}
    \text{\cref*{c5_U_Us_Ue} holds} \sepc \Rightarrow \sepc \rho \text{ pure} \;\; \forall \ket{\psi_{\tn{in}}}.
\end{equation}
The opposite implication is however not true:
\begin{equation}
    \text{\cref*{c5_U_Us_Ue} holds} \sepc \nLeftarrow \sepc \rho \text{ pure} \;\; \forall \ket{\psi_{\tn{in}}}.
\end{equation}
For example, taking both the system and environment to be qubits and $U$ to be a SWAP gate gives $\rho = \ket{\mathrm{e}_0}\bra{\mathrm{e}_0}$ for every input state. No fixed unitary $U_{\tn{sys}}$ can map every input state to the same output state, so \cref{c5_U_Us_Ue} cannot hold.

On the other hand, there is another implication of \cref{c5_U_Us_Ue} holding, namely that there is a measurement basis for the environment which always results in a predictable value:
\begin{equation}
    \text{\cref*{c5_U_Us_Ue} holds} \sepc \Rightarrow \sepc \Pr\lp(k) = \delta_{kk'} \;\; \forall \ket{\psi_{\tn{in}}},
\end{equation}
for some $k'$. In other words, the environment is left in a fixed pure state independent of $\ket{\psi_{\tn{in}}}$ (although it can depend on $U$ and $\ket{\mathrm{e}_0}$). And this implication \emph{is} reversible, but only in finite dimensions! I.e., 
\begin{equation}\label{c5_U_tot_dec_from_Pr_k}
    \text{\cref*{c5_U_Us_Ue} holds} \sepc \stackrel{\mathclap{\mbox{\tn{\scriptsize{fin.\ dim.}}}}}{\Leftarrow} \sepc \Pr\lp(k) = \delta_{kk'} \;\; \forall \ket{\psi_{\tn{in}}}.
\end{equation}
The reason is that a predictable environment output for every input implies a quantum operation that can be represented by a single Kraus operator satisfying $\adj{E}_{k'} \padj{E}_{k'} = \identity$. I.e., $E_{k'}$ is isometric. In finite dimensions this means that $E_{k'}$ must be unitary as well, but this is in general not true for infinite dimensions. 

Thus, in infinite dimensions, a larger set of possible time evolutions is possible when interacting with an environment than for closed systems, even when the environment is left in a predictable state. And it is not needed that the environment is infinite-dimensional; it is enough that the system is. 

This is precisely what happens in \cref{c5_U_W}. An initial state $\ket{0} \otimes \ket{\mathrm{e}}$ can represent a source initialized in an excited state with the other modes being vacuum. The time evolution $U$ then has the source decay to its ground state $\ket{\mathrm{g}}$ according to
\begin{equation}
    U \seph \lp\big(\ket{0} \otimes \ket{\mathrm{e}}) = W \ket{0} \otimes \ket{\mathrm{g}}.
\end{equation}
The output is a product state, meaning that there is no entanglement between the source and system. In fact, the source is left in a predictable state $\ket{\mathrm{g}}$, so that any subsequent measurement of the source in the $\{\ket{\mathrm{g}}, \ket{\mathrm{e}}\}$ basis always gives the outcome $\mathrm{g}$. This means that the operator $W$ acting on the system space must be isometric, but for the infinite-dimensional Hilbert space of a quantum field, $W$ does not have to be unitary. 

To sum up, we have seen that in order to construct strictly localized states, we may look for Licht operators $W$ satisfying \cref{c5_licht_W_phi} or \cref{c5_licht_W_A}. A strictly localized state is then given by $W \ket{0}$. Alternatively, we may look for unitary operators satisfying \cref{c5_Ud_phi_U} or \cref{c5_Ud_A_U}. In that case, a possibly mixed strictly localized state is given by tracing out the environment (if any) after letting $U$ act on $\ket{0}$ and some initial state of the environment (if any). The reason for writing \enquote{if any} is that although Licht operators are not \emph{necessarily} unitary, they \emph{can} be; if $W$ is itself unitary, we may take it directly as our $U$ without needing to extend the Hilbert space with a source space.

\subsection{Discussion of strictly localized states}
\label{c5_sec_discussion_localized_states}
We now discuss some properties of strictly localized states, compare them with their nonrelativistic counterparts, and consider the localization of post-measurement states of measuring local observables.

\subsubsection{Some properties of strictly localized states}
\label{c5_sec_some_properties_localized_states}
An important property stated in \cite{knight1961} is that strictly localized states cannot be finite superpositions of number states:

\begin{theorem}\label{c5_thm_knight_inf_photon_numbers}\textup{(Knight's theorem).}
    Let $\mathcal{O}$ be a region whose complement $\compl{\mathcal{O}}$ contains a nonempty open subregion. A nonvacuum pure state (of a hermitian scalar field or the electromagnetic field) strictly localized to $\mathcal{O}$ contains terms with arbitrarily high particle numbers.    
\end{theorem}

In particular, single particles or number states cannot be strictly localized. Knight argued for \cref{c5_thm_knight_inf_photon_numbers} using diagrams and analyticity [see also \cite{bievre2006}]. We give a simple proof based on the Reeh--Schlieder theorem in \cref{c5_sec_proof_knight}. We also have the observation \cite{gulla2023,rustad2023}:

\begin{remark}\label{c5_rem_localized_states_not_dense}\remheadp{}
    The states strictly localized to $\mathcal{O}$ are not dense in the Hilbert space.
\end{remark}

To see this, consider a single-photon state with a nonzero tail outside $\mathcal O$. There is then a bounded local observable outside $\mathcal{O}$ whose expectation value differs from its vacuum value by some nonzero amount. Since expectation values of bounded operators vary continuously with the state, this remains true for all sufficiently nearby states. Thus no state in some neighborhood of the single-photon state is strictly localized to $\mathcal O$.

Still, it turns out that any state in the Hilbert space can be approximated arbitrarily well by a superposition of strictly localized states \cite[Ch.~V.5]{haag1996}. This is not surprising in light of the Reeh--Schlieder theorem, and does not contradict strict localization because of \cref{c5_rem_licht_operators_superposition}.

In \cref{c7_sec_examples_of_localized_states} we give several examples of strictly localized states. One natural example that should be mentioned already is states generated by the Licht operator $W=\exp\lp(i\phi_f)$, where $\phi_f$ is a smeared field of the form \eqref{c4_phi_f_bounded_algebra}. As discussed in detail in \cref{c7_sec_coherent_states}, this gives a coherent state strictly localized to the support region of $f\lp(x)$ in addition to the forward and backward lightcones. Such coherent states illustrate  \cref{c5_rem_licht_operators_comp,c5_rem_licht_operators_superposition,c5_thm_knight_inf_photon_numbers} well: First, a composition of such Licht operators gives an exponential with the superposed fields in the exponent, which is clearly a new Licht operator, exactly in line with \cref{c5_rem_licht_operators_comp}. Second, it is well known that a superposition of two distinct coherent states is generally not a coherent state. And clearly, according to \cref{c5_rem_licht_operators_superposition}, such superpositions are generally not strictly localized either. Finally, the particle number of a coherent state has a Poisson number distribution up to infinity, consistent with \cref{c5_thm_knight_inf_photon_numbers}.

Coherent states give valuable intuition about strictly localized states. Given that classical fields trivially can be strictly localized, it is no surprise that the coherent states can be as well, as they constitute the link between quantum fields and classical fields. We keep them in mind when discussing certain differences between localization in nonrelativistic quantum mechanics and quantum field theory.

\makeshortsubsubsection[0.97]
\subsubsection{Difference from nonrelativistic quantum mechanics}
\label{c5_sec_diff_nonrel_qm}
In nonrelativistic quantum mechanics the position eigenstates in a given region of three-dimensional space form a basis of strictly localized states. Moreover, for two disjoint spatial regions, the corresponding subspaces of localized states are orthogonal.

In quantum field theory we should not expect this situation to persist \cite{falcone2024}. Taking the strictly localized coherent states as examples, we know that superposition is a highly nontrivial operation [see, e.g., \cite[Ch.~7]{gerry2004}], generally resulting in nonclassical states. Thus, although it is somewhat inconvenient that superpositions of strictly localized states are not generally strictly localized, this is not a problem for the theory. As argued by \cite{wallace2006}, there are several properties of quantum states that are not preserved under superposition, such as being energy eigenstates or being entangled. 

Still, it may appear surprising that two states strictly localized to disjoint regions of space at some fixed time $t=0$ are not necessarily orthogonal. Some authors even argue that this is a deficiency of Knight's definition of strict localization \cite{fleming2000}. But why would we expect such states to be orthogonal? Perhaps because in nonrelativistic quantum mechanics, states localized to disjoint regions can be perfectly distinguished by an ideal position measurement and therefore must belong to orthogonal subspaces.

In general, two orthogonal global states can be perfectly distinguished by some global measurement. However, to determine whether two states can be distinguished within a given region, one must consider measurements local to that region. In quantum field theory, projectors local to spacelike separated regions cannot in general be mutually orthogonal, as discussed in \cref{c4_sec_connection_to_malaments_theorem}. Thus, the connection between spatial separation and orthogonality familiar from nonrelativistic quantum mechanics does not carry over to quantum field theory.

This nonorthogonality is well illustrated by coherent states. If two coherent states describe weak pulses in disjoint regions, both states are close to vacuum and therefore have a large mutual overlap. But this does not imply any overlap of the regions in which the corresponding field disturbances are localized.

The distinction between global states and local measurement statistics can be illustrated further by a simple analogy from nonrelativistic quantum mechanics \cite{redhead1995,wallace2006}. It is well known that the vacuum state in quantum field theory is entangled. Let the Bell state
\begin{equation}
    \ket{\Omega} = \frac{\ket{00}+\ket{11}}{\sqrt{2}} 
\end{equation}
be regarded as a \enquote{vacuum state}. Here, for example, $\ket{00} \equiv \ket{0_A}\otimes\ket{0_B}$ is a state with qubits $\ket{0_A}$ and $\ket{0_B}$ in Alice's and Bob's subsystems, respectively. Furthermore, for local unitary operations $U_A$ at Alice's side, the states $U_A\ket{\Omega}$ can be considered strictly localized to Alice in the sense that all of Bob's measurement statistics agree with those of $\ket{\Omega}$.

As an analogy of the Reeh--Schlieder theorem, we now realize that Alice may create any of the four Bell states from $\ket{\Omega}$ by applying the Pauli matrices locally. Thus she may create a complete basis of the total Hilbert space. For example, if Alice applies the Pauli $X$ matrix, $X_A$, the total state transforms according to
\begin{equation}
    \ket{\Omega} = \frac{\ket{00}+\ket{11}}{\sqrt{2}} \mapsto X_A \ket{\Omega}  = \frac{\ket{10}+\ket{01}}{\sqrt{2}}. 
\end{equation}
Now, if Bob applies the Pauli $X_B$ matrix on his side instead, the resulting state is the same! In other words, their inner product $\bra{\Omega}X_A X_B \ket{\Omega} = 1$. This provides an analogy of the situation we have encountered above; strictly localized states are not necessarily orthogonal, even though they are created in disjoint regions in space.

\subsubsection{Post-measurement states of local observables}
\label{c5_sec_post_measurement_states_of_local_measurements}
We now turn from local unitary operations to local measurements on vacuum and ask whether the resulting states are strictly localized. Here we must distinguish between states conditioned on a particular measurement outcome and states averaged over all outcomes, with the latter generally requiring the mixed-state description introduced in \cref{c5_sec_mixed_localised_states}. From \cref{c4_rem_local_projectors_inf_dimensional}, it follows that a rank-one projector, in particular a projector onto a strictly localized pure state, is not a local operator. This suggests that different post-measurement states associated with a local observable may not be strictly localized \cite{falcone2024}. We first verify this result explicitly, then discuss implications of the result.

For simplicity, we consider a scalar field. As a concrete local observable, consider the smeared field \eqref{c4_phi_f_bounded_algebra}, which can be rewritten to
\begin{equation}\label{c5_smeared_field}
    \phi_f = \frac{1}{\sqrt 2}\lp\big(\padj{a}_1 + a_1^\dagger), \quad a_1^\dagger = \mint{\dd[3]k} \frac{f\lp(\vect k,\omega)}{\sqrt{\omega}}  \adj{a}_{\vect{k}},
\end{equation}
where
\begin{equation}
    f\lp(\vect k,\omega) = \frac{1}{\lp(2\pi)^{3/2}}  \mint{\dd t} \! \mint{\dd[3] x} f\lp(\vect x,t) e^{i\omega t - i\vect k\cdot\vect x}.
\end{equation}
In \cref{c5_smeared_field}, $\omega$ is given by the dispersion relation. With a suitable normalization of the smearing function $f\lp(\vect x,t)$, we obtain $\comm[\big]{\padj{a}_1}{\adj{a}_1} = 1$. Then $\phi_f$ is isomorphic to the position operator of a quantum mechanical harmonic oscillator. The observable can then be spectrally decomposed as \citetext{\citealp[App.~4]{barnett1997}; \citealp{gulla2023}}
\begin{equation}
    \phi_f = \mintn{\dd X} X \seph P_X,
\end{equation}
with
\begin{equation}\label{c5_P_X}
    P_X = \ket{X_1}\bra{X_1} \otimes \identity_2 \otimes \identity_3 \otimes \dotsb,
\end{equation}
where
\begin{equation}
    \ket{X_1} = \pi^{-1/4} e^{- X^2\! / 2} e^{- a_1^{\dagger2}\! / 2 + \sqrt{2} \seph X \adj{a}_1}  \ket{0_1}
\end{equation}
are eigenvectors of $\phi_f$ such that 
\begin{equation}
    \phi_f  \ket{X_1} = X  \ket{X_1}.
\end{equation}
These continuous-spectrum expressions are understood as the ideal limit of a measurement with sufficiently fine but finite resolution. In \cref{c5_P_X} we have written out explicitly the identity projectors of modes 2, 3, etc.; this is not done in \cref{c5_smeared_field} although they are implicitly present.

Suppose for example that the original state before measuring $\phi_f$ is vacuum $\ket{0}$. The probability density for each outcome $X$ is then
\begin{equation}
    p\lp(X) = \qexv{0}{P_X} = \frac{1}{\sqrt{\pi}} e^{- X^2}\!.
\end{equation}
Assuming the Lüders state-update rule, the conditional post-measurement state corresponding to each outcome $X$ is given by applying the spectral projector and normalizing:
\begin{equation}
    \frac{P_X  \ket{0}}{\norm{P_X  \ket{0}}} = \ket{X_1} \otimes \ket{0_2} \otimes \ket{0_3} \otimes \dotsb \equiv \ket{X}.
\end{equation}
If $f\lp(\vect x,t)$ is supported in some region of spacetime $\mathcal O$, $\phi_f$ is local to $\mathcal O$. A natural question is then: do we expect the post-measurement state $\ket{X}$ to be strictly localized to $\mathcal O$?

For simplicity, we let the smearing region $\mathcal{O}$ be a region in three-dimensional space for $t=0$, such that we can write $f\lp(\vect k,\omega) = f\lp(\vect k)$. To find out whether $\ket{X}$ is strictly localized to $\mathcal{O}$, consider measuring the field $\phi\lp(\vect x)$ at some point $\vect x$ in 3D space for $t=0$. Decompose
\begin{equation}
    \phi\lp(\vect x) = \frac{1}{\sqrt 2} \lp[ \Phi_1\lp(\vect x) a_1 + \Phi_1^*\lp(\vect x) a_1^\dagger ] + \phi_\perp(\vect x),
\end{equation}
where $\phi_\perp(\vect x)$ contains terms linear in $a_n$ and $a_n^\dagger$ for $n\geq 2$, involving no $a_1$ nor $a_1^\dagger$. Inserting \cref{c1_phi_x} on the left-hand side, and taking the commutator with $a_1^\dagger$, we have
\begin{equation}\label{c5_Phi1}
    \Phi_1\lp(\vect x) = \frac{1}{\lp(2\pi)^{3/2}} \mint{\dd[3]k} \frac{f\lp(\vect k)}{\omega} e^{i\vect k\cdot\vect x},
\end{equation}
which is real since $f\lp(\vect x)$ is. Thus,
\begin{equation}\label{c5_phi_Phi_1}
    \phi\lp(\vect x) = \phi_f \Phi_1\lp(\vect x) + \phi_\perp(\vect x).
\end{equation}
The expectation value is now obtained straightforwardly:
\begin{equation}\label{c5_qexv_X1_phi}
    \qexv{X}{\phi\lp(\vect x)} = X \Phi_1\lp(\vect x).
\end{equation}

If the factor $1/\omega$ were not present in the Fourier integral \eqref{c5_Phi1}, we would get $\Phi_1\lp(\vect x) = 0$ outside the support of $f\lp(\vect x)$. However, the factor $1/\omega$ leads to a nonlocal convolution in space, which makes $\Phi_1\lp(\vect x) \neq 0$ in general. Thus in general, $\qexv{X}{\phi\lp(\vect x)} \neq 0$, whereas $\qexv{0}{\phi\lp(\vect x)} = 0$. This demonstrates the fact that the Lüders conditional post-measurement state $\ket{X}$ following a measurement of a local observable in $\mathcal{O}$ is not strictly localized \cite{rustad2026}!

This does however not pose any problems with causality, because the nonlocal nature of the post-measurement state was calculated based on knowing the outcome $X$ of the measurement. Observers elsewhere do not know this result, so they must calculate the nonselective post-measurement state averaged over all outcome values $X$:
\begin{equation}\label{c5_dens_op_post_measur}
    \rho = \mintn{\dd X} p\lp(X)  \ket{X}\bra{X} = \mintn{\dd X} \frac{1}{\sqrt{\pi}} e^{- X^2}  \ket{X}\bra{X}.
\end{equation}
Recall that these expressions are understood as the ideal limit of a measurement with sufficiently fine but finite resolution $\Delta X$, so that the integrals over outcomes are shorthand for sums over normalized finite-resolution post-measurement states. The expectation value of the average state \eqref{c5_dens_op_post_measur} when measuring $\phi\lp(\vect x)$ is
\begin{equation}\label{c5_ave_qexv_phi}
\begin{aligned}
    \tr\lp[\rho \phi\lp(\vect x)] &= \mintn{\dd X} \frac{1}{\sqrt{\pi}} e^{- X^2} \qexv{X}{\phi\lp(\vect x)} \\
    &= \Phi_1\lp(\vect x) \mintn{\dd X} \frac{1}{\sqrt{\pi}} X e^{- X^2}  = 0.
\end{aligned}
\end{equation}
Thus, even though the post-measurement state conditioned on the outcome is nonlocal, the nonselective post-measurement state can be strictly localized \cite{fewster2020,rustad2026}.

The field observable $\phi\lp(\vect x)$ is however not so useful for discussing causality, since it leads to the vanishing result \eqref{c5_ave_qexv_phi} for all $\vect x$. We therefore consider the normal-ordered energy density:
\begin{equation}
    \norder{T^{00}} = \norder{ \frac{1}{2}m^2\phi^2 + \frac{1}{2}\lp(\nabla\phi)^2 + \frac{1}{2}\dot\phi^2}.
\end{equation}
Similarly to \cref{c5_phi_Phi_1} we can expand $\dot\phi\lp(\vect x)$:
\begin{equation}
    \dot\phi\lp(\vect x) = \pi_f f\lp(\vect x) + \dot\phi_\perp(\vect x),
\end{equation}
where
\begin{equation}\label{c5_dotphi_Phi_1}
    \pi_f = \frac{i}{\sqrt 2}\lp\big(a_1^\dagger-\padj{a}_1).
\end{equation}
Using \cref{c5_phi_Phi_1,c5_dotphi_Phi_1} we find
\begin{align}
   \qexv{X}{\norder{T^{00}\lp(\vect x)}} &= \lp(X^2-\frac{1}{2}) \lp[ \frac{1}{2}m^2\Phi_1^2\lp(\vect x) + \frac{1}{2}\lp\big(\nabla\Phi_1\lp(\vect x))^2 ] \nonumber \\
   & + \frac{1}{2}f^2\lp(\vect x) \qexv{X}{\lp(\pi_f^2-\frac{1}{2})}. \label{c5_qexv_phi_2}
\end{align}
The last term contains the expectation value of the squared momentum operator, which quantifies the momentum uncertainty in a position eigenstate. Accordingly this expectation value diverges as we approach the unnormalizable state $\ket{X}$.

The result \eqref{c5_qexv_phi_2} may appear surprising. For $\vect x$ outside $\mathcal{O}$, the last term vanishes, such that we can have negative (normal-ordered) energy density when $X^2-1/2$ is negative. Negative energy density is however quite common in quantum field theory \cite{epstein1965}; it arises from normal ordering, involving the subtraction of a commutator. For example, while $\phi^2$ certainly is positive, $\norder{\phi^2}$ is not. Physically, we may think of negative energy density as a result of suppressing the quantum field fluctuations below the vacuum value. In this case this occurs because the measurement postselects vacuum fluctuations for which the energy density is smaller than the vacuum value.

While the energy density for $\ket{X}$ can be different from the vacuum value outside $\mathcal{O}$, the nonselective post-measurement state, as described by the density operator \eqref{c5_dens_op_post_measur}, combines both positive and negative energy densities, giving zero average value: If we calculate the averaged counterpart of \cref{c5_qexv_phi_2} for $\vect x$ outside $\mathcal{O}$, similarly to \cref{c5_ave_qexv_phi}, we find
\begin{equation}\label{c5_ave_qexv_n:phi_2}
\begin{aligned}
    \tr\lp(\rho \, \norder{T^{00}\lp(\vect x)}) &= \lp[ \frac{1}{2}m^2\Phi_1^2\lp(\vect x) + \frac{1}{2}\lp\big(\nabla\Phi_1\lp(\vect x))^2 ] \\
    & \cdot \mintn{\dd X} \frac{1}{\sqrt{\pi}} \lp(X^2-\frac{1}{2}) e^{- X^2}  = 0.
\end{aligned}
\end{equation}
Again, this is consistent with causality; a nonzero value would mean that the post-measurement state is immediately detectable far away from the measurement.

\subsection{Proof of Knight's theorem}
\label{c5_sec_proof_knight}
We now give a proof of Knight’s theorem, which states that (nonvacuum) strictly localized pure states must contain terms with arbitrarily high particle numbers. The proof is simple, since the analyticity argument is already built into the Reeh--Schlieder theorem, on which it rests. Moreover, since the Reeh--Schlieder theorem applies both to the hermitian Klein--Gordon field and to the electromagnetic field, both types of particles, scalar bosons and photons, are covered. We use the notation $\phi_f$ for the smeared field, indicating a scalar field. The same argument applies to the electromagnetic field when $\phi_f$ is replaced by the appropriately smeared gauge-invariant electromagnetic field operators introduced in \cref{c4_sec_polynomial_local_operators}.

The proof needs a small adaptation of the Reeh--Schlieder theorem. Let $\ket{\chi_N}$ be an $N$-particle state, and $\mathcal{D}$ be any nonempty open region of spacetime. Then the Reeh--Schlieder theorem (\cref{c4_sec_the_reeh_schlieder_theorem}) for the polynomial local algebra tells us that there exists an operator $L$ local to $\mathcal{D}$ such that $\bra{\chi_N} L \ket{0} \approx 1$. Expressing $L$ as a finite sum of finite products of smeared fields, this implies that 
\begin{equation}\label{c5_reeh_schlieder_M}
    \bra{\chi_N} \phi_{f_1}\cdots\phi_{f_M} \ket{0} \neq 0,
\end{equation}
for some $M\geq N$, some real test functions $f_m\lp(x)$ supported in $\mathcal{D}$, and smeared fields defined by \cref{c4_phi_f_pol_algebra}. We now expand the product of fields using Wick's theorem:
\begin{align}
    & \phi_{f_1}\cdots\phi_{f_M} = \norder{\phi_{f_1}\cdots\phi_{f_M}} \\
    & + \sum\text{contractions $\cdot$ $\norder{\text{products of fewer fields}}$}. \nonumber 
\end{align}
After acting on the vacuum state and projecting onto $\bra{\chi_N}$, only terms with $N$ uncontracted fields remain. By the condition \eqref{c5_reeh_schlieder_M} at least one of these terms must be nonzero. Relabeling such that the surviving fields have labels $1$, $\dotsc$, $N$, we have
\begin{equation}\label{c5_reeh_schlieder_N}
    \bra{\chi_N} \norder{\phi_{f_1}\cdots\phi_{f_N}} \ket{0} \neq 0.
\end{equation}
Put $f\lp(x) = \alpha_1 f_1\lp(x) + \dotsb + \alpha_N f_N\lp(x)$, where $\alpha_1$, $\dotsc$, $\alpha_N$ are real. Then $\bra{\chi_N} \norder{\phi_{f}^N} \ket{0}$ is a polynomial in $\alpha_1$, $\dotsc$, $\alpha_N$, whose coefficient of $\alpha_1 \cdots \alpha_N$ is proportional to the left-hand side of condition \eqref{c5_reeh_schlieder_N}. Thus the polynomial is not identically zero. We have arrived at a useful consequence of the Reeh--Schlieder theorem:

\begin{remark}\label{c5_rem_reeh_schlieder_chi_N_norder_phi_N}\remheadp{}
    Let $\mathcal{D}$ be any nonempty open region of spacetime, and $\ket{\chi_N}$ be any $N$-particle state. For some real smearing function $f\lp(x)$ supported in $\mathcal{D}$ we have
    \begin{equation}\label{c5_reeh_schlieder_phiN}
        \bra{\chi_N} \norder{\phi_f^N} \ket{0} \neq 0.
    \end{equation}
\end{remark}

We are now ready to prove Knight's theorem. A special case of Knight's theorem is that single-particle states cannot be strictly localized. Since that case is particularly simple, we begin there. The following subsection can be skipped by readers only interested in the general case.

\subsubsection{Proof that single particles cannot be strictly localized}
\label{c5_sec_proof_single_particles_not_localized}
Let $\ket{\chi}$ be a single-particle state of a neutral scalar boson or a photon. Assume, for contradiction, that $\ket{\chi}$ is strictly localized to some region $\mathcal O$, and let $\mathcal{D}$ be a nonempty open region contained in $\compl{\mathcal O}$. From the Reeh--Schlieder theorem in the form \eqref{c5_reeh_schlieder_N} or \cref{c5_rem_reeh_schlieder_chi_N_norder_phi_N}, there is a smearing function $f\lp(x)$ supported in $\mathcal{D}$ such that
\begin{equation}\label{c5_reeh_schlieder_1}
    \bra{\chi} \phi_{f} \ket{0} \neq 0.
\end{equation}
By a suitable normalization of $f\lp(x)$, the smeared field \eqref{c4_phi_f_pol_algebra} can be written in the form $\phi_f = a + a^\dagger$, for some annihilation operator $a$ with $\comm{a}{a^\dagger} = 1$. Since $\ket{\chi}$ is assumed to be strictly localized to $\mathcal O$, we have
\begin{align}
    0 &= \qexv{0}{\norder{\phi_f^2}} = \qexv{\chi}{\norder{\phi_f^2}} = 2 \qexv{\chi}{a^\dagger a} \\ &= 2 \bra{\chi} a^\dagger \ket{0}\bra{0} \seph a \ket{\chi}
    = 2 \abs{\bra{\chi} a^\dagger \ket{0}}^2 = 2 \abs{\bra{\chi} \phi_f \ket{0}}^2, \nonumber
\end{align}
which contradicts condition \eqref{c5_reeh_schlieder_1}. Thus single particles (scalar bosons, photons) cannot be strictly localized.

We observe that the nonlocalizability of single-particle states can almost be regarded as a corollary of the Reeh--Schlieder theorem. In \cref{a1_sec_proof_single_photons_cannot_be_localized} we give an independent proof from first principles based on the electromagnetic energy density.

\subsubsection{Proof that finite particle-number superpositions cannot be strictly localized}
\label{c5_sec_general_proof_knight}
We now consider the general case, that no superposition of states up to $N$ particles can be strictly localized. Let $\ket{\chi}$ be a nonvacuum state strictly localized to $\mathcal O$, and let $\mathcal{D}$ be a nonempty open region contained in $\compl{\mathcal O}$. Suppose for contradiction that $\ket{\chi}$ has a highest nonzero multiparticle component:
\begin{equation} \label{c5_chi_N_particle_cutoff}
    \ket{\chi} = \ket{\chi_0} + \dotsb + \ket{\chi_N}, \quad \ket{\chi_N} \neq 0, \sepc N \geq 1.
\end{equation}
Due to strict localization, we have
\begin{equation}\label{c5_chi_strict_loc}
    \qexv{\chi}{\norder{\phi_f^{2N}}} = \qexv{0}{\norder{\phi_f^{2N}}} = 0.
\end{equation}
Choose $f\lp(x)$ to satisfy condition \eqref{c5_reeh_schlieder_phiN} for the highest nonzero component $\ket{\chi_N}$. With a suitable normalization of $f\lp(x)$ we can write  $\phi_f = a + a^\dagger$ for some annihilation operator $a$. Since every term in $\norder{\phi_f^{2N}}$ contains a total of $2N$ creation and annihilation operators, and since $\ket{\chi}$ contains no more than $N$ particles, only the term with $N$ creation operators and $N$ annihilation operators can contribute:
\begin{align}
    0 &= \qexv{\chi}{\norder{\phi_f^{2N}}} = \qexv{\chi_N}{\norder{\phi_f^{2N}}} \nonumber\\
    &= \binom{2N}{N} \qexv{\chi_N}{a^{\dagger N}a^N} = \binom{2N}{N} \abs{\bra{\chi_N} a^{\dagger N} \ket{0}}^2 \nonumber\\
    &= \binom{2N}{N} \abs{\bra{\chi_N} \norder{\phi_f^N} \ket{0}}^2.
\end{align}
Combining with \cref{c5_rem_reeh_schlieder_chi_N_norder_phi_N} we have a contradiction; thus there is no $N$-particle cutoff in \cref{c5_chi_N_particle_cutoff}.


%% file: 6_localization_and_causality.tex


\newcommand{\FigSixOne}{%
    \begin{figure*}
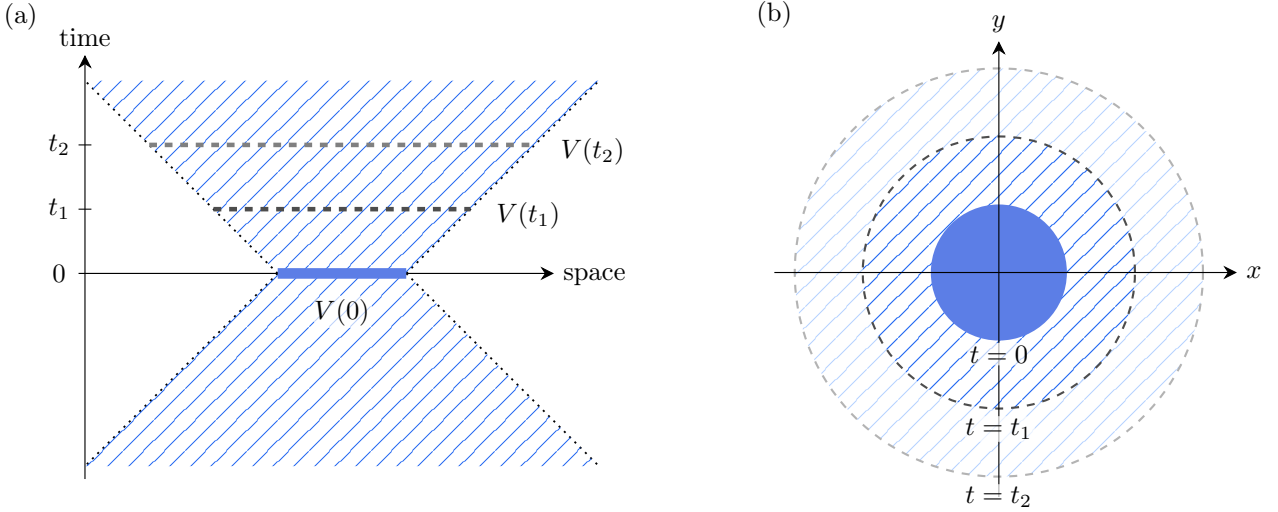
%
    \hspace*{-10mm}%
    \raisebox{4mm}{%
        \iftoggle{FigFromPDF}{%
            \includegraphics{fig6-1a.pdf}%
        }{%
            \tikzsetnextfilename{fig6-1a}%
            \begin{tikzpicture}[scale=0.85]%
                \node at (-5.0, 4) {(a)};%
                \draw[thick, dotted, line width=0.7pt] (-1, 0) -- (-4,  3);%
                \draw[thick, dotted, line width=0.7pt] ( 1, 0) -- ( 4,  3);%
                \draw[thick, dotted, line width=0.7pt] (-1, 0) -- (-4, -3);%
                \draw[thick, dotted, line width=0.7pt] ( 1, 0) -- ( 4, -3);%
                \fill[myhatch, hatch phase=8pt, pattern color=MyBlueMC] (-4,  3) -- (-1, 0) -- (-4, -3) -- (4, -3) -- (1, 0) -- (4, 3);%
                \draw[line width=.6mm, black!50, dash pattern={on 2.92pt off 2.92pt}] (-3, 2) -- (3, 2);%
                \draw[line width=.6mm, black!70, dash pattern={on 2.95pt off 2.95pt}] (-2, 1) -- (2, 1);%
                \draw[axis_arrow] (-4.1,  0.0) -- ( 3.3, 0.0) node[right, yshift=-.4mm] {space};%
                \draw[axis_arrow] (-4.0, -3.2) -- (-4.0, 3.4) node[above] {time};%
                \draw[line width=1.4mm, MyBlueM] (1, 0) -- (-1, 0);%
                \draw (-4.1, 1.0) -- (-3.9, 1.0);%
                \draw (-4.1, 2.0) -- (-3.9, 2.0);%
                \node at (-4.4,  0) {$0$};%
                \node at (-4.4,  1) {$t_1$};%
                \node at (-4.4,  2) {$t_2$};%
                \node[fill=white, fill opacity=0.7, text opacity=1, rounded corners=5pt, inner sep=1pt] at (0.0,  -0.6) {$V\lp(0)$};%
                \node at (2.9,  0.9) {$V\lp(t_1)$};%
                \node at (3.9,  1.9) {$V\lp(t_2)$};%
            \end{tikzpicture}%
        }%
    }%
    \hspace*{15mm}%
    \raisebox{0mm}{%
        \iftoggle{FigFromPDF}{%
            \includegraphics{fig6-1b.pdf}%
        }{%
            \tikzsetnextfilename{fig6-1b}%
            \begin{tikzpicture}[scale=0.9]%
                \node at (-3.3, 3.8) {(b)};%
                \draw[thick, dashed, black!30, myhatch, pattern color=MyBlueML] (0, 0) circle (3.0);%
                \draw[thick, dashed, black!70, myhatch, pattern color=MyBlueMC] (0, 0) circle (2.0);%
                \fill[MyBlueM] (0, 0) circle (1.0);%
                \draw[axis_arrow] (-3.3,  0.0) -- (3.5, 0.0) node[right] {$x$};%
                \draw[axis_arrow] ( 0.0, -3.3) -- (0.0, 3.4) node[above] {$y$};%
                \node[fill=white, fill opacity=0.7, text opacity=1, rounded corners=4pt, inner sep=1pt] at (0, -1.2) {$t = 0$};%
                \node[fill=white, fill opacity=0.7, text opacity=1, rounded corners=4pt, inner sep=1pt] at (0, -2.3) {$t = t_1$};%
                \node[fill=white, fill opacity=0.7, text opacity=1, rounded corners=4pt, inner sep=1pt] at (0, -3.3) {$t = t_2$};%
            \end{tikzpicture}%
        }%
    }%
    \caption{\label{c6_fig_full_light_cone}Visualization of the localization region \eqref{c6_G_light_cone}. At any time $t$, the set $\mathcal{O}$ determines the current spatial localization region $V\lp(t)$. In this case, $\mathcal{O}$ is the lightcone of $V\lp(0)$. (a) Spacetime diagram of $\mathcal{O}$. (b) 2D illustration of the spatial localization region $V\lp(t)$ for some different times $t$.}%
    \end{figure*}%
}

\newcommand{\FigSixTwo}{%
    \begin{figure}%
    \hspace*{2mm}%
    \raisebox{0mm}{%
        \iftoggle{FigFromPDF}{%
            \includegraphics{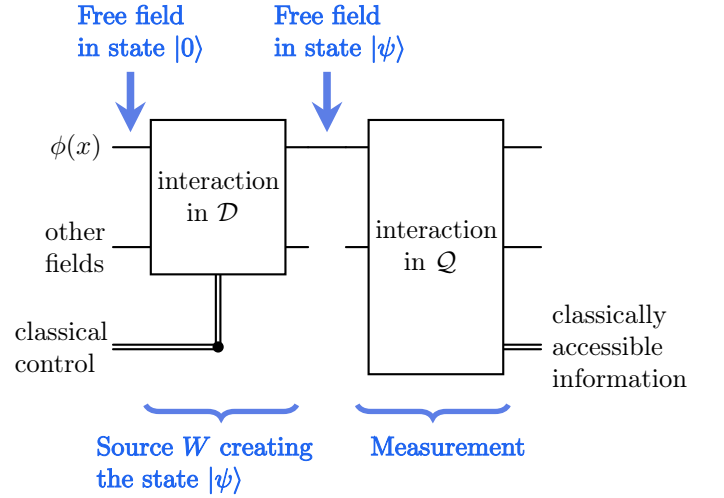}%
        }{%
            \clap{%
                \tikzsetnextfilename{fig6-2}%
                \begin{tikzpicture}%
                    \node{%
                        \begin{quantikz}[column sep=0.5cm, row sep=0.6cm]%
                            \lstick{$\phi\lp(x)$} %
                            & %
                            \gate[2]{%
                                \begin{minipage}{15mm}%
                                    \centering\vspace{3mm}\hspace{-1.5mm}%
                                    \smash{\parbox{14.5mm}{\centering interaction\\in $\mathcal{D}$}}%
                                    \vspace{1.4mm}%
                                \end{minipage}%
                            } %
                            &[-2mm] %
                            & %
                            &[-2mm] %
                            \gate[3]{%
                                \begin{minipage}{15mm}%
                                    \centering\vspace{3mm}\hspace{-1.5mm}%
                                    \smash{\parbox{14.5mm}{\centering interaction\\in $\mathcal{Q}$}}%
                                    \vspace{1.4mm}%
                                \end{minipage}%
                            } %
                            & %
                            \\ %
                            %
                            \lstick[label style={text width=14mm}]{%
                                \hspace*{-1mm}%
                                \rlap{\begin{minipage}{14mm}{\begin{flushleft}other\\fields\end{flushleft}}\end{minipage}}%
                            } %
                            & %
                            & %
                            & %
                            \wireoverride{n} %
                            & %
                            & %
                            \\ %
                            %
                            \lstick[label style={text width=14mm}]{%
                                \hspace*{-9mm}%
                                \rlap{\begin{minipage}{14mm}{\begin{flushleft}classical\\control\end{flushleft}}\end{minipage}}%
                            } %
                            \setwiretype{c} %
                            & %
                            \ctrl[vertical wire=c]{-1} %
                            & %
                            \setwiretype{n} %
                            & %
                            & %
                            & %
                            \setwiretype{c} %
                            \rstick[label style={text width=20mm}]{%
                                \vspace{-5.0mm}%
                                \begin{flushleft}classically\\accessible\\information\end{flushleft}%
                            }%
                        \end{quantikz}%
                    };%
                    \draw[big_arrow] (-2.9, 2.4) to (-2.9, 1.6);%
                    \node at (-2.6, 2.9) [color=MyBlueMC] {%
                        \parbox{20mm}{\begin{flushleft}\sboldtext{Free field\\in state $\ket{0}$}\end{flushleft}}%
                    };%
                    \draw[big_arrow] (-0.3, 2.4) to (-0.3, 1.6);%
                    \node at (0.0, 2.9) [color=MyBlueMC] {%
                        \parbox{20mm}{\begin{flushleft}\sboldtext{Free field\\in state $\ket{\psi}$}\end{flushleft}}%
                    };%
                    \draw [decorate, decoration={brace, amplitude=5pt, mirror, raise=4ex}, MyBlueM, line width=0.7mm]%
                        (-2.7, -1.4) to (-0.7, -1.4);%
                    \node at (-1.6, -2.805) [color=MyBlueMC] {%
                        \parbox{35mm}{\begin{flushleft}\sboldtext{Source $W$ creating\\the state $\ket{\psi}$}\end{flushleft}}%
                    };%
                    \draw [decorate, decoration={brace, amplitude=5pt, mirror, raise=4ex}, MyBlueM, line width=0.7mm]%
                        (0.1, -1.4) to (2.1, -1.4);%
                    \node at (1.3, -2.55) [color=MyBlueMC] {\sboldtext{Measurement}};%
                \end{tikzpicture}%
            }%
        }%
    }%
    \caption{\label{c6_fig_physical_process_state_creation_and_measurement}A quantum circuit model of a physical process with a source $W$ located in some spacetime region $\mathcal{D}$ that produces a state $\ket{\psi}$ of the field $\phi\lp(x)$, followed by a measurement of $\phi\lp(x)$ in some region $\mathcal{Q}$.}%
    \end{figure}%
}

\newcommand{\FigSixThree}{%
    \begin{figure}%
    \hspace*{0mm}%
    \raisebox{0mm}{%
        \iftoggle{FigFromPDF}{%
            \includegraphics{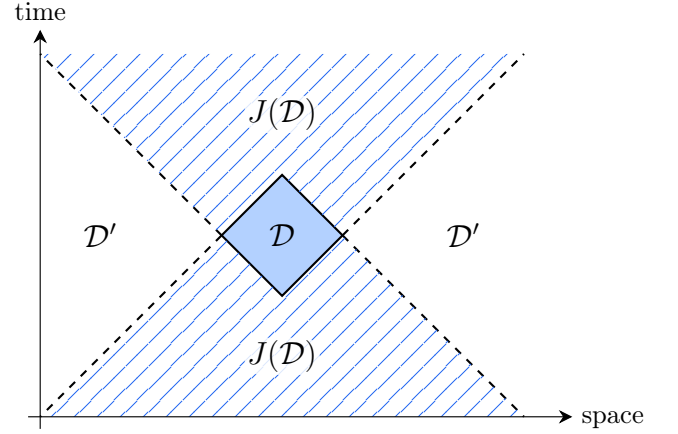}%
        }{%
            \tikzsetnextfilename{fig6-3}%
            \begin{tikzpicture}[scale=0.8]%
                \fill[MyBlueML] (1, 0) -- (0, 1) -- (-1, 0) -- (0, -1) -- (1, 0);%
                \draw[thick]   (1, 0) -- (0, 1) -- (-1, 0) -- (0, -1) -- (1, 0);%
                \draw[dashed, thick] ( 1, 0) -- ( 4,  3);%
                \draw[dashed, thick] ( 1, 0) -- ( 4, -3);%
                \draw[dashed, thick] (-1, 0) -- (-4,  3);%
                \draw[dashed, thick] (-1, 0) -- (-4, -3);%
                \node at (0, 0) {\large$\mathcal{D}$};%
                \fill[thick, myhatch, hatch distance=6.1pt, hatch phase=-3.5pt, pattern color=MyBlueMC] (-4,  3) -- (-1, 0) -- (0,  1) -- (1, 0) -- (4,  3);%
                \fill[thick, myhatch, hatch distance=6.1pt, hatch phase=-3.5pt, pattern color=MyBlueMC] (-4, -3) -- (-1, 0) -- (0, -1) -- (1, 0) -- (4, -3);%
                \node[fill=white, fill opacity=0.85, text opacity=1, rounded corners=6pt, inner sep=1pt] at ( 0,  2) {\large$J\lp(\mathcal{D})$};%
                \node[fill=white, fill opacity=0.85, text opacity=1, rounded corners=6pt, inner sep=1pt] at ( 0, -2) {\large$J\lp(\mathcal{D})$};%
                \node at ( 3,  0) {\large$\splike{\mathcal{D}}$};%
                \node at (-3,  0) {\large$\splike{\mathcal{D}}$};%
                \draw[axis_arrow] (-4.2, -3.0) -- ( 4.8, -3.0) node[right, yshift=-.4mm] {space};%
                \draw[axis_arrow] (-4.0, -3.2) -- (-4.0,  3.4) node[above] {time};%
            \end{tikzpicture}%
        }%
    }%
    \caption{\label{c6_fig_causal_diamonds}Let $\mathcal{D}$ for simplicity be a bounded spacetime diamond. Denote its lightcone by $J\lp(\mathcal{D})$. The complement of the lightcone $\compl{J\lp(\mathcal{D})}$ is equal to the spacelike region $\splike{\mathcal{D}}$ of $\mathcal{D}$.}%
    \end{figure}%
}

\newcommand{\FigSixFour}{%
    \begin{figure}%
    \hspace*{0mm}%
    \raisebox{0mm}{%
        \iftoggle{FigFromPDF}{%
            \includegraphics{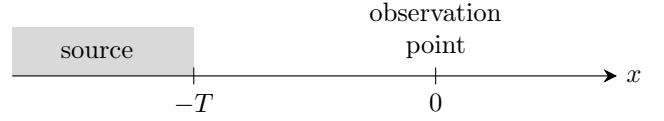}%
        }{%
            \tikzsetnextfilename{fig6-4}%
            \begin{tikzpicture}[scale=1.6]%
                    \fill [black!15] (0, 0.4) rectangle (1.5, 0);%
                    \draw (0.7, 0.07) node[above] {source};%
                    \draw[axis_arrow] (0, 0) -- (5, 0) node[right] {$x$};%
                    \draw (1.5, 0.07) -- (1.5, -0.07) node[below] {$-T$};%
                    \draw (3.5, 0.07) -- (3.5, -0.07) node[below] {$0$};%
                    \draw (3.5, 0.07) node[above] {\parbox{20mm}{\centering observation\\point}};%
                \end{tikzpicture}%
        }%
    }%
    \caption{\label{c6_fig_1d_on_demand_source}Let a 1D source be located in the region $x < -T$, and consider local observables at the observation point $x = 0$. The source is turned on at $t = -T$, which means that any measurement at times $t < 0$ must have the same outcome probabilities as in vacuum.}%
    \end{figure}%
}

\iftoggle{ExportTikz}{%
    \FigSixOne
    \FigSixTwo
    \FigSixThree
    \FigSixFour
     
}{}

\section{Localization and causality}
\label{c6_sec_localization_and_causality}
One of the main motivations for the theory of strictly localized states is to be able to describe localized \emph{information} in quantum field theory. And specifically, we hope that such states could represent localized information that propagates in a causal way, such as in the Fermi problem. However, a priori there is no reason to assume that the dynamics of strictly localized states is in fact causal. Moreover, what is meant by causality in quantum field theory in the first place is not always trivial. In this section, we explore these questions.

Note that the connection between strict localization and information is not limited to causal propagation. The same notion that a state agrees with the vacuum for all measurements outside a specified region has appeared under different names in entropy and black-hole physics. So-called \enquote{Vacuum-bounded} mixed states have been used in studies of localized entropy and black-hole evaporation \cite{olum1997,olum1998}, and \enquote{Vacuum-outside-$R$} states have been used in connection with entropy bounds \cite{page2008}. More recently, the same localization condition was used to derive a Bekenstein-type bound in quantum field theory \cite{longo2025}.

In this review, we focus on causal propagation. First, we briefly discuss relativistic causality in quantum field theory, and provide a precise formulation in terms of external, classical control and probabilities of local measurements. We then look at how the dynamics of strictly localized states is contained in the definition of the state $\ket{\psi}$ and of the spacetime region $\mathcal{O}$. This analysis shows that strictly localized states indeed always represent causally propagating, localized information in quantum field theory. Finally, we look at how strictly localized states are physically generated by localized on-demand sources.

\subsection{Relativistic causality}
\label{c6_sec_relativistic_causality}
Despite its pivotal importance for physical theories, relativistic causality is a concept for which it can be difficult to find an exact mathematical statement. It is easy to get an intuitive idea of this concept, but lack of precise language can quickly lead to confusion. This is especially true when it is combined with the measurement rules of quantum mechanics, and even more so with quantum field theory. For example, as mentioned in \cref{c5_sec_what_is_a_localized_state}, there has been much discussion about causality in combination with measurement probabilities not vanishing (\cref{c4_cor_local_probabilities_never_vanish}). The goal of this section is to formulate a meaningful, precise statement of relativistic causality for quantum field theory.

\makeshortsubsubsection[0.97]
\subsubsection{Information propagation from external interactions}
\label{c6_sec_information_propagation_from_external_interactions}
Causality is the principle that an effect cannot occur before its cause. With the framework of special relativity, this principle must hold in every inertial frame, so information cannot propagate faster than the speed of light $c$.

Information is in general some feature of some physical theory, such as of quantum states in quantum field theory. However not all features represent information. What we consider information in this context is any feature that is \emph{observable} and \emph{classical}. Observable means anything that is not dependent on a redundant degree of freedom in the theory, such as gauge choices, a global quantum phase, etc. Classical refers to features that with high accuracy behave according to classical information theory, i.e., being stable, copyable, etc. This means that quantum states are not themselves considered information in this context, which makes sense since global states change instantaneously over spacelike distances in the standard collapse description of Bell-type experiments. On the other hand, the result of a quantum measurement does represent information.

Second, to meaningfully talk about information propagation, we need a mechanism for controlling where some information initially is localized. A way to do this reliably is to make a division between some primary system of interest and some elements that can be considered external. That is, since analyzing causality in fully closed systems can be difficult, we can rather analyze the flow of information that enters the primary system through its interaction with these external elements. Furthermore, we assume that the external interaction can be controlled classically. This way of keeping track of when information enters the system is a safe way to avoid misunderstandings about where the information is initially localized. For example it shows that the information propagation in an optical pulse happens with the velocity of the wavefront and not with the group velocity. 

Relativistic causality is then the statement that an external, classical control in some spacetime region $\mathcal{D}$ cannot lead to any information in the primary system \emph{changing} outside the lightcone of $\mathcal{D}$. The importance of focusing on change is reminiscent of the discussion in \cref{c4_sec_interpreting_local_measurement_probabilities} about how to interpret probabilities of local quantum measurements, but it is actually also present classically. For instance, if we perturb a classical point charge at some time, the change in the electric field propagates outwards within the lightcone. This does however not mean that the field outside the lightcone must be zero, only that the outside field must be \emph{unchanged} by the perturbation.

\subsubsection{Quantum randomness}
\label{c6_sec_quantum_randomness}
In quantum theories, there is one more nuance to consider, namely that the physical theory is not able to predict all the information. This happens in the standard Copenhagen formulation of quantum mechanics, where information coming from the outcome of a measurement process is determined stochastically. That is, the physical theory predicts a probability for each outcome, but then the actual result is randomly selected within that probability distribution. In light of this, some authors have introduced a distinction between so-called \enquote{strong vs.\ weak} causality \cite{hegerfeldt1998a}: strong causality being that information is limited by the speed of light in any individual experiment, and weak causality being that this only holds \emph{on average} over many repeated experiments. 

However, in our view, such a distinction does not appear to be relevant. After all, in the standard formulation of quantum mechanics, individual measurement outcomes are generally random, while the theory predicts their probability distribution. Therefore the difference between predicted probabilities and actual observed outcomes is purely random. In particular, this means that the only thing we can meaningfully say changes \emph{due to} the classical control is the measurement probability. This is what we refer to as causality. This does not mean that causality only applies on average. For example, if a classical particle travels a distance $r$ with a velocity drawn from a probability distribution, causality means that the travel time is at least $r$ in every individual trial.

\subsubsection{Relativistic causality for quantum field theory}
\label{c6_sec_statement_of_relativistic_causality_for_quantum_field_theory}
Putting it all together, we thus get the following formulation of relativistic causality in quantum field theory:

\begin{remark}\label{c6_def_causality_quantum_field_theory}\remheadp{}
    Relativistic causality in quantum field theory dictates that an external, classical control in some spacetime region $\mathcal{D}$ cannot affect the probabilities of local measurement outcomes in any spacelike separated region $\splike{\mathcal{D}}$. This localized control can for example be switching on a localized interaction or source, locally triggering a measurement, or similar.
\end{remark}

This formulation also makes the role of strictly localized states in quantum field theory clear. According to \cref{c5_def_localized_states}, states strictly localized to a spacetime region $\mathcal{O}$ give the same expectation value as vacuum for any local measurement outside $\mathcal{O}$. Thus, strictly localized states are states that can carry localized information. According to \cref{c5_thm_knight_inf_photon_numbers} or \cref{c5_sec_proof_single_particles_not_localized}, the fundamental single-particle excitations of the quantum fields cannot themselves represent such localized information. Instead, Fock states must combine in nontrivial ways to do so, making quantum field theory much more interesting.

\subsection{Propagation of strictly localized states}
\label{c6_sec_propagation_of_localized_states}
Thus a strictly localized state $\ket{\psi}$ can in some sense represent information that is localized to some spacetime region $\mathcal{O}$. But causality is all about the \emph{propagation} of localized information. We might picture strictly localized states as representing the change to the quantum field  due to some local, external interaction, which gradually propagates outwards with time. But how is this intuitive picture reflected in the state definition $\ket{\psi}$ and the spacetime region $\mathcal{O}$? 

\FigSixOne

\subsubsection{Dynamics}
\label{c6_sec_dynamics}
As a start, note that the time evolution is somewhat hidden by our use of the Heisenberg picture. This picture is very useful given that we work with free fields, since the states $\ket{\psi}$ and localization region $\mathcal{O}$ appear as fixed quantities. There is however dynamics contained in these quantities. For instance if we calculate the expectation value of a local observable, e.g., $\qexv{\psi}{\phi\lp(x)}$, at some spacetime point $x$, this will be some function $f\lp(x)$. The dynamics of the state is contained within $\ket{\psi}$ in the sense that the particular function $f\lp(x)$ is determined by $\ket{\psi}$.

Similarly, the region $\mathcal{O}$ that $\ket{\psi}$ is strictly localized to is written as a single quantity and therefore appears to be a static region. However, $\mathcal{O}$ is a region of space\emph{time}, meaning that a specification of $\mathcal{O}$ describes a region of space for each time. For instance imagine that we know that the state $\ket{\psi}$ is strictly localized to the spacetime region
\begin{equation}\label{c6_G_light_cone}
    \mathcal{O} = \set{\lp(c t, \vect{x})^\mu}{\abs{\vect{x}} \leq r_0 + c \abs{t}, \; t \in \reals},
\end{equation}
for some radius $r_0 > 0$. This region is illustrated in \cref{c6_fig_full_light_cone}. For each point in time $t$, label the current region of space $\ket{\psi}$ is strictly localized to by $V\lp(t)$. Clearly, $V\lp(0)$ is a ball of radius $r_0$, and the radius expands linearly by $c \abs{t}$. I.e., $\mathcal{O}$ is the lightcone of a three-dimensional ball. If we want to know what region of space $\ket{\psi}$ is strictly localized to at some other, later time $t_1 > 0$, this is evident from the description of the region $\mathcal{O}$. In this case it would be the spatial volume $V\lp(t_1) = \set{\vect{x}}{\abs{\vect{x}} \leq r_0 + c t_1}$.

\subsubsection{Possible localization regions \texorpdfstring{$\mathcal{O}$}{O}}
\label{c6_sec_possible_localization_regions_G}
In view of the lightcone region  \eqref{c6_G_light_cone}, one may ask whether this is a typical localization region for strictly localized states. Of course the actual region $\mathcal{O}$ will depend on the specific state $\ket{\psi}$ in question, but is \emph{any} region $\mathcal{O}$ possible or are there some universal principles that must apply? This question was analyzed in \cite{knight1961}, and two principles were identified. First, Knight showed that for any strictly localized nonvacuum state $\ket{\psi}$, the localization region $\mathcal{O}$ cannot be bounded in time. I.e., for all times $t \in \lp(- \infty, \infty)$, there must exist a point $\lp(c t, \vect{x})^\mu \in \mathcal{O}$ for some $\vect{x}$. This is quite logical, because we are working with a free theory, and thus the state $\ket{\psi}$ cannot simply disappear; it must at all times be localized to \emph{somewhere}. 

 Second, Knight showed that if a state $\ket{\psi}$ is strictly localized to some spatial volume $V$ at a given time, then it remains strictly localized to the lightcone of $V$. This is good news. One of the main goals of the theory of strictly localized states is to be able to describe localized information in quantum field theory. Knight's result shows that information embedded in states that are strictly localized \emph{stays} localized and propagates at maximally the speed of light. This therefore establishes a strong internal consistency of the theory, in that not only do there exist \emph{some} strictly localized states that satisfy this causal behavior, but in fact \emph{all} strictly localized states do.

\subsection{On-demand sources}
\label{c6_sec_on_demand_sources}
We can then turn to another question: how are strictly localized states produced in the first place? When discussing the physical process of measurement in \cref{c3_sec_what_is_a_measurement}, we assumed that a quantum field of interest $\phi\lp(x)$ starts out in some state $\ket{\psi}$. We further assumed that this state was sufficiently stable and isolated that we can consider $\phi\lp(x)$ to be free of interactions initially. But how does the field attain the state $\ket{\psi}$?

\subsubsection{What is a localized source?}
\label{c6_sec_what_is_a_localized_source}
The physical process we have in mind consists of two steps, depicted in \cref{c6_fig_physical_process_state_creation_and_measurement}. As in \cref{c3_sec_what_is_a_measurement}, we consider some quantum field $\phi\lp(x)$ that is initially not interacting strongly with any other fields. We can therefore with good accuracy consider it initially a free field with no interactions. It is then useful to be in the Heisenberg picture, where the free time evolution is carried by the field operator $\phi\lp(x)$. We further assume that the field starts out in the (time-independent) vacuum state $\ket{0}$. 

Then in some time interval, $\phi\lp(x)$ interacts with some other fields. In line with the discussion of causality in \cref{c6_sec_relativistic_causality}, we assume that this interaction is controlled by some external, classical input. This means that there is some external element that has the effect of (classically) turning on or off the interaction between $\phi\lp(x)$ and the other quantum fields. In typical experiments, this classical control will for example be an experimentalist pushing a button. We label the spacetime region where this controlled interaction takes place as $\mathcal{D}$. We do not want to assume anything more than necessary about the time evolution of this interaction in $\mathcal{D}$. The only thing we assume is that it leaves the field $\phi\lp(x)$ in some pure state $\ket{\psi}$ (after tracing out other degrees of freedom). During the time where the interaction is non-negligible, we go to the Schrödinger picture, where the unknown time evolution is carried by the state of the combined system. After the interaction, we trace out the other fields to obtain the reduced state of the field $\phi\lp(x)$. This will generally result in a mixed state, but we limit our analysis to situations where it happens that the reduced state is pure, which we denote by $\ket{\psi}$.

\FigSixTwo

We then assume that the interaction gets weaker again after some time, and that the field is again sufficiently stable and isolated that we can consider $\phi\lp(x)$ a free field. It is then convenient to go back to the Heisenberg picture, where the state $\ket{\psi}$ is again time independent. We interpret the classically controlled interaction in $\mathcal{D}$ as a localized source creating the state $\ket{\psi}$ for the field $\phi\lp(x)$. This is how we arrive at the initial configuration of \cref{c3_fig_measurement_process}. A measurement then happens in some spacetime region $\mathcal{Q}$ according to the physical process described in \cref{c3_sec_what_is_a_measurement}.

\subsubsection{The need for a classical control}
\label{c6_sec_the_need_for_a_classical_control}
In general, an interaction that slowly becomes negligible sufficiently far from $\mathcal{D}$ is sufficient for considering a free field theory far away from $\mathcal{D}$. However, to analyze information flow, we need to identify where information enters the system. In particular, according to the discussion in \cref{c6_sec_relativistic_causality}, to talk about relativistic causality, the information entering the system needs to be present \emph{only} in a sharply defined region $\mathcal{D}$. That is the role of the classical control input in \cref{c6_fig_physical_process_state_creation_and_measurement}: assuming that the controlled interaction is confined to a spacetime region $\mathcal{D}$ ensures that information may enter the system in a sharply defined spacetime region. 

On the other hand, on the measurement side, relativistic causality for \cref{c6_fig_physical_process_state_creation_and_measurement} concerns any information that can be obtained \emph{at least} by local measurements in $\mathcal{Q}$. That is why we do not need to assume a classical control on the measurement side of \cref{c6_fig_physical_process_state_creation_and_measurement}. Of course the measurement may be a controlled one, as we can take the classical control as part of the \enquote{other fields}. However, our model where the measurement has no explicit classical control is in fact more general, as it also encompasses measurement processes that we have no control over.

Note also that if we want to implement the standard causality argument of two spacelike separated measurements not influencing each other, these measurements cannot both be uncontrolled; that would be too general. After all, \cref{c6_def_causality_quantum_field_theory} calls for a classical control, meaning that at least one of the measurements has to be controlled. This also makes sense intuitively, because the argument for causality relies on the possibility of transmitting information by \emph{choosing} to measure or not to measure. Clearly such a choice requires at least one of the measurements to be controlled, but not both. We see that the model in \cref{c6_fig_physical_process_state_creation_and_measurement} can also describe these types of situations by considering the interaction in $\mathcal{D}$ as the controlled measurement.

\subsubsection{Causality of localized sources}
\label{c6_sec_causality_of_localized_sources}
In summary, the model in \cref{c6_fig_physical_process_state_creation_and_measurement} is captured by our formulation of relativistic causality in \cref{c6_def_causality_quantum_field_theory}. Therefore, the source located in $\mathcal{D}$ can only lead to observable effects within the lightcone of $\mathcal{D}$, which we label $J\lp(\mathcal{D})$ (see \cref{c6_fig_causal_diamonds}). This means that the produced state $\ket{\psi}$ must give the same measurement statistics as the vacuum for all physically realizable local measurements in $\splike{\mathcal{D}}$. In line with the inner and outer sets introduced in \cref{c3_sec_physically_realizable_local_measurements}, a sufficient condition for this is that the produced state $\ket{\psi}$ is strictly localized to $J\lp(\mathcal{D})$. Here we choose to specialize to the case where all observables in the bounded local algebra do in fact correspond to physically realizable local measurements, so that this condition is also necessary. Under this assumption, any pure state $\ket{\psi}$ produced on demand from vacuum by a source in $\mathcal{D}$ must be strictly localized to $J\lp(\mathcal{D})$.

Other arguments can also lead to the same conclusion. For instance, it would suffice that the physically realizable measurements define the same class of strictly localized states as the full local algebra. Alternatively, the conclusion can also follow directly from assumptions about the source dynamics. For a source modeled by local interactions confined to $\mathcal{D}$, spacelike commutativity motivates requiring that its evolution commute with the field in $\splike{\mathcal{D}}$. This holds for instance when the source operation belongs to the corresponding local algebra in $\mathcal{D}$. As discussed in \cref{c5_sec_finding_licht_operators,c5_sec_unitary_operators}, this commutation property, imposed on an isometry $W$ or a field-source unitary $U$, already ensures strict localization of the produced state, independently of whether every observable in the local algebra is physically realizable.

In this setting, we can therefore apply Licht's theorem (\cref{c5_thm_lichts_theorem}) to characterize the states produced by localized on-demand sources:

\begin{remark}\remheadp{}
    Any source localized to some spacetime region $\mathcal{D}$ that produces a pure state from vacuum is characterized by a Licht operator $W$ associated with $J\lp(\mathcal{D})$. 
\end{remark}

\FigSixThree

How can we find Licht operators associated with $J\lp(\mathcal{D})$? By \cref{c5_def_licht_operators}, an operator $W$ is a Licht operator associated with $J\lp(\mathcal{D})$ if and only if $\adj{W} W = \identity$ and
\begin{equation}\label{c6_comm_W_phi_0}
    \comm{W}{L} = 0, \sepc \forall L \in \balg{\splike{\mathcal{D}}}.
\end{equation}
Looking at this equation, it is actually easy to see one place where we might find such an operator $W$, namely, in the local algebra $\balg{\mathcal{D}}$. Due to the spacelike commutativity of the local algebras, any operator local to $\mathcal{D}$ commutes with all operators local to $\splike{\mathcal{D}}$ and therefore satisfies \cref{c6_comm_W_phi_0}. Thus, if we can find an operator $W \in \balg{\mathcal{D}}$ such that $\adj{W} W = \identity$, then $W$ is a Licht operator associated with $J\lp(\mathcal{D})$. In summary:

\begin{remark}\label{c6_rem_psi_produced_G}\remheadp{}
    For any spacetime region $\mathcal{D}$,
    \begin{equation}\label{c6_psi_produced_G}
    \begin{aligned}
        &\tn{A source in $\mathcal{D}$ producing a pure state $\ket{\psi}$} \\
        &\qquad \Rightarrow \sepc \tn{$\ket{\psi}$ only detectable in $J\lp(\mathcal{D})$} \\
        &\qquad \Leftrightarrow \sepc \tn{$\ket{\psi}$ locally equivalent to $\ket{0}$ in $\splike{\mathcal{D}}$} \\
        &\qquad \Leftrightarrow \sepc \tn{$\ket{\psi}$ strictly localized to $J\lp(\mathcal{D})$} \\
        &\qquad \Leftrightarrow \sepc \comm{W}{L} = 0, \; \forall L \in \balg{\splike{\mathcal{D}}} \\
        &\qquad \Leftarrow \sepc W \in \balg{\mathcal{D}},
    \end{aligned}
    \end{equation}
    with $\ket{\psi} = W  \ket{0}$ and $\adj{W} W = \identity$.
\end{remark}

\subsubsection{Haag duality}
\label{c6_sec_haag_duality}
Note that the last arrow in \cref{c6_rem_psi_produced_G} not necessarily being bidirectional is related to an interesting concept called Haag duality. To satisfy the commutation condition \eqref{c6_comm_W_phi_0}, we seek operators $W$ that commute with all $L \in \balg{\splike{\mathcal{D}}}$. As we have seen, operators $W \in \balg{\mathcal{D}}$ do the job. One might then wonder: are the operators in $\balg{\mathcal{D}}$ the \emph{only} bounded operators that do the job? 

This is generally a question known as Haag duality \citetext{\citealp[Ch.~1.3]{horuzhy1990}; \citealp{garbarz2022}}. It was first proposed in \cite{haag1962,haag1963}, and it has been proved to hold for certain situations, such as for free scalar fields where the region $\mathcal{D}$ is a spacetime diamond \cite{araki1964,osterwalder1973,leyland1978,hislop1986}. For our purposes, we are however content with seeing that $W \in \balg{\mathcal{D}}$ is sufficient for \cref{c6_comm_W_phi_0}, and we will not look into the necessity further here.

\subsubsection{Sources triggered on demand}
\label{c6_sec_sources_triggered_on_demand}
To summarize, in \cref{c6_sec_what_is_a_localized_source} and \cref{c6_fig_physical_process_state_creation_and_measurement} we described a physical process that can realize a source located in a spacetime region $\mathcal{D}$. Furthermore, we have seen in \cref{c6_sec_causality_of_localized_sources} that we can look for operators representing such sources among isometric operators $W \in \balg{\mathcal{D}}$. But one question remains: does a process such as the one described in \cref{c6_fig_physical_process_state_creation_and_measurement} actually occur in practice, experimentally? It is not hard to construct a system that interacts with some field $\phi\lp(x)$ and that is sensitive to some external, classical input. But for our model, we also need the external input to control the interaction within some sharply defined spacetime region $\mathcal{D}$.

Such external controls are in general a somewhat difficult subject in physics. We can typically take any part of the system that is sufficiently independent, such as an experimentalist, and designate it as an external element. However, how could we show that the experimentalist's input is itself localized in space and time? We might consider using the same argument where the experimentalist is also modeled as a quantum source $W$ according to \cref{c6_fig_physical_process_state_creation_and_measurement}, but this would then require \emph{another} external, localized input to be present. Said in another way, the model in \cref{c6_fig_physical_process_state_creation_and_measurement} connects localized, controlled interactions to strictly localized states, but it cannot itself show that localized signals exist to begin with; such an argument would be circular. 

There seems to be one obvious way to motivate why such a localized classical input should exist, and that is \emph{free will}. Concretely, we can assume that there are external actors, e.g., experimentalists, that can make a decision freely and locally within a sharply defined spacetime region $\mathcal{D}$. That is, they can make a decision that is not determined by any other sources and that is made \emph{here and now}. Such a decision can therefore serve as the localized, external, classical input.

In this way, free will provides an argument for why the model in \cref{c6_fig_physical_process_state_creation_and_measurement} can be realized experimentally. In fact, it means that this model is actually quite general, and applies to a wide range of experiments. For example, in any experiment where we decide to push a button to generate a pulse by an optical source, the resulting photonic state $\ket{\psi}$ must by our assumptions be strictly localized to the lightcone of this controlled interaction. We designate such locally triggered sources as \emph{on-demand} sources, to highlight that they operate at the request of an external, classical control. In addition, on-demand sources are assumed to be reliable, in the sense that we do not allow post-selecting away any trials (as this would need a different analysis). 

\FigSixFour

Refs.~\cite{gulla2021,gulla2023} consider such an on-demand setup with a 1D optical source (see \cref{c6_fig_1d_on_demand_source}) located in the half-space $x < - T$, for some constant $T > 0$. The source is assumed to be switched on at time $t = - T$ by a free choice of an experimentalist, producing a pure state $\ket{\psi}$. By causality, any local measurement at some observation point made before the signal has had time to propagate there must give the same outcome probabilities as in vacuum. The possible optical states $\ket{\psi}$ satisfying this requirement are investigated, with particular emphasis on how closely they can approximate single photons.

On-demand sources are not only important for quantum information technology; they are also useful for theoretical analyses of information propagation in quantum optical systems. Since the produced state is strictly localized and therefore has a sharp leading edge, the propagation of information can be tracked sharply by the time-dependent expectation values of local observables. In this way one gets definite answers to, e.g., questions surrounding the tunneling times of states close to single photons \cite{gulla2021a}.

\subsubsection{The role of free will}
\label{c6_sec_the_role_of_free_will}
Note that in the above discussion, the need to invoke free will to realize the source model might seem unnecessary. However, without it there does not seem to be any reason why sharply localized information should exist in the universe to begin with. Indeed, what if all quantum states, including the ones governing the experimentalist, had infinite tails in space and time \cite{hegerfeldt1998}? In such a world, there might be no sharply defined cause and effect anymore, and the source setup in our analysis might never occur. Instead, events and interactions could be smeared out in time, with small probability tails stretching off to the infinite past and into the infinite future. Such a world model would seemingly lead to no inconsistencies. 

Yet, we think there are good arguments against such a viewpoint. First, the Paley--Wiener limitation does not apply to all quantum states; for instance coherent states can be strictly localized \cite{knight1961,bialynicki-birula1998} (see \cref{c7_sec_coherent_states}) and can therefore in principle be generated on demand. In light of this, it is somewhat arbitrary to abandon sharp causality purely because one type of state cannot be strictly localized. Second, if the world were nonlocal, why would it appear to be so local? Even if the fundamental interactions were nonlocal and predetermined by infinite tails, these microscopic interactions would have to combine in a way to produce our macroscopic experience of free choice, at least as an illusion. 

Ultimately though, the question of whether sharp causality and free will exist at a fundamental level is hard to resolve by current physics, and we therefore leave it aside. The theory of local measurements and strictly localized states makes it seem at least \emph{possible} to incorporate free, local decisions with quantum field theory. We therefore take as an assumption in our analysis that on-demand sources exist.


%% file: 7_examples_of_localized_states.tex


\newcommand{\FigSevenOne}{%
    \begin{figure}%
    \hspace*{0mm}%
    \raisebox{0mm}{%
        \iftoggle{FigFromPDF}{%
            \includegraphics{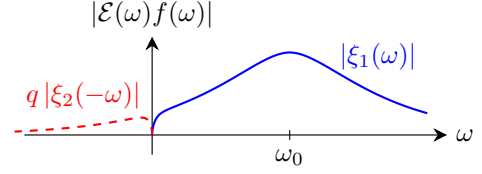}%
        }{%
            \tikzsetnextfilename{fig7-1}%
            \begin{tikzpicture}[domain=-1.3:3,scale=1.3]%
                \draw[axis_arrow] (-1.3,0) -- (3,0) node[right] {$\omega$};%
                \draw[axis_arrow] (0,-0.2) -- (0,1) node[above] {$\abs{\mathcal E\lp(\omega) f\lp(\omega)}$};%
                \draw[blue,thick,smooth] plot[domain=0:4,samples=2001,scale=0.7] (\x,{1.2*sqrt(abs(\x)/sqrt(\x*\x+0.01))*exp(-1*(\x-2)^2/(1+abs(\x-2)^(1.001)))});%
                \draw[red,dashed,thick,smooth] plot[domain=0:-2,samples=1001,scale=0.7] (\x,{1.2*sqrt(abs(\x)/sqrt(\x*\x+0.01))*exp(-1*(\x-2)^2/(1+abs(\x-2)^(1.001)))});%
                \draw[color=red] (-0.7,0.4) node {$q \abs{\xi_2\lp(-\omega)}$};%
                \draw[color=blue] (2.3,0.8) node {$\abs{\xi_1\lp(\omega)}$};%
                \draw (1.4,0.05) -- (1.4,-0.05) node[below] {$\omega_0$};%
            \end{tikzpicture}%
        }%
    }%
    \caption{\label{c7_fig_fomega}Decomposing the positive and negative spectrum of a typical smearing function into modes $\xi_1\lp(\omega)$ and $\xi_2\lp(\omega)$, respectively.}%
    \end{figure}%
}

\newcommand{\FigSevenTwo}{%
    \begin{figure}%
    \hspace*{0mm}%
    \raisebox{0mm}{%
        \iftoggle{FigFromPDF}{%
            \includegraphics{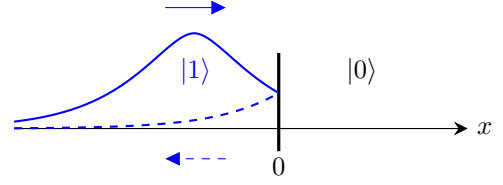}%
        }{%
            \tikzsetnextfilename{fig7-2}%
            \begin{tikzpicture}[scale=1.0]%
                \draw[axis_arrow] (-3.5,0) -- (2.5,0) node[right] {$x$};%
                \draw[very thick] (0,-0.3) -- (0,1);%
                \draw[blue,thick,smooth] plot[domain=-5:0,samples=200,scale=0.7] (\x,{1.8*exp(-(\x+1.6)^2/(1+abs(\x+1.6)^(1.001)))});%
                \draw[blue,arrows={-Stealth[inset=0pt,angle=50:6pt]}] (-1.5, 1.6) -- (-0.7, 1.6);%
                \node[blue] at (-1.1,1.5/2) {$\ket{1}$};%
                \node[black] at (1.1,1.5/2) {$\ket{0}$};%
                \draw[blue,thick,dashed,smooth] plot[domain=-5:0,samples=200,scale=0.7](\x,{1.8*exp(-(-\x+1.6)^2/(1+abs(-\x+1.6)^(1.001)))});%
                \draw[blue,dashed,arrows={-Stealth[inset=0pt,angle=50:6pt]}] (-0.7, -0.4) -- (-1.5, -0.4);%
                \draw (0,-0.5) node {$0$};%
            \end{tikzpicture}%
        }%
    }%
    \caption{\label{c7_fig_truncated_photon}\cite{rukan2026}. Schematic snapshot of an incident photon being reflected by a mirror at $x=0$ (black vertical line), shortly before the removal of the mirror. The solid blue line is the expected energy density of the forward-propagating incident part; the dashed blue line is the backward-propagating reflected part. The mirror is subsequently removed during a small interval $-T<t<0$.}%
    \end{figure}%
}

\newcommand{\FigSevenThree}{%
    \begin{figure}%
    \hspace*{0mm}%
    \raisebox{0mm}{%
        \iftoggle{FigFromPDF}{%
            \includegraphics{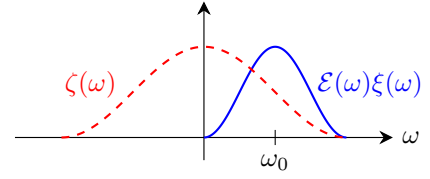}%
        }{%
            \tikzsetnextfilename{fig7-3}%
            \begin{tikzpicture}[scale=1.0]%
                \draw[axis_arrow] (-2.5,0) -- (2.5,0) node[right] {$\omega$};%
                \draw[axis_arrow] (0,-0.3) -- (0,1.8);%
                \draw[blue,thick,smooth] plot[domain=0:2*pi,samples=100,scale=0.3] (\x,{2*(1-cos((180/pi)*\x))});%
                \draw[red,thick,smooth,dashed] plot[domain=-2*pi:2*pi,samples=100,scale=0.3] (\x,{2*(1+cos((90/pi)*\x))});%
                \draw (pi*0.3,0.1) -- (pi*0.3,-0.1) node[below] {$\omega_0$};%
                \draw[red] (-1.5,0.7) node {$\zeta\lp(\omega)$};%
                \draw[blue] (2.2,0.7) node {$\mathcal E\lp(\omega)\xi\lp(\omega)$};%
            \end{tikzpicture}%
        }%
    }%
    \caption{\label{c7_fig_weakly_localized_photon_spectrum}The spectrum $\mathcal E\lp(\omega)\xi\lp(\omega)$ characterizing the photon, and the spectrum $\zeta\lp(\omega)$ of the classical pulse.}%
    \end{figure}%
}

\newcommand{\FigSevenFour}{%
    \begin{figure}%
    \hspace*{0mm}%
    \raisebox{0mm}{%
        \iftoggle{FigFromPDF}{%
            \includegraphics{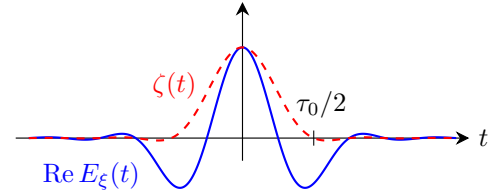}%
        }{%
            \tikzsetnextfilename{fig7-4}%
            \begin{tikzpicture}[scale=1.0]%
                \draw[axis_arrow] (-3,0) -- (3,0) node[right] {$t$};%
                \draw[axis_arrow] (0,-0.3) -- (0,1.8);%
                \draw[blue,thick,smooth] plot[domain=-3*pi:3*pi,samples=200,scale=0.3] (\x,{2*sin(2*\x*180/pi)*(1/\x - 1/(2*\x-2*pi) - 1/(2*\x+2*pi))});%
                \draw[red,thick,smooth,dashed] plot[domain=-3*pi:3*pi,samples=200,scale=0.3] (\x,{2*sin(2*\x*180/pi)*(1/\x - 1/(2*\x-pi) - 1/(2*\x+pi))});%
                \draw (pi*0.3,-0.1) -- (pi*0.3,0.1);%
                \draw (pi*0.3+0.1,0.15) node[above] {$\tau_0/2$};%
                \draw[red] (-0.9,0.7) node {$\zeta\lp(t)$};%
                \draw[blue] (-2,-0.5) node {$\Re E_{\xi}\lp(t)$};%
            \end{tikzpicture}%
        }%
    }%
    \caption{\label{c7_fig_weakly_localized_photon_field}The photon pulse as described by $\Re E_\xi\lp(t)$, and the classical pulse field $\zeta\lp(t)$.}
    \end{figure}%
}

\iftoggle{ExportTikz}{%
    \FigSevenOne
    \FigSevenTwo
    \FigSevenThree
    \FigSevenFour
     
}{}

\section{Examples of localized states}
\label{c7_sec_examples_of_localized_states}
In this section we discuss several examples of strictly localized states. The simplest way to construct strictly localized states is usually by finding the associated Licht operator $W$. This is done with the help of \hyperref[c5_rem_licht_operators_phi]{Remarks \ref*{c5_rem_licht_operators_phi}} or \ref{c5_rem_licht_operators_A}: we must find an operator $W$ that commutes with the field outside the localization region and is an isometry, $W^\dagger W = 1$. 

In \cref{c7_sec_general_methods_for_constructing_licht_operators} we start by discussing general methods for constructing Licht operators. We then use these methods to establish different families of strictly localized states. In \cref{c7_sec_coherent_states,c7_sec_squeezed_states} we consider coherent states and squeezed states, respectively. These can be generated by unitary Licht operators. We then introduce $\ket{\eta_{1,2}}$ states in \cref{c7_sec_eta12} and similar entangled states in \cref{c7_sec_other_states}, which are close to number states and Bell number states, respectively. These states can be constructed by polar decomposition of local operators. A completely different method of constructing strictly localized states is by sending non-localized states through a time-dependent optical shutter. For example (\cref{c7_sec_truncated_photon}), we may obtain strictly localized states close to a single photon by truncating the leading tail in this way. In \cref{c7_sec_localized_modes} we consider localized modes, and find that an excitation in a localized mode is locally equivalent to the corresponding vacuum, but not strictly localized. Finally, in \cref{c7_sec_weakly_localized_photons}, we consider weakly localized photons and discuss constraints on the pulse widths.

\subsection{General methods for constructing strictly localized states}
\label{c7_sec_general_methods_for_constructing_licht_operators}
Let $\mathcal{D}$ be any bounded spacetime region (see \cref{c6_fig_causal_diamonds}). In line with \cref{c6_rem_psi_produced_G} we want to find a state strictly localized to the lightcone of $\mathcal{D}$, which we can do by finding a Licht operator associated with $J\lp(\mathcal{D})$. This can for example be done by finding a Licht operator local to $\mathcal{D}$, i.e., an operator $W \in \balg{\mathcal{D}}$ such that $\adj{W} W = \identity$. Explicitly constructing operators in the local algebra $\balg{\mathcal{D}}$ is easy: we can for example just take a smeared field
\begin{equation}\label{c7_phi_f}
    \phi_f = \int_{\mathcal{D}} \sepia \dd[4] x \sepib f\lp(x) \phi\lp(x),
\end{equation}
with $f\lp(x)$ real, and truncate its spectrum to some bounded interval. The difficult part is finding such operators that are isometric, $\adj{W} W = \identity$. 

One method for achieving this suggested by \cite{knight1961} is  to take
\begin{equation}
\label{c7_W_eiL}
    W = e^{iL}, \sepc L \text{ local and self-adjoint.}
\end{equation}
For example, $L$ can be taken to be a smeared field or a product of two smeared fields. This procedure leads to strictly localized coherent states (\cref{c7_sec_coherent_states}) and strictly localized squeezed states (\cref{c7_sec_squeezed_states}), respectively.

A natural question is whether all strictly localized pure states can be generated by operators of the form \eqref{c7_W_eiL}, a conjecture put forth by \cite{knight1961}. The answer turns out to be no. The limitation of the method \eqref{c7_W_eiL} is that such operators will always be unitary. As will be demonstrated in \cref{c7_sec_eta12}, there are several examples of interesting states resulting from Licht operators that are isometric but not unitary. By uniqueness of the Licht operator associated with a given region, these states cannot also be generated by a unitary Licht operator associated with that region. Obtaining nonunitary Licht operators can be done by polar decomposition. 

\begin{theorem}\label{c7_thm_polar_decomposition}\textup{(Polar decomposition of unbounded operators).\footnote{\label{c7_foot_polar_decomposition}\cite{schmuedgen2012}, Theorem~7.2 and subsequent discussion.}}
    On a Hilbert space $H$, let $L$ be a closed (possibly unbounded but densely defined) operator with trivial kernel:
    \begin{equation}\label{c7_ker_L_0}
        \ker\lp(L) = \lp{0}.
    \end{equation}
    The operator $L$ can then be decomposed as
    \begin{equation}\label{c7_polar_decomp}
        L = W \abs{L} = W \sqrt{\adj{L} L},
    \end{equation}
    where $\abs{L}$ is positive and self-adjoint, and we have that $\ker\lp(\abs{L}) = \ker\lp(L) = 0$. The operator $W$ is an isometry on $H$, i.e., $\adj{W} W = \identity$. Note that $W$ need not be unitary. 
\end{theorem}

For $L$ satisfying the conditions of \cref{c7_thm_polar_decomposition}, the operator $W$ can be written explicitly as
\begin{equation}\label{c7_V_L}
    W = L \abs{L}^{-1} = L \frac{1}{\sqrt{\adj{L} L}}.
\end{equation}
Here the reciprocal square root is defined by applying the spectral theorem (\cref{c4_thm_spectral}) to the positive self-adjoint operator $\adj{L} L$. We can also verify that it is isometric since
\begin{equation}
    \adj{W} W = \frac{1}{\sqrt{\adj{L} L}} \adj{L} L \frac{1}{\sqrt{\adj{L} L}} = \identity.
\end{equation}
We can use this method to reduce the problem of finding isometric local operators to instead finding local operators with a trivial kernel, i.e., local operators without 0 as an eigenvalue---a much easier task. 

\begin{remark}\label{c7_rem_local_polar_decomposition}\textup{(Polar decomposition of local operators).\footnote{\label{c7_foot_local_polar_decomposition}\cite{fewster2019}, Sec.~6 and \cite{lledo2009}, Proposition~3.1 and Sec.~4.1.3.}}
    For some bounded region $\mathcal{D}$, let $L$ be a local operator in $\balg{\mathcal{D}}$ or a smeared single field of the form \eqref{c4_phi_f_bounded_algebra} or \eqref{c4_A_f_bounded_algebra} local to $\mathcal{D}$, with a possibly complex smearing function. Furthermore, let $\ker\lp(L) = \lp{0}$. Then the polar decomposition \eqref{c7_polar_decomp} of $L$ yields a $W \in \balg{\mathcal{D}}$ with $W^\dagger W = \identity$, meaning that $W$ is a Licht operator local to $\mathcal{D}$. 
\end{remark}

Note that compositions of Licht operators give new Licht operators (\cref{c5_rem_licht_operators_comp}). Therefore the families of strictly localized states in this section can be combined. For example, we can obtain strictly localized displaced squeezed states by combining Licht operators from \cref{c7_sec_coherent_states,c7_sec_squeezed_states}. 

For the examples in the following sections, we simplify to 1D by specializing to sources emitting only forward-propagating light of a single linear polarization. We therefore use the description introduced in \cref{c1_sec_notation}, writing the vector potential at a fixed observation point, taken to be $x=0$, as a scalar function of time:
\begin{equation}\label{c7_A_t}
    A\lp(t) = \minti{\dd\omega} \mathcal{A}\lp(\omega) \lp[a\lp(\omega) e^{- i \omega t} + \adj{a}\sephn\lp(\omega) e^{i \omega t}],
\end{equation}
where $\mathcal{A}\lp(\omega) = 1/\sqrt{\lp(2\pi)\lp(2\omega)}$, and $a\lp(\omega)$ is a ladder operator corresponding to the single-polarization forward-moving modes. Any photonic state of these modes can be expressed as functions of $a\lp(\omega)$ and $\adj{a} \sephn \lp(\omega)$. For example a single-photon state $\ket{1}$ of these modes can always be written as
\begin{equation}\label{c7_1}
    \ket{1} = \minti{\dd \omega} \xi\lp(\omega) \adj{a} \sephn \lp(\omega)  \ket{0},
\end{equation}
for some function $\xi\lp(\omega)$ in the mode space $L^2\sephn\lp(0, \infty)$. This description can represent, e.g., single-polarization optical signals traveling in one direction through an optical fiber. 

For forward propagation, the time dependence at a fixed position can equivalently be viewed as a dependence on $-x$ at $t=0$. Distinct arguments then correspond to spacelike separated points, where the full field commutes with itself. However, since $A\lp(t)$ contains only the forward-propagating part of the field, its commutator $\comm{A\lp(t)}{A\lp(t')}$ need not vanish for $t'\neq t$. Indeed, a straightforward calculation shows that $\comm{A\lp(t)}{A\lp(t')} \neq 0$ in general, but fortunately,
\begin{equation}\label{c7_A_E_comm}
    \comm{A\lp(t)}{E\lp(t')} = -\frac{i}{2} \delta\lp(t-t'),
\end{equation}
where
\begin{equation}\label{c7_E_t}
\begin{aligned}
    E\lp(t) 
    &= -\frac{dA\lp(t)}{dt} \\
    &= \minti{\dd\omega} \lp[\mathcal{E}\lp(\omega) a\lp(\omega) e^{- i \omega t} + \mathcal{E}^*\sephn\lp(\omega) \adj{a}\sephn\lp(\omega) e^{i \omega t}]
\end{aligned}
\end{equation}
is the forward-propagating electric field, with $\mathcal E\lp(\omega) = i\omega\mathcal A\lp(\omega)$. We therefore construct Licht operators by smearing the electric field
\begin{equation}\label{c7_E_f}
    E_f = \frac{1}{\sqrt{2\pi}} \mintn{dt} f\lp(t) E\lp(t).
\end{equation}
If the smearing function $f\lp(t)$ has support only in some interval $\mathcal{O}$, we obtain $\comm{A\lp(t)}{E_f} = 0$ for $t \in \compl{\mathcal{O}}$.

It is sometimes convenient to work with a countable basis, here referred to as a basis of \emph{pulse modes}. Define an orthonormal set of pulse modes $\xi_i\lp(\omega)$ spanning the function space $L^2\lp(0, \infty)$:
\begin{subequations}\label{c7_xi_basis}
\begin{align}
    \minti{\dd \omega} \xi_i^*\lp(\omega) \xi_j\lp(\omega) &= \delta_{ij}, \label{c7_xi_orthog} \\
    \sum_i \xi_i^*\lp(\omega) \xi_i\lp(\omega') &= \delta\lp(\omega - \omega').
\end{align}
\end{subequations}
We can then define pulse-mode ladder operators by
\begin{equation}\label{c7_a_i_def}
    a_i^\dagger = \minti{\dd \omega} \xi_i\lp(\omega) a^\dagger\sephn\lp(\omega).
\end{equation}
From $\comm{a\lp(\omega)}{a^\dagger\sephn\lp(\omega')} = \delta\lp(\omega-\omega')$ and \cref{c7_xi_orthog} they satisfy
\begin{equation}\label{eq:a_n_comm}
    \comm[\big]{\padj{a}_i}{a_j^\dagger} = \delta_{ij}.
\end{equation}
The pulse modes $\xi_i\lp(\omega)$ are clearly not associated with energy eigenstates; the energy content is given by both the particle number and also the spectrum itself.

Using the complete set $\xi_i\lp(\omega)$, we can now rewrite the first term of \cref{c7_A_t}, i.e., the positive-frequency part of $A\lp(t)$:
\begin{equation}\label{eq:E_pos_i}
    A^+\lp(t) = \sum_i A_i\lp(t) a_i,
\end{equation}
where
\begin{equation}\label{eq:A_i_def}
    A_i\lp(t) = \minti{\dd \omega} \mathcal{A}\lp(\omega) \xi_i\lp(\omega) e^{-i\omega t}.
\end{equation}
In the following sections we explore different families of states localized to some bounded or unbounded interval $\mathcal{O}$ in time.

\subsection{Coherent states}
\label{c7_sec_coherent_states}
The fact that coherent states can be strictly localized was apparent already from the original work \cite{knight1961}; related localization properties were also discussed in \cite{bialynicki-birula1998}. Let $f\lp(t)$ be a real smearing function supported in $\mathcal{O}$, i.e., $f\lp(t)=0$ for $t \in \compl{\mathcal{O}}$. From \cref{c7_E_f} we obtain
\begin{equation}\label{c7_E_f2}
    E_f = -i\alpha a_1^\dagger + i\alpha^* a_1,
\end{equation}
where
\begin{equation}
    \alpha a_1^\dagger = \minti{\dd\omega} \omega\mathcal A^*\sephn\lp(\omega)f\lp(\omega)a^\dagger\sephn\lp(\omega),
\end{equation}
and $f\lp(\omega)$ is the Fourier transform of $f\lp(t)$. The subscript 1 is used to indicate that we take the first mode $\xi_1\lp(\omega)$ in the pulse-mode basis \eqref{c7_xi_basis} to be given by\footnote{To obtain a square-integrable $\xi_1\lp(\omega)$, we need some weak assumption on $f\lp(\omega)$ or $f\lp(t)$; for example, that both $f\lp(t)$ and its derivative are in $L^2$.}
\begin{equation}\label{c7_xi1}
    \alpha\xi_1\lp(\omega) = \omega\mathcal A^*\sephn\lp(\omega)f\lp(\omega)
\end{equation}
for some normalization constant $\alpha$. By absorbing a complex phase factor into $\xi_1\lp(\omega)$, we may assume that $\alpha>0$.

We now set
\begin{equation}\label{c7_W_coh}
    W = \exp\lp(iE_f) = \exp\lp(\alpha a_1^\dagger - \alpha a_1), \sepc \alpha > 0,
\end{equation}
which makes $W$ a unitary Licht operator local to $\mathcal{O}$. We recognize $W$ as the coherent state displacement operator $W \equiv D_1\lp(\alpha)$ of pulse mode 1. Thus
\begin{equation}\label{c7_coh_state}
    \ket{\alpha_{1}} = D_1\lp(\alpha) \ket{0}
\end{equation}
is a coherent state strictly localized to $\mathcal{O}$. The localization is valid for any $\alpha > 0$, as can be realized by rescaling the smearing function $f\lp(t)$. In the limit $\alpha\to 0$ the state tends to the vacuum state.

It is interesting how the coherent state restores the negative frequencies such that it can be strictly localized. For operators $A$ and $B$ and a scalar $\lambda$, we have
\begin{equation}\label{c7_BCHexp}
    e^{\lambda B}A e^{-\lambda B} = A + \lambda \comm{B}{A} + \frac{\lambda^2}{2!} \comm[\big]{B}{\comm{B}{A}} + \dotsb.
\end{equation}
Using this identity and $D_1\lp(\alpha) = \exp\lp(iE_f)$ we find from \cref{c7_E_f,c7_A_E_comm} that
\begin{equation}\label{c7_coh_E_transf}
\begin{aligned}
    D_1^\dagger\lp(\alpha) A\lp(t) D_1\lp(\alpha) &= A\lp(t) + \frac{1}{2\sqrt{2\pi}} f\lp(t).
\end{aligned}
\end{equation}
The expected field is therefore
\begin{equation}\label{eq:coh_A_expv}
     \qexv{\alpha_1}{A\lp(t)} = \frac{1}{2\sqrt{2\pi}} f\lp(t) = \frac{1}{4\pi} \mintn{\dd \omega} f\lp(\omega) e^{- i \omega t},
\end{equation}
which is a Fourier integral over all frequencies. Since $f\lp(t)=0$ outside $\mathcal{O}$, the expectation value \eqref{eq:coh_A_expv} vanishes there as well, as required from the fact that the state \eqref{c7_coh_state} is strictly localized to $\mathcal{O}$. We can also find, e.g., the expected (normal-ordered) energy density, which is proportional to $\alpha^2$ but also dependent on the spectrum of the mode $\xi_1\lp(\omega)$. As expected, the energy density vanishes outside $\mathcal{O}$.

\subsection{Squeezed states}
\label{c7_sec_squeezed_states}
Similarly to \cref{c7_W_coh}, we can generate Licht operators from a quadratic smeared field,
\begin{equation}\label{c7_W_Ef2}
    W = \exp\lp(iE_f^2),
\end{equation}
or, more generally, from a smeared product of two fields,
\begin{equation}\label{c7_W_fst}
    W = \exp\lp(\frac{i}{2\pi} \int \sepia ds \sepia \int \sepia dt \sepib f\lp(s,t) E\lp(s)E\lp(t)),
\end{equation}
where $f\lp(t)$ is real and vanishes outside $\mathcal O$, while $f^*\sephn\lp(t,s)=f\lp(s,t)$ and $f\lp(s,t)$ vanishes unless $s,t\in\mathcal O$. Clearly the operators \eqref{c7_W_Ef2} and \eqref{c7_W_fst} are unitary and commute with $A\lp(t)$ for $t \in \compl{\mathcal{O}}$, so the resulting states $W\ket{0}$ are strictly localized to $\mathcal{O}$.

We now consider \cref{c7_W_Ef2} in more detail. By substituting \cref{c7_E_f2}, we can write
\begin{equation}
    W = \exp\lp(-i\alpha^2 (a_1^2 + a_1^{\dagger 2} - 2 a_1^\dagger \padj{a}_1 - 1)). 
\end{equation}
The exponential can be disentangled using the technique in \cite{truax1985}, giving
\begin{equation}
    W =  \exp\lp(\mu a_1^{\dagger 2}/2) \exp\lp(\nu (a_1^\dagger \padj{a}_1+1/2)) \exp\lp\Big(\mu a_1^2/2), 
\end{equation}
for
\begin{subequations}
\begin{align}
    \mu &= \frac{-2i\alpha^2}{1-2i\alpha^2}, \label{c7_mu_alpha} \\
    \nu &= -\ln\lp(1-2i\alpha^2).    
\end{align}
\end{subequations}
When the Licht operator $W$ acts on the vacuum, it produces the strictly localized state
\begin{equation}
    W\ket{0} = \exp\lp\big(\nu/2) \exp\lp(\mu a_1^{\dagger 2}/2) \ket{0}.
\end{equation}

This is a squeezed state; up to a phase factor it is also produced as $T_1\lp(\zeta)\ket{0}$ by the unitary squeeze operator \cite{truax1985} 
\begin{equation}
    T_1\lp(\zeta) = \exp\lp( \zeta^* a_1^2/2 - \zeta a_1^{\dagger 2}/2 ).
\end{equation}
The connection between the complex parameters is formulated in polar form:
\begin{equation}\label{c7_zeta_mu}
    \zeta = -\frac{\mu}{\abs{\mu}} \arctanh\abs{\mu}. 
\end{equation}
In other words, the squeezed vacuum $T_1\lp(\zeta)\ket{0}$ is strictly localized for the values of $\zeta$ given by \cref{c7_zeta_mu,c7_mu_alpha}, for any $\alpha>0$.

Note that although $W$ and $T_1\lp(\zeta)$ are both unitary and produce the same state from vacuum, they are not equal, even up to a phase factor. Due to uniqueness of the Licht operator (\cref{c5_thm_lichts_theorem}), $T_1\lp(\zeta)$ cannot be a Licht operator since $W$ is. Therefore $T_1\lp(\zeta)$ does not commute with $A\lp(t)$ outside $\mathcal{O}$.

We can also construct strictly localized, multimode states. In \cref{c7_sec_truncated_photon} we discuss a strictly localized, multimode squeezed state obtained in a different way.

\subsection{\texorpdfstring{\boldmath$\ket{\eta_{1,2}}$}{|eta1,2>} states close to single photons}
\label{c7_sec_eta12}
So far we have discussed strictly localized states generated by unitary Licht operators of the form $W=\exp\lp(iL)$ where $L$ is a self-adjoint operator, linear (\cref{c7_sec_coherent_states}) or quadratic (\cref{c7_sec_squeezed_states}) in the field in $\mathcal{O}$. We could continue this strategy, letting $L$ be cubic or even an $n$th-order polynomial in the field. Still the method is somewhat limited in that it gives only unitary Licht operators. We now turn to nonunitary isometries $W$, satisfying $W^\dagger W=\identity$ but not $WW^\dagger=\identity$.

In particular we construct a class of strictly localized states close to single photons. These states will be denoted $\ket{\eta_{1,2}}$. The subscripts 1 and 2 refer to two pulse modes, and $\eta$ is a parameter which decides how close to a single photon the state is (for $\eta\to 0$ the state approaches a single photon, but the limit is unattainable). We know that even though these states may have a single-photon component close to one, being strictly localized they must contain terms with arbitrarily high photon numbers, in accordance with \cref{c5_thm_knight_inf_photon_numbers}.

One motivation for studying these states is that single-photon states are quite central in quantum information technology. An on-demand source that is triggered at some time must by causality (under some weak assumptions) produce a strictly localized state (\cref{c6_sec_on_demand_sources}). Thus, on-demand photonic sources do not produce exact single photons, and it is of interest to consider the necessary deviation. 

To construct a strictly localized state close to a single photon, we start with a smeared field $E_f$ for a smearing function $f\lp(t)$ supported in some interval $\mathcal{O}$. To use polar decomposition (\cref{c7_thm_polar_decomposition}), we are not looking for an observable, so we take $f\lp(t)$ complex so that in general $\adj{E}_f \neq E_f$. We will establish that 0 is not in the spectrum of $\abs{E_f}$ for a broad range of smearing functions $f\lp(t)$. Then we obtain from \cref{c7_V_L} a Licht operator
\begin{equation}\label{c7_W_E_f}
    W = E_f \frac{1}{\sqrt{\adj{E}_f \padj{E}_f}}.
\end{equation}

For reasons that become clear shortly, we normalize $f\lp(t)$ such that
\begin{equation}\label{c7_normalization_f}
    \minti{\dd\omega} \abs{\mathcal{E}\lp(\omega)}^2 \abs{f\lp(\omega)}^2 = 1,
\end{equation}
and adjust the phase $f\lp(t) \mapsto e^{i\theta} f\lp(t)$ for some real constant $\theta$ such that
\begin{equation}
    \minti{\dd\omega} \abs{\mathcal{E}\lp(\omega)}^2 f\lp(\omega) f\lp(- \omega) \geq 0.
\end{equation}
For any smeared operator $E_f$ of the form \eqref{c7_E_f}, we can write
\begin{equation}\label{c7_E_f_a1_a2}
    E_f = a_1^\dagger + p a_1 + q a_2,
\end{equation}
for some constants $p,q \geq 0$. The intuition behind this is that we can let the positive-frequency spectrum be the mode $\xi_1\lp(\omega)$. Then the negative-frequency spectrum will contain some part along $\xi_1\lp(\omega)$ and some orthogonal part $\xi_2\lp(\omega)$. Finally, the freedom in the phases of $\xi_1\lp(\omega)$ and $\xi_2\lp(\omega)$ allows us to eliminate any phases for $p$ and $q$. The details of showing \cref{c7_E_f_a1_a2} are to rewrite \cref{c7_E_f} as
\begin{equation}\label{E_f_F}
    E_f = \minti{\dd \omega} \mathcal{E}\lp(\omega) f\lp(-\omega) a\lp(\omega) + \minti{\dd \omega} \mathcal{E}^*\sephn\lp(\omega) f\lp(\omega) a^\dagger\sephn\lp(\omega)
\end{equation}
and then define
\begin{subequations}\label{c7_xi1_xi2_p_q}
\begin{align}
    \xi_1\lp(\omega) &= \mathcal{E}^*\sephn\lp(\omega) f\lp(\omega), \\
    p &= \minti{\dd\omega} \abs{\mathcal{E}\lp(\omega)}^2 f\lp(\omega) f\lp(- \omega), \label{c7_p}\\
    \xi_{\perp}\lp(\omega) &= \mathcal{E}^*\sephn\lp(\omega) f^*\sephn\lp(- \omega) - p \xi_1\lp(\omega), \label{c7_xi_perp}\\
    q^2 &= \minti{\dd\omega} \abs{\xi_{\perp}\lp(\omega)}^2, \label{c7_q}\\
    \xi_2\lp(\omega) &= \frac{1}{q} \xi_{\perp}\lp(\omega), \sepc \text{if } q\neq 0.
\end{align}
\end{subequations}

Already now, we see how the Licht operator \eqref{c7_W_E_f} can give a strictly localized state close to a single photon. By picking a smearing function with spectrum mainly for $\omega>0$, we will obtain $p,q \ll 1$ such that $E_f$ is roughly equal to $a_1^\dagger$. The resulting state $W \ket{0}$ will therefore be roughly the single photon $a_1^\dagger \ket{0}$. It is not possible to obtain equality, since given $f\lp(\omega) = 0$ for $\omega<0$ the Paley--Wiener criterion (\cref{c2_thm_paley_wiener}) forbids $f\lp(t) = 0$ outside $\mathcal{O}$.

Another interesting observation is that the only way to obtain $q=0$ is by having $p=1$. Assuming $q=0$, \cref{c7_xi1_xi2_p_q} means that $f^*\sephn\lp(-\omega) = p f\lp(\omega)$. Define $g\lp(t) = f^*\sephn\lp(t) - p f\lp(t)$ which is supported in $\mathcal{O}$. The Fourier transform is $g\lp(\omega) = f^*\sephn\lp(-\omega) - p f\lp(\omega) = 0$ for $\omega>0$. By the Paley--Wiener criterion this is impossible unless $g\lp(t)=0$, and therefore $p=1$ and $f\lp(t)$ is real. For this special case, $E_f$ is hermitian, so the construction \eqref{c7_W_E_f} does not yield a nonunitary isometry as wanted (however the construction in \cref{c7_sec_coherent_states,c7_sec_squeezed_states} can be used, leading to coherent states, squeezed states, etc.). Excluding this case, we must have $q>0$, so the strictly localized state made from the smeared field will be expressed from two modes.

To rewrite the Licht operator \eqref{c7_W_E_f} in a more useful form, we aim to find a unitary Bogoliubov transformation $U$ such that
\begin{equation}
    E_f = a_1^\dagger + p a_1 + q a_2 = \text{const.} \cdot U a_1^\dagger \adj{U}. \label{c7_bog_U}
\end{equation}
If this can be done, $\adj{E}_f\padj{E}_f \propto U \lp\big(1 + \adj{a}_1\padj{a}_1) \adj{U}$ does not have 0 in its spectrum. Furthermore, the Licht operator \eqref{c7_W_E_f} can be expressed
\begin{equation}
    W = U \adj{a}_1 \adj{U} \frac{1}{\sqrt{U \padj{a}_1 \adj{a}_1 \adj{U}}} = U \adj{a}_1 \frac{1}{\sqrt{\padj{a}_1 \adj{a}_1}} \adj{U} = U \seph \adj{\wt{a}}_1 \adj{U},
\end{equation}
where
\begin{equation}
    \adj{\wt{a}}_1 = \adj{a}_1 \frac{1}{\sqrt{\padj{a}_1 \adj{a}_1}} = \sum_n \ket{{n + 1}_1} \bra{n_1} \label{c7_at}
\end{equation}
is the right shift operator on mode 1, i.e., a ladder operator on the states $\ket{n_1}$ without the usual normalization constant $\sqrt{n + 1}$:
\begin{equation}
    \adj{\wt{a}}_1  \ket{n_1} = \ket{{n + 1}_1}.
\end{equation}

To determine the required Bogoliubov transformation $U$, we require
\begin{equation}
    p^2 + q^2 < 1, \label{c7_cond_bog}
\end{equation}
and determine the constant in \cref{c7_bog_U} such that the canonical commutator relation is preserved ($\comm*{\padj{b}_1}{\adj{b}_1} = 1$):
\begin{equation}
    \adj{b}_1 \equiv U a_1^\dagger \adj{U} = \frac{a_1^\dagger + p a_1 + q a_2}{\sqrt{1-p^2-q^2}}. \label{c7_bog_U_b}
\end{equation}
Take $U=TS$, where
\begin{subequations}
\begin{align}
    T &= e^{\rho a_1^2/2 - \rho a_1^{\dagger2}/2}, \label{c7_Tdef}\\
    S &= e^{\gamma a_1 a_2 - \gamma a_1^\dagger a_2^\dagger} \label{c7_Sdef}
\end{align}
\end{subequations}
are the one- and two-mode squeeze operators respectively \cite{schumaker1985}. The intuition is that $T$ may produce the $a_1$ term, while $S$ gives $a_2$. Using the formula \eqref{c7_BCHexp} we find that \cref{c7_bog_U_b} is satisfied when
\begin{subequations}
\begin{align}
    \tanh\rho &= p, \\
    \tanh\gamma &= \frac{q}{\sqrt{1-p^2}}.
\end{align}
\end{subequations}

We are now ready to interpret the constructed localized state
\begin{equation}\label{c7_eta12}
    W\ket{0}, \sepc W = TS \seph \adj{\wt{a}}_1\adj{S}\adj{T}.
\end{equation}
Clearly, when $p\ll 1$ and $q\ll 1$, the squeeze operators $S$ and $T$ are close to identity, such that the resulting state $W\ket{0}$ is close to a single photon in mode 1. To approach this situation, we pick a smearing function $f\lp(t)$ with most of its spectrum $f\lp(\omega)$ for $\omega>0$. This can be achieved if $f\lp(t)$ is a pulse with typical variation scale $\tau$ for the envelope much longer than the inverse carrier (central) frequency $\omega_0$, i.e., $\omega_0 \tau \gg 1$. We may for example take
\begin{equation}
    f\lp(t) = w\lp(t) \exp\lp(-i\omega_0 t),
\end{equation}
for some real $w\lp(t)$ which is slowly varying over $2\pi/\omega_0$, but still only supported in the interval $\mathcal{O}$. This is only possible if the localization region $\mathcal{O}$ is large compared with $2\pi/\omega_0$.

To get a feeling for what this strictly localized state looks like, we consider the special case $p=0$. It is not entirely obvious that $p=0$ is possible; however, if $p\neq 0$ for a given smearing function $f\lp(t)$, one can obtain $p=0$ by a small modification of $f\lp(t)$ [see the Supplemental Material for \cite{gulla2021}]. With $p=0$, the strictly localized state $W \ket{0}$ takes the form
\begin{equation}\label{c7_eta12_ort}
    \ket{\eta_{1,2}} = S \seph \adj{\wt{a}}_1 \adj{S} \ket{0} = c_1  \ket{1_1 \: 0_2} + c_2  \ket{2_1 \: 1_2} + c_3  \ket{3_1 \: 2_2} + \dotsb
\end{equation}
for some coefficients $c_1, \ c_2, \ \dotsc$, obtained by expanding the exponential in $S$. The subscripts 1 and 2 in the kets refer to the modes $\xi_1\lp(\omega)$ and $\xi_2\lp(\omega)$, which in this case are given by [see \cref{c7_xi1_xi2_p_q}]:
\begin{subequations}\label{c7_xi1_xi2_q}
\begin{alignat}{2}
    \xi_1\lp(\omega) &= \mathcal{E}^*\sephn\lp(\omega) f\lp(\omega), \sepc &&\omega>0, \\
    \xi_2\lp(\omega) &= \frac{1}{q} \mathcal{E}^*\sephn\lp(\omega) f^*\sephn\lp(- \omega), \sepc &&\omega>0.
\end{alignat}
\end{subequations}
Intuitively, the state \eqref{c7_eta12_ort} is strictly localized since the two modes make up a combined spectrum from $-\infty$ to $\infty$ (see \cref{c7_fig_fomega}). The mode $\xi_1\lp(\omega)$ is obtained from the positive spectrum, which is the main part of $f\lp(\omega)$, while the mode $\xi_2\lp(\omega)$ is obtained from the small negative spectrum tail of $f\lp(\omega)$, after normalization. 

The parameter that decides the fidelity to a single photon is the amount of negative frequencies in \cref{c7_fig_fomega}. Whenever $\mathcal E\lp(\omega)$ is used for negative frequencies, we use the extension $\mathcal{E}\lp(-\omega)= \mathcal{E}^*\sephn\lp(\omega)$, as in \cref{c1_qexv_alpha_xi_E_t}. Let $\eta$ be the fraction of the squared norm for the negative part of the spectrum:
\begin{equation}
    \eta = \frac{\mintm{\dd\omega} \abs{\mathcal E\lp(\omega)f\lp(\omega)}^2}{\mintn{\dd\omega} \abs{\mathcal E\lp(\omega)f\lp(\omega)}^2} = \frac{q^2}{1+q^2}.
\end{equation}
If $\eta \ll 1$ we have $\eta \approx q^2 \approx \gamma^2$. Expanding \cref{c7_Sdef}, we find that the fidelity between $\ket{\eta_{1,2}}$ and the single-photon part is
\begin{equation}
\begin{aligned}    
    F &\equiv \abs{\braket{1_1 \: 0_2}{\eta_{1,2}}} = \abs{c_1} = 1 - \frac{3 - 2 \sqrt{2}}{2} \eta + \order{\eta^2} \\
    &\approx 1 - 0.09 \eta. 
\end{aligned}
\end{equation}
While the limit $\eta\to 0$ is unattainable, one may come as close as one wishes when the smearing function envelope varies sufficiently slowly over a cycle $2\pi/\omega_0$.

\FigSevenOne

In Refs.~\cite{gulla2021} and \cite{gulla2023}\footnote{In these references a different method is used to construct the $\ket{\eta_{1,2}}$ state, where the starting point is a so-called seed function $G\lp(\omega)$, related to our smearing function $f\lp(t)$ by $G\lp(\omega) \propto \sqrt{-i\omega} f\lp(\omega)$. A three-dimensional version of the state is given in \cite{ryen2022}.}, the $\ket{\eta_{1,2}}$ state is used to determine lower bounds for how close to a single-photon target one can get with an on-demand source (see \cref{c6_sec_localization_and_causality} for a discussion on the connection between causality and localization). These references also give upper bounds for the fidelity. Depending on the target, the upper bounds depend on the amount of desired (unphysical) negative frequencies of the target pulse, or size of the (noncausal) negative time tail of a target photon. The limitations for realizing strictly localized states close to single photons on demand can be summarized as follows:

\begin{remark}\label{c7_rem_loc_states_few_cycles}\remheadp{}
There are two distinct restrictions for ultrashort photon pulses. First, the absence of negative frequencies places a lower bound on the duration of a single-photon pulse measured in optical cycles (\cref{c7_sec_weakly_localized_photons}). Second, a strictly localized state can never be an exact single photon. This departure can be made arbitrarily small in the many-cycle limit, but becomes appreciable for localization to only a few optical cycles \cite{gulla2021,gulla2023}.
\end{remark}

We have thus seen that some interesting localized pure states, such as $\ket{\eta_{1,2}}$, can only be generated by Licht operators $W$ that are isometric but not unitary. The Licht operator in \cref{c7_eta12} is of this category since the shift operator $\wt{a}$ satisfies $\wt{a} \seph \adj{\wt{a}} = \identity$, but $\adj{\wt{a}} \wt{a} \neq \identity$. It is interesting to note that the so-called Wold decomposition \cite{neumann1930,wold1954,sarkar2014} characterizes isometries by decomposing them into unitary and shift components.

\subsection{Other strictly localized states}
\label{c7_sec_other_states}
In the previous section, we focused on strictly localized states close to single photons. Since products of Licht operators are also Licht operators, $W^n \ket{0}$ will be a strictly localized state close to an $n$-photon state, for $W$ given by \cref{c7_eta12}. We may also produce strictly localized states by compositions of this and the other Licht operators seen so far.

It is tempting to try to construct strictly localized states by superposition to obtain strictly localized entangled states. However, in light of \cref{c5_rem_licht_operators_superposition} this is not straightforward. Instead, entangled states can be produced by sending $\ket{\eta_{1,2}}$ through a beamsplitter. Since $\ket{\eta_{1,2}}$ is close to a single photon, the output of a 50\% beamsplitter will be close to the single-photon output $\lp\big(\ket{01} + \ket{10})/\sqrt{2}$, where, e.g., $\ket{01}$ represents vacuum in one output port and a single photon in the other.

Another way of producing an entangled, strictly localized state is by superposition of local operators \emph{before} inserting the result into \cref{c7_V_L}. For example, we may take $L = E^1_f + E^2_f$ for orthogonal polarizations 1 and 2:
\begin{equation}\label{c7_W_phi_f_phi_g}
    W = \lp\big(E^1_f+E^2_f) \frac{1}{\sqrt{\lp\big(E^{1\dagger}_f+E^{2\dagger}_f)\lp\big(E^1_f+E^2_f)}}.
\end{equation}
Define a unitary operator $U = U^1 \otimes U^2$, where the unitaries $U^i = T^i S^i$ are composed of one-mode ($T^i$) and two-mode ($S^i$) squeeze operators for each polarization $i$, as defined in \cref{c7_sec_eta12}. We can write
\begin{equation}\label{c7_W_phi_f_phi_g_U}
    W = U \lp\big(a^{1\dagger}_1 + a^{2\dagger}_1) \frac{1}{\sqrt{\lp\big(a^{1}_1 + a^{2}_1)\lp\big(a^{1\dagger}_1 + a^{2\dagger}_1)}} U^\dagger,
\end{equation}
where, again, superscripts refer to polarization. 

By picking a smearing function as in \cref{c7_sec_eta12}, the operator $U$ can be close to identity, so the produced state can be close to a Bell state of the two polarization modes:
\begin{equation}
    W\ket{0} \approx \frac{a^{1\dagger}_1 + a^{2\dagger}_1}{\sqrt 2} \ket{0}.
\end{equation}
Here we have inserted a factor $1/\sqrt 2$ both in the numerator and denominator of \cref{c7_W_phi_f_phi_g_U}, to express $W$ with the normalized ladder operator $\lp\big(a^{1\dagger}_1 + a^{2\dagger}_1)/\sqrt{2}$.

\subsection{Truncated photons}
\label{c7_sec_truncated_photon}
In \cref{c7_sec_eta12} we discussed strictly localized states that could have been produced by photonic on-demand sources. Another natural way of creating strictly localized states close to single photons is to use a time-dependent optical shutter to truncate an incoming photon into a strictly localized state by removing, e.g., a part of the leading tail \cite{rukan2026}.

A time-dependent shutter can be realized as a time-dependent mirror, reflecting the leading tail of the photon, while letting the remaining part through (see \cref{c7_fig_truncated_photon}). We assume that the mirror reflectivity is (approximately) unity for $t<-T$ and zero for $t>0$, so $T$ is the time needed to turn off the mirror. Systems with explicit time dependence break time-translation symmetry, and are therefore not necessarily energy conserving. Indeed, quantum optical analyses of such systems \cite{moore1970,dewitt1975,fulling1976,yablonovitch1989,dodonov1993,law1994,cirone1999,braunstein2005,foo2020} lead to Bogoliubov transformations, creating new photons from vacuum induced by the external control.

\FigSevenTwo

If the initial state with the mirror present is vacuum, the state after the mirror has been removed (for $t>0$) can be written in the form \cite{rukan2026}
\begin{equation}\label{c7_shutter_new_vacuum}
    \mathcal{N}\exp\left[-\frac{1}{2}\mintn{dk} \! \mintn{dk'}  K\lp(k,k')a^{\dagger}\sephn\lp(k)a^{\dagger}\sephn\lp(k')\right]\ket{0}.
\end{equation}
Here, $\mathcal N$ is a normalization constant, $K\lp(k,k')$ is a function determined by the detailed time dependence of the mirror, $a\lp(k)$ is the annihilation operator as a function of one-dimensional wavenumber $k$, and $\ket{0}$ is the vacuum state associated with forward- and backward-propagating modes. The state \eqref{c7_shutter_new_vacuum} describes the production of photons due to the time dependence of the mirror. Given the form of the exponential, this is a multimode-squeezed vacuum state. When only interested in the forward-propagating modes, one must trace out the backward-propagating modes, giving rise to a mixed state. Consistent with causality, this state is strictly localized to $x<t+T$. For the gradual-removal scheme studied by \cite{rukan2026}, the expected number of produced photons has an upper bound that tends to zero as $T \to \infty$.

Returning to the original single-photon input, the truncated photon state is determined by first calculating the total state for $t>0$, including the backward-propagating modes (with negative $k$'s) \cite{rukan2026}:
\begin{equation}\label{c7_truncated_photon}
    \mathcal{N} a_{\zeta}^{\dag} \exp\left[-\frac{1}{2}\mintn{dk} \! \mintn{dk'} K\lp(k,k')a^{\dagger}\sephn\lp(k)a^{\dagger}\sephn\lp(k')\right] \ket{0}.
\end{equation}
Here
\begin{equation}
    a_\zeta^\dagger = \mintn{dk} \zeta\lp(k)a^\dagger\sephn\lp(k),
\end{equation}
where $\zeta\lp(k)$ is determined by the incoming photon spectrum and the time dependence of the mirror. The state \eqref{c7_truncated_photon} can be seen as a photon-added version of the state \eqref{c7_shutter_new_vacuum}; however, the spectrum $\zeta\lp(k)$ has a complicated relation to the original photon spectrum $\xi\lp(k)$.

By tracing out the backward-propagating modes from the state \eqref{c7_truncated_photon}, we are left with the truncated photon state. This state is complicated and contains terms with arbitrarily high photon numbers, in accordance with the mixed-state extension of Knight's theorem \cite{rustad2026}. However, considering forward-propagating modes, the truncated photon state is locally equivalent to the original photon for $x<t$, and locally equivalent to vacuum for $x>t+T$ (see \cref{c5_def_local_equivalence})\footnote{The equivalence to vacuum ahead of the transition region is approximate for nonzero initial mirror transmissivity, but can be made arbitrarily accurate by reducing this transmissivity.}.

\subsection{Localized modes}
\label{c7_sec_localized_modes}
Localization has also been studied from the perspective of field modes in, e.g., \cite{shirokov1980,colosi2008,vazquez2014,su2016a,foo2020,onoe2022,camblong2024}. This can be done, for example, by smearing the field with test functions supported in a bounded region, or by constructing mode functions using coordinates adapted to spacetime domains such as the interior and exterior of a causal diamond. While the implementations differ in detail (some select only a finite set of localized modes as an effective subsystem, whereas others introduce a complete localized mode basis), the common structure is that ladder operators corresponding to the new modes $\padj{b}_i,b_i^\dagger$ are linear combinations of the ladder operators corresponding to the standard global modes $\padj{a}_j,a_j^\dagger$. This defines a Bogoliubov transformation. The Bogoliubov transformation need not be unitarily implementable, in which case the corresponding Fock representations are unitarily inequivalent. 
Excitations built from the $b_i$ modes can be locally equivalent to the corresponding \enquote{new vacuum} $\ket{0^b}$, defined by $b_i\ket{0^b}=0$, outside their localization region, without being locally equivalent to the physical vacuum $\ket{0}$.

Below we consider a simple version of such a Bogoliubov transformation in which only a single localized mode is constructed. In this case the transformation is unitarily implementable by a two-mode squeeze operator. For simplicity, we stick to the one-dimensional, forward-propagating situation. The following analysis will depend on that in \cref{c7_sec_eta12}, where we start with a smearing function $f\lp(t)$ supported in some interval $\mathcal{O}$, and from its positive and negative frequency spectra identify modes $\xi_1\lp(\omega)$ and $\xi_2\lp(\omega)$ according to \cref{c7_xi1_xi2_q}. Again we assume $p=0$ and $q=\tanh\gamma$, for some $\gamma>0$.

Consider the corresponding mode set $A_i\lp(t)$ in \cref{eq:A_i_def}. It turns out that we can construct a localized mode $B_1\lp(t)$ supported in $\mathcal{O}$ and a (delocalized) mode $B_2\lp(t)$ using the Bogoliubov transformation in \cref{c7_sec_eta12}:
\begin{subequations}\label{c7_bog_U_p0}
\begin{align}
    \adj{b}_1 &= \frac{a_1^\dagger + q a_2}{\sqrt{1-q^2}},  \\
    \adj{b}_2 &= \frac{a_2^\dagger + q a_1}{\sqrt{1-q^2}}.  
\end{align}   
\end{subequations}
This leads to the new modes
\begin{subequations}\label{c7_bog_U_p0_modes}
\begin{align}
    B_1\lp(t) &= \frac{A_1\lp(t) - q A_2^*\lp(t)}{\sqrt{1-q^2}},  \label{c7_bog_B_1} \\
    B_2\lp(t) &= \frac{A_2\lp(t) - q A_1^*\lp(t)}{\sqrt{1-q^2}},  
\end{align}   
\end{subequations}
such that the field $A\lp(t)$ can be expressed with either set:
\begin{subequations}
\begin{align}
    A\lp(t) &= \sum_i \lp[A_i\lp(t) a_i + A_i^*\sephn\lp(t) \adj{a}_i]  \label{c7_A_Ai} \\
    &= \sum_i \lp[B_i\lp(t) b_i + B_i^*\sephn\lp(t) \adj{b}_i]. \label{c7_A_Bi}
\end{align}
\end{subequations}
We have let all modes $i\geq 3$ be unchanged, $B_i\lp(t) = A_i\lp(t)$. To verify that $B_1\lp(t)$ is supported in $\mathcal{O}$, we calculate \cref{c7_bog_B_1} after inserting \cref{eq:A_i_def,c7_xi1_xi2_q}:
\begin{equation}
    B_1\lp(t) = - \frac{i}{2\sqrt{2\pi}} \sqrt\frac{1}{1-q^2} f\lp(t).
\end{equation}

We have two different vacua, $\ket{0}$ and $\ket{0^b}$, defined by 
\begin{subequations}
\begin{align}
    a_i \ket{0} = 0, \label{c7_def_a_vac} \\
    b_i \ket{0^b} = 0, \label{c7_def_b_vac}   
\end{align}    
\end{subequations}
for all $i$. From \cref{c7_bog_U_b} we get
\begin{equation}
    b_1 = S a_1 S^\dagger,
\end{equation}
since $T=\identity$ for $p=0$. The same relation applies to mode 2. In other words, the Bogoliubov transformation is implemented by the two-mode squeeze operator \eqref{c7_Sdef}, and we find
\begin{equation}
    \ket{0^b} = S \ket{0}.
\end{equation}

Clearly, even though the mode $B_1\lp(t)$ is supported in $\mathcal{O}$, the (unnormalized) state $b_1^\dagger \ket{0}$ is not strictly localized, as it corresponds to a single photon $a_1^\dagger \ket{0}$. An interesting question is whether the state
\begin{equation} \label{c7_loc_mode_ex}
    b_1^\dagger \ket{0^b}
\end{equation}
is strictly localized to $\mathcal{O}$. As shown in \cite{vazquez2014}, the answer is that it is locally equivalent to $\ket{0^b}$ in $\compl{\mathcal{O}}$. However, it is not locally equivalent to $\ket{0}$. Indeed, considering an operator $L$ local to $\compl{\mathcal{O}}$, we have
\begin{align} \label{c7_loc_eq_b1_ex}
    \qexv{0^b}{b_1 L b_1^\dagger} = \qexv{0^b}{L},    
\end{align}
since $\comm{L}{b_1} = 0$. This commutator vanishes because the field operators generating $L$ have the form \eqref{c7_A_Bi}, and $B_1\lp(t)=0$ in $\compl{\mathcal{O}}$. The local equivalence \eqref{c7_loc_eq_b1_ex} does not mean that the state \eqref{c7_loc_mode_ex} is strictly localized, since
\begin{equation}
    \qexv{0^b}{L} \neq \qexv{0}{L}
\end{equation}
in general. This can be demonstrated by substituting $E^2\lp(t)$ for $L$, and calculating the difference between the two expectation values for $t\in\compl{\mathcal{O}}$, noting that $B_1\lp(t)=0$ there.

Note that $\mathcal O$ is a time interval at a fixed observation point; propagation is already included through the dependence on $t-x$. Thus the failure of strict localization is not simply due to an excitation propagating away. On the other hand,
\begin{equation}
    b_1^\dagger \frac{1}{\sqrt{\padj{b}_1b_1^\dagger}} \ket{0}
\end{equation}
\emph{is} strictly localized, and can be close to a single photon; in fact it is the state obtained by polar decomposition and discussed in \cref{c7_sec_eta12}.

In the spirit of relativistic quantum information, it is interesting to calculate the reduced density operator corresponding to the localized mode $B_1\lp(t)$, in the global vacuum $\ket{0}$. We express $\ket{0} = S^\dagger \ket{0^b}$, where 
\begin{equation}
    S = e^{\gamma a_1a_2 - \gamma a_1^\dagger a_2^\dagger} = e^{\gamma b_1b_2 - \gamma b_1^\dagger b_2^\dagger}.
\end{equation}
From the form of the exponent of $S$, the operator $S^\dagger$ creates the same number of excitations in modes $b_1$ and $b_2$. Writing a general such state, and demanding\linebreak\textit{} $(b_1-qb_2^\dagger)\ket{0}=0$ [resulting from \cref{c7_def_a_vac}], we find that $\ket{0}$ can be expressed in the form
\begin{equation}\label{c7_vac_b1}
    \ket{0} = \sqrt{1-q^2} \sum_n q^n \ket{n^{b_1}}\ket{n^{b_2}}.
\end{equation}
Here, e.g., $\ket{n^{b_1}}$ means $n$ excitations of the mode $b_1$, and we have omitted the modes $b_i$ for $i\geq 3$. Clearly, the vacuum state, expressed in the form \eqref{c7_vac_b1}, is an entangled state. If we trace out the modes $i\geq 2$, we obtain the density operator
\begin{equation} \label{c7_rho1_b1}
    \rho_1 = \sum_n \frac{\expval{n}^n}{(1+\expval{n})^{n+1}} \ket{n^{b_1}}\bra{n^{b_1}},
\end{equation}
with
\begin{equation}
    \expval{n} = \frac{q^2}{1-q^2}.
\end{equation}
The state \eqref{c7_rho1_b1} has thermal excitation number statistics with mean $\expval{n}$.

Two comments are in order. First, even though the mode $B_1\lp(t)$ is supported in $\mathcal{O}$, the local algebra is larger than that generated by the component $B_1\lp(t)b_1 + B_1^*\lp(t)b_1^\dagger$ of the field. This is because the other modes $B_i\lp(t)$ have supports overlapping with $\mathcal{O}$ as well.

Second, as demonstrated in \cite{onoe2022}, for a short-duration measurement of the vacuum $\ket{0}$, virtual particles are transferred to the detector as real excitations. We illustrate this feature through the excitation-number statistics of a localized mode. The observable $\padj{E}_f E_f^\dagger$ made from the smeared electric field \eqref{c7_E_f} can be expressed [see \cref{c7_E_f_a1_a2}]
\begin{equation}
    \padj{E}_f E_f^\dagger = \lp(1-q^2) b_1^\dagger \padj{b}_1, 
\end{equation}
which is proportional to the number operator $b_1^\dagger \padj{b}_1$. For the vacuum $\ket{0}$, the corresponding excitation-number statistics are therefore the thermal statistics in \cref{c7_rho1_b1}. Thus the local observable $\padj{E}_fE_f^\dagger$ assigns a thermally distributed \enquote{excitation number} to the vacuum. For $q$ close to 1, the mean number $\expval{n}$ gets large. As discussed in \cref{c7_sec_eta12} this occurs when the width of $f\lp(t)$ is small compared to the inverse carrier frequency.

\subsection{Weakly localized photons: Bounds for duration \texorpdfstring{\boldmath$\cdot$}{\textperiodcentered} bandwidth}
\label{c7_sec_weakly_localized_photons}
We have seen that single photons cannot be strictly localized (\cref{c5_thm_knight_inf_photon_numbers}), but that we can come close to single photons with certain strictly localized states (\cref{c7_sec_eta12}). In the situations where strict localization is not dictated by, e.g., causality (\cref{c6_sec_on_demand_sources}), there is another possibility, namely to consider weakly localized photons (\cref{c5_sec_strictly_localized_pure_states}). 

For the conventional case in quantum optics, the bandwidth is much less than the central frequency. One can then derive Heisenberg uncertainty relations, or duration-bandwidth constraints [see, e.g., \cite{averchenko2020} and references therein], from the properties of the Fourier transform, similar to the case with classical signals. It is then quite common to discuss single photons with pulse envelopes given by, e.g., Gaussians or Hermite--Gaussians.

Lifting the assumption of small bandwidths compared to the central frequency, we noted already in \cref{c1_sec_invitation} that the normal-ordered electric energy density of a single photon $\ket{1_\xi}$ has the form
\begin{equation}\label{c7_qexv_1_xi_E_t}
    \frac{1}{2} \qexv{1_\xi}{\norder{E^2\lp(t)}} = \abs{E_{\xi}\lp(t)}^2,
\end{equation}
where
\begin{equation}\label{c7_E_xi_t}
    E_{\xi}\lp(t) = \minti{\dd \omega} \mathcal{E}\lp(\omega) \xi\lp(\omega) e^{- i \omega t}.
\end{equation}
Due to the lower limit 0 in the Fourier integral \eqref{c7_E_xi_t}, the Paley--Wiener criterion (\cref{c2_thm_paley_wiener}) dictates that the falloff with time must be slower than exponential. Thus gaussian pulse envelopes are not possible.

The restriction to positive frequencies in \cref{c7_E_xi_t} has consequences beyond the asymptotic falloff constraint described by the Paley--Wiener criterion. More specifically, we will discuss how the pulse width of the weakly localized photon is constrained from below. The treatment is illustrated by an example.

To get a short pulse, we need a wide spectrum (see \cref{c7_fig_weakly_localized_photon_spectrum}). We therefore consider an example where the photon spectrum in \cref{c7_E_xi_t} is a \enquote{raised cosine} (ignoring normalization):
\begin{equation}
    \mathcal{E}\lp(\omega)\xi\lp(\omega) = 
    \begin{cases}
        1-\cos\lp(\frac{\pi\omega}{\omega_0}), & 0\leq\omega \leq 2\omega_0, \\
        0, & \omega > 2\omega_0,
    \end{cases}
\end{equation}
leading to
\begin{equation}
    \Re E_{\xi}\lp(t) = \sin\lp(2\omega_0 t) \lp(\frac{1}{t} - \frac{1}{2t-\tau_0} - \frac{1}{2t+\tau_0}),
\end{equation}
where $\tau_0 = 2\pi/\omega_0$. To get a fair comparison, we let a classical pulse $\zeta\lp(t)$ have the same maximum frequency $2\omega_0$. However, classical pulses may contain negative frequencies as well, so in order to get a short pulse we use the entire band from $-2\omega_0$ to $2\omega_0$:
\begin{subequations}
\begin{align}
    \zeta\lp(\omega) &= 
    \begin{cases}
        1+\cos\lp(\frac{\pi\omega}{2\omega_0}), & \abs{\omega} \leq 2\omega_0 \\
        0, & \abs{\omega} > 2\omega_0,
    \end{cases} \\
    \zeta\lp(t) &= \sin\lp(2\omega_0 t) \lp(\frac{1}{t} - \frac{1}{2t-\tau_0/2} - \frac{1}{2t+\tau_0/2}).
\end{align}
\end{subequations}
Here $\zeta\lp(t)$ is the inverse Fourier transform of $\zeta\lp(\omega)$, however normalized such that the value for $t=0$ matches $\Re E_\xi\lp(0)$. The two spectra and pulses are plotted in \cref{c7_fig_weakly_localized_photon_spectrum,c7_fig_weakly_localized_photon_field}.

We are now ready to compare\footnote{One may wonder why we compare the classical field $\zeta\lp(t)$ with $\Re E_{\xi}\lp(t)$, given that only the absolute value is present in \cref{c7_qexv_1_xi_E_t}. It could be natural to consider the expectation value of the field itself; however this expectation value vanishes. We may therefore instead consider, e.g., the two-point correlation function, which can be evaluated to $\qexv{1_\xi}{\norder{E\lp(\tau)E\lp(t)}} = 2\Re E_{\xi}^*\lp(\tau)E_{\xi}\lp(t)$. Picking a fixed $\tau$ such that $E_{\xi}^*\lp(\tau)$ is real, we obtain a result proportional to $\Re E_{\xi}\lp(t)$.} the pulse widths of weakly localized photons and classical pulses, beginning with their absolute widths. For classical pulses, the widths are constrained by the standard Fourier transform property (or Heisenberg uncertainty), $\Delta\omega\Delta\tau \gtrsim 1/2$, where $\Delta\omega$ is the bandwidth and $\Delta \tau$ is the duration. 

\FigSevenThree

\FigSevenFour

For the weakly localized photon, the bandwidth is half the bandwidth of the classical pulse, even though the maximum frequency is the same. This is because the negative frequencies are absent. The result is a pulse width that is twice that of the classical pulse.

However, the comparison changes more substantially when we instead consider normalized pulse widths, i.e., pulse widths measured in number of cycles. Since the classical pulse may have its spectral content centered about any frequency, even $\omega=0$, there is no lower bound for the normalized pulse width. For the example in \cref{c7_fig_weakly_localized_photon_field}, the field $\zeta\lp(t)$ has no carrier oscillations. In contrast, since the photon spectrum contains only positive frequencies, its characteristic \enquote{carrier} frequency $\omega_0$, defined for instance by the center of its spectrum, must be positive. We therefore get a rough lower bound of order unity for the normalized width of the weakly localized photon,
\begin{equation}\label{c7_norm_pulse_width_bound}
    \omega_0 \Delta\tau \gtrsim 1,    
\end{equation}
where the precise numerical constant depends on the chosen definition of pulse width [see also \cite{brunetti2002}]. Therefore, a single-photon pulse with no carrier oscillations, such as the classical field in \cref{c7_fig_weakly_localized_photon_field}, is impossible.


%% file: 8_conclusion.tex

\iftoggle{ExportTikz}{%
     
}{}

\section{Conclusion}
\label{c8_sec_conclusion}
We conclude this review by answering some questions that have been treated in the previous sections.

\notocsubsubsection{Relativistic particles cannot be strictly localized because they do not have a suitable position operator/\allowbreak{}position eigenstate/\allowbreak{}wavefunction/\allowbreak{}spatial number operator/\allowbreak{}etc.?}
There could be many ways to define what it means to be localized. The reason why the listed concepts are useful to begin with is that in nonrelativistic quantum mechanics they correspond to some local measurement scheme. The question to ask is therefore whether these concepts still correspond to anything locally measurable in quantum field theory. Although they can sometimes be well-approximated by measurements, their usual nonrelativistic interpretation does not necessarily carry over to exact local measurements in quantum field theory. Therefore, localization in quantum field theory does not seem to depend fundamentally on the various properties of these concepts. This is discussed in \cref{c3_sec_what_does_it_mean_for_a_particle_to_be_localized}.

\notocsubsubsection{What is locally measurable?}
In quantum field theory it is common to postulate that the field itself, and superpositions of products of the field in a region $\mathcal{O}$, are locally measurable in $\mathcal{O}$. The precise statement is formulated in terms of (gauge-invariant) smeared-field observables in \emph{local operator algebras} (\cref{c4_sec_local_operators}). The physical realizability of local measurements in quantum field theory is still debated (\cref{c3_sec_physically_realizable_local_measurements}), so we take a pragmatic, conservative approach: When arguing that something \emph{does not} hold for a local measurement, we use a safe characterization with the criteria failing for some pointwise local field observable such as the energy density. When arguing that something holds for \emph{all} local measurements, we use a stringent characterization where the criteria must hold for all observables in the local operator algebra.

\notocsubsubsection{What does it mean for relativistic particles to be strictly localized?}
We find that the appropriate way to formalize strict localization in quantum field theory is through local measurement. This approach connects localization to measurable outcomes in spacetime and to relativistic causality. In line with \cite{knight1961}, a state is said to be strictly localized to some spacetime region $\mathcal{O}$ if it is indistinguishable from vacuum everywhere outside $\mathcal{O}$; i.e., it has the same expectation value as vacuum for all local measurements outside $\mathcal{O}$. We define this in \cref{c5_sec_localized_states}.

\notocsubsubsection{Photons that are maximally localized have an energy density that falls off asymptotically as \texorpdfstring{$\sim \bm{r^{-7}}$}{\~r-7} with radial distance \texorpdfstring{$\bm{r}$}{r}?}
This is false, as \cite{bialynicki-birula1998} demonstrates a single-photon state with a certain spherical symmetry whose expected energy density falls off asymptotically as $\sim \exp\lp(- \gamma \sqrt{r})$ for some $\gamma > 0$. The same analysis also shows that the expected energy density for single photons with this symmetry must fall off asymptotically slower than $\exp\lp(-\gamma r)$ for all $\gamma > 0$.

\notocsubsubsection{Which photonic states can be strictly localized?}
Single photons cannot be strictly localized, as their expected energy density cannot be confined to a bounded spatial region. This has been shown by \cite{bialynicki-birula2009}. The fact also follows from Knight's theorem (\cref{c5_thm_knight_inf_photon_numbers}), which can easily be established for scalar bosons and photons from the Reeh--Schlieder theorem (\cref{c5_sec_proof_knight}). Certain aspects of strict localization of coherent states have been studied in \cite{knight1961,bialynicki-birula1998}. We show in \cref{c7_sec_coherent_states} that coherent optical states can be strictly localized, since their expectation value may equal that for vacuum for all local observables outside a sharp region. This makes sense because coherent states behave as the classical solutions. In \cref{c7_sec_examples_of_localized_states} we go through several other examples of optical states that are strictly localized, including states that approximate single photons and (entangled) Bell states.

\notocsubsubsection{Photons cannot be strictly localized because they have zero mass?}
The mass of the particle does not directly matter, as for instance the proofs in \cref{c5_sec_proof_single_particles_not_localized} or \cref{a1_sec_proof_single_photons_cannot_be_localized} showing that single particles cannot be strictly localized work with and without mass. The main change is that for massive particles there is a regime where one might use nonrelativistic quantum mechanics with decent accuracy. It is true that in nonrelativistic quantum mechanics, there is less of a localization problem. However, applying nonrelativistic quantum mechanics will nevertheless always be an approximation.

\notocsubsubsection{Photons cannot be strictly localized because of the factor $1/\sqrt{2 \omega_{\vect{k}}}$ in the $A^\mu$-field?}
The fact that photons cannot be strictly localized comes from the positive-frequency dynamics. This was already evident in the 1D special case in \cref{c1_sec_invitation}. The positive-frequency dynamics is also seen as the key ingredient of the Reeh--Schlieder theorem on which Knight's theorem rests. The factor $1/\sqrt{2 \omega_{\vect{k}}}$ is not the reason for the lack of localizability (\cref{c5_sec_reason_photon_cannot_be_localized}). The proof in \cref{a1_sec_proof_single_photons_cannot_be_localized} works with any such factor; the reason is that the square root can be canceled by a corresponding square root in the mode spectrum.

\notocsubsubsection{Photonic states being strictly localized is not compatible with relativistic causality due to Hegerfeldt's/Malament's theorem?}
This is false. The way to reconcile strict localization with causality in line with Hegerfeldt's and Malament's theorems is argued in detail in \cref{c2_sec_hegerfeldts_theorem,c2_sec_interpreting_measurement_probabilities_in_quantum_field_theory,c4_sec_connection_to_malaments_theorem}.

\notocsubsubsection{For a notion of localization to make sense, states localized to spacelike separated regions must necessarily be (close to) orthogonal?}
This is false, as discussed in \cref{c5_sec_diff_nonrel_qm}. For example, two weak coherent states describing pulses in disjoint regions have a large overlap due to their common vacuum component.

\notocsubsubsection{What are the implications of the Reeh--Schlieder theorem for local measurement and the state space?}
The Reeh--Schlieder theorem (\cref{c4_sec_the_reeh_schlieder_theorem}) demonstrates that any state in the Fock space can be approximated, to any precision, when a local operator acts on the vacuum. Regarding the operator as a Kraus operator, we may thus produce, with nonzero probability, a photon wavepacket concentrated near the Moon instantaneously by some local operation on Earth. This does not violate causality: the remotely changed state is conditional on a particular local outcome (which occurs with extremely low probability), while the complete, nonselective local operation cannot change measurement statistics in a spacelike separated region. The Reeh--Schlieder theorem has some important implications, including the fact that the probability of any possible local measurement outcome in vacuum cannot vanish (\cref{c4_sec_probabilities_of_local_measurements_do_not_vanish}). Furthermore, the Fock space does not factor into spaces associated with complementary regions of spacetime (\cref{c5_sec_hilbert_space_does_not_factor}), making the notion of strict localization nontrivial.

\notocsubsubsection{Is there a Heisenberg uncertainty relation for photons?}
As in nonrelativistic quantum mechanics, the usual Fourier transform uncertainty relation applies in quantum field theory. For photons, it constrains the bandwidth $\cdot$ pulse length product. However, the absence of negative frequencies imposes an additional constraint, which becomes important when the bandwidth is comparable to the central (carrier) frequency. This has the effect that the normalized pulse width (pulse width measured in number of cycles) must roughly be larger than $\sim 1$ (\cref{c7_sec_weakly_localized_photons}). This is not the case for classical signals, which can have any normalized pulse width.

\notocsubsubsection{Are modes associated with bounded regions of spacetime related to strictly localized states?}
One may introduce a complete set of modes, where one mode is supported in some bounded region $\mathcal{O}$. Let $\ket{0^b}$ be the vacuum state with respect to these modes, while $\ket{0}$ is the usual vacuum state (associated with global modes). A single excitation of the localized mode will be a state that is locally equivalent to $\ket{0^b}$ outside $\mathcal{O}$. It is however not locally equivalent to $\ket{0}$; thus localization of the mode does not by itself imply strict localization of the state. This is demonstrated in \cref{c7_sec_localized_modes}.


%% file: 9_appendix.tex

\iftoggle{ExportTikz}{%
     
}{}

\section{Independent proof that single photons cannot be strictly localized}
\label{a1_sec_proof_single_photons_cannot_be_localized}
In the main text we stated the result that single photons cannot be strictly localized (\cref{c5_thm_knight_inf_photon_numbers}), which was proved straightforwardly from the Reeh--Schlieder theorem in \cref{c5_sec_proof_single_particles_not_localized}. Here we give a self-contained proof, from first principles, of the lack of localizability of single photons, based directly on the electromagnetic energy density. There are a few assumptions made in this proof, namely that the photon is physical (two spatial, transverse polarization directions) and that it has finite energy.

The method of the proof is to show that the expectation value of the normal-ordered electromagnetic energy density, which is a local observable, cannot vanish in an appropriate region. To start, we show in \cref{c5_sec_decoupling_polarization_degrees_of_freedom} that the polarization degrees of freedom for physical photons decouple, and it is sufficient to consider one helicity component at a time. This argument is based on the analysis in \cite{bialynicki-birula1998}. In \cref{c5_sec_what_localization_regions_are_impossible} we then discuss what types of localization regions we can consider, e.g., bounded regions, half-infinite regions, etc. We then conclude the proof in \cref{c5_sec_proof_that_eq_c5_Phi_norm_x_0_is_impossible} by showing that the spatial localization region of a single photon cannot be bounded, or even a half-space. 

The proof is similar to the analysis in \cite{knight1961}, which shows that single particles of a hermitian scalar field cannot be strictly localized. After reducing to one polarization direction, the proof uses exactly the same key ingredients as Knight: positive energy and Fourier analyticity. Another proof showing that single photons cannot be strictly localized can be found in \cite{bialynicki-birula2009}, generalizing the earlier analysis of uniform localization in \cite{bialynicki-birula1998} to arbitrary physical photons. This proof is based on a different technique than Knight's and our proof, instead showing that two complementary observables, such as the electric and magnetic energy densities, cannot simultaneously vanish.

\subsection{Decoupling polarization degrees of freedom}
\label{c5_sec_decoupling_polarization_degrees_of_freedom}
We consider physical photons, i.e., photons consisting of up to two polarization directions $s$. Any physical single-photon state is of the form
\begin{equation}\label{c5_1_photon}
    \ket{1} = \sum_{s = 1}^2 \mint{\dd[3] k} \xi_s\lp(\vect{k}) \adj{a}_{\vect{k}s}  \ket{0},
\end{equation}
where $\xi_s\lp(\vect{k})$ are two arbitrary functions in $L^2\sephn\lp(\reals^3)$ and $\adj{a}_{\vect{k}s}$ is a single-mode creation operator satisfying $\comm[\big]{\padj{a}_{\vect{k}s}}{\adj{a}_{\vect{k}'\!s'}} = \delta\lp(\vect{k} - \vect{k}') \delta_{ss'}$. In Coulomb gauge, the 3D vector potential is
\begin{equation}\label{c5_A_x}
    \vect{A}\lp(x) = \sum_{s = 1}^2 \mintf{\frac{\dd[3] k}{\lp(2 \pi)^{3/2}}} \frac{1}{\sqrt{2 \omega_{\vect{k}}}} \lp(\vect{e}_{\vect{k}s} a_{\vect{k}s} e^{- i k \cdot x} + \vect{e}_{\vect{k}s}^{\vphantom{\dagger}*} \adj{a}_{\vect{k}s} e^{i k \cdot x}).
\end{equation}
Here $k^0 = \omega_{\vect{k}} = \abs{\vect{k}}$ and $\vect{e}_{\vect{k}s}$ are (potentially complex) transverse polarization vectors satisfying
\begin{equation}\label{c5_polarization_vectors}
    \vect{k} \cdot \vect{e}_{\vect{k}s} = 0, \sepc \vect{e}_{\vect{k}s}^* \cdot \vect{e}_{\vect{k}s'}^{} = \delta_{ss'}, \sepc s, s' = 1, 2.
\end{equation}

One convenient choice of polarization vectors is right and left circular ones. If $\vect{e}_{\vect{k}1}$ and $\vect{e}_{\vect{k}2}$ are real, linear polarization vectors, we can construct complex, circular ones as
\begin{equation}
    \vect{e}_{\vect{k}\pm} = \frac{1}{\sqrt{2}} \lp(\vect{e}_{\vect{k}1} \pm i \vect{e}_{\vect{k}2}),
\end{equation}
which still satisfy \cref{c5_polarization_vectors}. With this choice of polarization vectors, it is convenient to introduce the Riemann--Silberstein field
\begin{equation}
    \vect{F}\lp(x) = \vect{E}\lp(x) + i \vect{B}\lp(x),
\end{equation}
based on the electric $\vect{E}\lp(x) = - \partial_t \vect{A}\lp(x)$ and magnetic $\vect{B}\lp(x) = \nabla \cross \vect{A}\lp(x)$ fields. Using \cref{c5_A_x} we can show that $\vect{F}\lp(x)$ takes the form
\begin{equation}\label{c5_R_x}
    \vect{F}\lp(x) = \mintf{\frac{\dd[3] k}{\lp(2 \pi)^{3/2}}} \sqrt{2 \omega_{\vect{k}}} \seph \vect{e}_{\vect{k}+} \lp(a_{\vect{k}+} e^{- i k \cdot x} + \adj{a}_{\vect{k}-} e^{i k \cdot x}).
\end{equation}
Here we have used that we may absorb phase factors into the ladder operators, e.g., $i a_{\vect{k}s} \mapsto a_{\vect{k}s}$, $- i \adj{a}_{\vect{k}s} \mapsto \adj{a}_{\vect{k}s}$.

Note that there is no sum over polarization directions $\sum_s$ in \cref{c5_R_x} anymore. This is possible because $\vect{F}\lp(x)$ is now a complex field, so that $\vect{F}\lp(x)$ and $\smash{\adj{\vect{F}}\sephn\lp(x)}$ without sums have the same number of degrees of freedom as the two real fields $\vect{E}\lp(x)$ and $\vect{B}\lp(x)$ with sums. Cleverly, this new complex field is defined in such a way that $\vect{F}\lp(x)$ represents purely right-handed light and $\smash{\adj{\vect{F}}\sephn\lp(x)}$ represents purely left-handed light. 

We want to show that the state $\ket{1}$ cannot be strictly localized, which can be done by showing that for all choices of $\xi_{+}\lp(\vect{k})$ and $\xi_{-}\lp(\vect{k})$, the expectation value of some local observable cannot vanish in any appropriate region. One local observable in electromagnetism is the normal-ordered energy density, which is proportional to
\begin{equation}
    \norder{\vect{E}^2\sephn\lp(x)} + \norder{\vect{B}^2\sephn\lp(x)} = \norder{\adj{\vect{F}}\sephn\lp(x) \cdot \vect{F}\lp(x)}.
\end{equation}
Calculating the expectation value of this observable for the state $\ket{1}$ in \cref{c5_1_photon} gives
\begin{equation}\label{c5_qexv_1_Rd_R}
    \qexv{1}{\norder{\adj{\vect{F}}\sephn\lp(x) \cdot \vect{F}\lp(x)}} = \abs{\vect{\Phi}_+\lp(x)}^2 + \abs{\vect{\Phi}_-\lp(x)}^2,
\end{equation}
where
\begin{equation}\label{c5_Phi_pm_x}
    \vect{\Phi}_{\pm}\lp(x) = \mintf{\frac{\dd[3] k}{\lp(2 \pi)^{3/2}}} \sqrt{2 \omega_{\vect{k}}} \seph \xi_{\pm}\lp(\vect{k}) \vect{e}_{\vect{k}\pm} e^{- i k \cdot x}.
\end{equation}
Since the energy density is a sum of two positive quantities, it can only vanish if both $\vect{\Phi}_{+}\lp(x)$ and $\vect{\Phi}_{-}\lp(x)$ vanish separately. Since $\vect{\Phi}_{+}\lp(x)$ only depends on $\xi_{+}\lp(\vect{k})$, and $\vect{\Phi}_{-}\lp(x)$ only depends on $\xi_{-}\lp(\vect{k})$, we can solve for $\xi_{+}\lp(\vect{k})$ and $\xi_{-}\lp(\vect{k})$ separately. 

Both cases are covered by answering whether there can exist a function $\xi\lp(\vect{k})$ such that
\begin{equation}\label{c5_Phi_x_0_G}
    \vect{\Phi}\lp(x) = 0 \alme, \sepc \forall x \in \compl{\mathcal{O}},
\end{equation}
for some appropriate region $\mathcal{O}$, for
\begin{subequations}\label{c5_Phi_x}
\begin{align}
    &\vect{\Phi}\lp(x) = \mintf{\frac{\dd[3] k}{\lp(2 \pi)^{3/2}}} \vect{A}\lp(\vect{k}) e^{- i \omega_{\vect{k}} x_0 + i \vect{k} \cdot \vect{x}}, \\
    &\vect{A}\lp(\vect{k}) = \mathcal{B}\lp(\omega_{\vect{k}}) \xi\lp(\vect{k}) \vect{e}_{\vect{k}}.
\end{align}
\end{subequations}
To be as general as possible, we here let $\vect{e}_{\vect{k}}$ be a transverse polarization vector and $\mathcal{B}\lp(\omega_{\vect{k}})$ be an arbitrary function of the frequency $\omega_{\vect{k}}$. For example a Klein--Gordon field has $\mathcal{B}\lp(\omega_{\vect{k}}) = 1 / \sqrt{2 \omega_{\vect{k}}}$, whereas $\vect{F}\lp(x)$ in \cref{c5_R_x} has $\mathcal{B}\lp(\omega_{\vect{k}}) = \sqrt{2 \omega_{\vect{k}}}$. In addition, we also let the field potentially be massive, by letting
\begin{equation}\label{c5_k0_omega_m}
    \omega_{\vect{k}} = \sqrt{\vect{k}^2 + m^2},
\end{equation}
for some $m \geq 0$ which may be zero. 

Note that the localization of a single particle of a hermitian scalar field $\phi\lp(x)$ can similarly be demonstrated to be impossible by considering \cref{c5_Phi_x_0_G}, only without the polarization vector $\vect{e}_{\vect{k}}$ in \cref{c5_Phi_x}. This is also what \cite{knight1961} considers. The remaining part of the proof is unaffected by the inclusion of this polarization vector. Thus, the proof is valid for (massless) physical photons as well as massive or massless, hermitian scalar bosons. 

The function $\xi\lp(\vect{k})$ must be in $L^2\sephn\lp(\reals^3)$ since $\ket{1}$ must be normalized. We make one assumption here, namely that the function $\mathcal{B}\lp(\omega_{\vect{k}}) \xi\lp(\vect{k})$ is in $L^2\sephn\lp(\reals^3)$ as well. This may not be true for certain special states $\ket{1}$; however, in the case of a photon such states must have infinite energy, as follows from \cref{c5_qexv_1_Rd_R,c5_Phi_pm_x}.

\subsection{Localization regions}
\label{c5_sec_what_localization_regions_are_impossible}
One question is what localization regions $\mathcal{O}$ we want to show are impossible for $\ket{1}$. For example we cannot rule out $\ket{1}$ being strictly localized to all of spacetime, although this is not what we usually mean by $\ket{1}$ being strictly localized. Here we ask whether there can exist some fixed region of space $V$, so that $\ket{1}$ is strictly localized to $V$ at two separate times, say at both $t = 0$ and $t = T > 0$. Note that this does not mean we assume that the spatial localization region of $\ket{1}$ stays constant with time. Rather, we assume that we can find some larger-than-necessary region $V$ that contains the localization regions at both times. For example if $\ket{1}$ is strictly localized to a region $V_0$ at $t = 0$ and to $V_1$ at $t = T$, we could take $V = V_0 \cup V_1$. 

The question is then what spatial regions $V$ we should consider. In the simplest case, we could assume that $V$ is bounded. Showing that no such $V$ can exist would imply that single photons must have a (detectable) tail to infinity in some direction during any time interval. This is what we would usually mean by a single photon not being strictly localized. However, we could also consider more general, infinite regions $V$. For example showing that $V$ cannot even be a half-infinite region would imply that single photons must have tails to infinity in several directions. In the most general case, we could consider regions $V$ that have support (almost) everywhere except some open, bounded region. This is the type of localization region shown to be impossible by the proof in the main text (\cref{c5_thm_knight_inf_photon_numbers}). 

Here we will not be quite as general, but we consider $V$ to be a half-space. Showing that $\vect{\Phi}\lp(x)$ cannot have support purely within such a region $V$ would imply that $\ket{1}$ cannot be strictly localized to any bounded or half-space region during any time interval. Thus assume for contradiction that $\vect{\Phi}\lp(x)$ can vanish outside some fixed half-space $V$ at both $t = 0$ and $t = T$. This means in particular that it is possible to choose coordinate axes so that, for example
\begin{equation}\label{c5_Phi_x_0}
    \vect{\Phi}\lp(x) = 0 \alme, \sepc \forall x_1, \; \forall x_2, \; \forall x_3 > 0, \; x_0 = 0, T.
\end{equation}
This means that the norm of $\vect{\Phi}\lp(x)$ must vanish in this region:
\begin{equation}\label{c5_Phi_norm_x_0}
    \int_{-\infty}^\infty \sepin \dd x_1 \int_{-\infty}^\infty \sepin \dd x_2 \int_0^\infty \sepii \dd x_3 \sepib \abs{\vect{\Phi}\lp(x)}^2 = 0, \sepc x_0 = 0, T.
\end{equation}

\subsection{Proof that Eq.~(\ref*{c5_Phi_norm_x_0}) is impossible}
\label{c5_sec_proof_that_eq_c5_Phi_norm_x_0_is_impossible}
From \cref{c5_Phi_x}, we see that $\vect{\Phi}\lp(x)$ is a 3D inverse Fourier transform $\vect{k} \mapsto \vect{x}$ of $\vect{A}\lp(\vect{k}) e^{- i \omega_{\vect{k}} x_0}$. We can expand this triple Fourier integral into each $k$ integral separately, and isolate the $k_3$ integral:
\begin{align}\label{c5_Phi_B}
    \vect{\Phi}\lp(x) &= \int \sepia \frac{\dd k_1}{\sqrt{2 \pi}} \int \sepia \frac{\dd k_2}{\sqrt{2 \pi}} \int \sepia \frac{\dd k_3}{\sqrt{2 \pi}} \sepib \vect{A}\lp(\vect{k}) e^{i k_1 x_1 + i k_2 x_2 + i k_3 x_3} \nonumber \\[-1mm]
    & \hspace{40.8mm} \cdot e^{- i \sqrt{\vect{k}^2 + m^2} \seph x_0} \\
    &= \int \sepia \frac{\dd k_1}{\sqrt{2 \pi}} \int \sepia \frac{\dd k_2}{\sqrt{2 \pi}} \sepib \vect{B}\lp(k_1, k_2; x_3, x_0) e^{i k_1 x_1} e^{i k_2 x_2}, \nonumber 
\end{align}
where
\begin{equation}\label{c5_B_k1_x2_t}
    \vect{B}\lp(k_1, k_2; x_3, x_0) = \int \sepia \frac{\dd k_3}{\sqrt{2 \pi}} \sepib \vect{A}\lp(\vect{k}) e^{i k_3 x_3 - i \sqrt{\vect{k}^2 + m^2}\seph x_0}.
\end{equation}

Since the Fourier transform is unitary, the fact that the norm of $\vect{\Phi}\lp(x)$ vanishes in $x_1$ and $x_2$ according to \cref{c5_Phi_norm_x_0} implies that the norm of $\vect{B}\lp(k_1, k_2; x_3, x_0)$ in $k_1$ and $k_2$ must vanish:
\begin{equation}
    \int_{-\infty}^\infty \sepin \dd k_1 \int_{-\infty}^\infty \sepin \dd k_2 \int_0^\infty \sepii \dd x_3 \sepib \abs{\vect{B}\lp(k_1, k_2; x_3, x_0)}^2 = 0, \sepc x_0 = 0, T.
\end{equation}
This in turn means that 
\begin{equation}
\begin{aligned}
    &\vect{B}\lp(k_1, k_2; x_3, x_0) = 0 \alme, \\[1mm]
    &\forall k_1, \; \forall k_2, \; \forall x_3 > 0, \; x_0 = 0, T.
\end{aligned}
\end{equation}

In other words, for $x_0 = 0$ and $x_0 = T$, there are (several values for) $k_1$ and $k_2$ such that $\vect{B}\lp(k_1, k_2; x_3, x_0)$ vanishes for (almost) all $x_3 > 0$. Thus by Titchmarsh's theorem \cite[Theorem~95]{titchmarsh1948}, the Fourier transform of $\vect{B}\lp(k_1, k_2; x_3, x_0)$ from $x_3 \mapsto k_3$ must be analytic in the upper half-plane of complex $k_3$. I.e.,
\begin{equation}\label{c5_A_k_analytic}
\begin{aligned}
    \vect{A}\lp(\vect{k}) e^{- i \sqrt{\vect{k}^2 + m^2}\seph x_0} \sepc &\tn{is analytic for } \Im\lp(k_3) > 0, \\
    & x_0 = 0, T,
\end{aligned}
\end{equation}
in the sense that it must hold for each vector component. For $x_0 = 0$, this means that $\vect{A}\lp(\vect{k})$ must be analytic for $\Im\lp(k_3) > 0$. For $x_0 = T$, it therefore means that $\rule{0pt}{15pt}\smash{e^{- i \sqrt{\vect{k}^2 + m^2}\seph T}}$ must be analytic for $\Im\lp(k_3) > 0$, except possibly at isolated zeros of $\vect{A}\lp(\vect{k})$. However this is impossible, since the complex square root in the exponent has branch points at
\begin{equation}
    k_3 = \pm i \sqrt{k_1^2 + k_2^2 + m^2}.
\end{equation}
Even if $m = 0$, the analyticity condition \eqref{c5_A_k_analytic} must hold for nonzero $k_1$ and $k_2$, meaning that the branch points are unavoidable. In conclusion we have shown that:

\begin{remark}\label{c5_rem_single_photons_cannot_be_localized}\textup{(Single photons cannot be strictly localized).}
    Physical single photons with finite energy cannot be strictly localized to any half-space (or bounded) spatial region during any time interval.
\end{remark}

Since the energy density \eqref{c5_qexv_1_Rd_R} is positive for any pure single-photon state, it follows that a mixed single-photon state cannot be strictly localized either.

\subsection{Fundamental reason photons cannot be strictly localized}
\label{c5_sec_reason_photon_cannot_be_localized}
From the proof we find that the fundamental reason that the photon cannot be strictly localized is the form of the dispersion relation \eqref{c5_k0_omega_m}, leading to a time-evolution factor that is nonanalytic in complex wavevector. The proof is particularly transparent in the 1D case in \cref{c1_sec_invitation}, where it is shown that the energy density is proportional to the squared modulus of a Fourier integral involving only positive frequencies. The Paley--Wiener criterion (\cref{c2_thm_paley_wiener}) then asserts that the photon must have infinite tails, which fall off more slowly than exponential.

A common misunderstanding is that the fundamental reason for the lack of localizability is the factor $1/\sqrt{2 \omega_{\vect{k}}}$ in the field expression \eqref{c5_A_x}, which after an inverse Fourier transform is nonlocal in space. However, in the proof we called this factor $\mathcal{B}\lp(\omega_{\vect{k}})$ and the detailed form of this function did not matter. The reason why this factor does not necessarily lead to nonlocality is that the square root can be canceled by a corresponding square root in the mode functions $\xi_\pm(\vect k)$.
